\documentclass[10pt,aps,prd,amsmath,amssymb,superscriptaddress,twocolumn,nofootinbib,showpacs,preprintnumbers,amsfonts, notitlepage]{revtex4-2}

\usepackage{hyperref,color,subfigure,cleveref,subfiles}
\usepackage{amsmath,amssymb,physics,mathtools}
\usepackage{lmodern,bm,upgreek}
\usepackage{graphicx,dcolumn,xcolor,xfrac,slashed}
\usepackage{multirow,xspace,tabularx}
\usepackage[shortlabels]{enumitem}
\newcolumntype{C}[1]{>{\centering\arraybackslash}p{#1}}

\hypersetup{
 colorlinks=true,
 allcolors=[RGB]{31 119 180}
}

\newcommand{\diff}{\textrm{d}}
\newcommand{\pom}{{I\!\!P}}
\newcommand{\m}{m}
\newcommand{\eV}{\ensuremath{{\mathrm{\,e\kern -0.1em V}}}\xspace}
\newcommand{\kev}{\ensuremath{{\mathrm{\,ke\kern -0.1em V}}}\xspace}
\newcommand{\mev}{\ensuremath{{\mathrm{\,Me\kern -0.1em V}}}\xspace}
\newcommand{\gev}{\ensuremath{{\mathrm{\,Ge\kern -0.1em V}}}\xspace}
\newcommand{\gevsq}{\ensuremath{{\mathrm{\,Ge\kern -0.1em V}^2}}\xspace}
\newcommand{\tev}{\ensuremath{{\mathrm{\,Te\kern -0.1em V}}}\xspace}
\newcommand{\nsgev}{\ensuremath{{\mathrm{Ge\kern -0.1em V}}}\xspace}
\newcommand{\nsmev}{\ensuremath{{\mathrm{Me\kern -0.1em V}}}\xspace}
\newcommand{\nskev}{\ensuremath{{\mathrm{ke\kern -0.1em V}}}\xspace}

\newcommand{\ceem}{Center for  Exploration  of  Energy  and  Matter,
Indiana  University,
Bloomington,  IN  47403,  USA}
\newcommand{\cern}{CERN, 1211 Geneva 23, Switzerland}
\newcommand{\icn}{Instituto de Ciencias Nucleares, 
Universidad Nacional Aut\'onoma de M\'exico, Ciudad de M\'exico 04510, Mexico}
\newcommand{\icsup}{Pedagogical University of Krakow, 30-084 Krak\'ow, Poland}
\newcommand{\indiana}{Department of Physics,
Indiana  University, Bloomington,  IN  47405,  USA}
\newcommand{\jlab}{Theory Center, Thomas  Jefferson  National  Accelerator  Facility, Newport  News,  VA  23606,  USA}
\newcommand{\rome}{INFN Sezione di Roma, Roma, I-00185, Italy}
\newcommand{\ub}{Departament de F\'isica Qu\`antica i Astrof\'isica and Institut de Ci\`encies del Cosmos, Universitat de Barcelona, Mart\'i i Franqu\`es 1, E08028, Spain}
\newcommand{\ucm}{Departamento de F\'isica Te\'orica, Universidad Complutense de Madrid and IPARCOS, 28040 Madrid, Spain}

\begin{document}

\title{\boldmath $\pi^-p\to\eta^{(\prime)}\, \pi^- p$ in the double-Regge region}
\preprint{JLAB-THY-21-3354}
\author{Ł.~Bibrzycki}
    \email{lukasz.bibrzycki@up.krakow.pl}
    \affiliation{\icsup}
    \affiliation{\indiana}
    \affiliation{\jlab}
\author{C.~Fern\'andez-Ram\'irez}
    \email{cesar.fernandez@nucleares.unam.mx}
    \affiliation{\icn}
\author{V.~Mathieu}
    \affiliation{\ub}
    \affiliation{\ucm}
\author{M.~Mikhasenko}
    \affiliation{\cern}
\author{M.~Albaladejo}
    \affiliation{\jlab}
\author{A.~N.~Hiller Blin}
    \affiliation{\jlab}
\author{A.~Pilloni}
    \affiliation{\rome}    
\author{A.~P.~Szczepaniak}
    \affiliation{\indiana}
    \affiliation{\jlab}
    \affiliation{\ceem}
\collaboration{Joint Physics Analysis Center}
\date{\today}

\begin{abstract}
The production of $\eta^{(\prime)}\pi$ pairs 
constitutes one of the golden channels 
to search for hybrid exotics,
with explicit gluonic degrees of freedom. 
Understanding
the dynamics and backgrounds associated to $\eta^{(\prime)}\pi$ 
production above the resonance region is required to impose additional constraints to the resonance extraction.
We consider the reaction $\pi^-p\to \eta^{(\prime)} \pi^- \,p$ measured by COMPASS.
We show that the data in $2.4 < m_{\eta^{(\prime)}\pi} < 3.0\gev$ can be described by 
amplitudes based on double-Regge exchanges.
The angular distribution of the meson pairs, in particular in the $\eta' \pi$ channel, can be attributed to flavor singlet exchanges, suggesting the
presence of a large gluon content that couples strongly to the produced mesons.
\end{abstract}

\maketitle

\section{Introduction}
Since the early days of the quark model, hadron spectroscopy has remained central to our understanding of QCD. High precision data on various reactions that have recently been collected from experiments at CERN, JLab, $B$- and charm factories have produced tantalizing evidence for the existence of exotic states that do not naturally fit within the quark model classification~\cite{Shepherd:2016dni,Klempt:2007cp,Guo:2017jvc}, {\it e.g.} pentaquark and tetraquark candidates~\cite{Esposito:2016noz,Olsen:2017bmm,Karliner:2017qhf}. The quantum numbers of some exotics are manifestly incompatible with a simple $q\bar q$ assignment. For example, states with $J^{PC}=1^{-+}$ have long been speculated to be hybrids, 
{\it i.e.} mesons where gluons play the role of constituents~\cite{Dudek:2011bn,Meyer:2015eta}.
The paucity of data and the need for a thorough partial wave analysis to disentangle resonance from nonresonant background can be a challenging endeavor. The COMPASS collaboration extracted the $\eta\pi$ and $\eta'\pi$ partial waves as a function of the invariant mass, $m_{\eta^{(\prime)}\pi} < 3.0\gev$ from the measurement of diffractive pion dissociation on a nucleon target at $191\gev$~\cite{Adolph:2014rpp}.
These odd waves carry exotic quantum numbers, $J^{PC} = 1^{-+}$, $3^{-+}$,... 
The key observations are that even waves are similar in both reactions, while the $P$-wave is significantly larger in $\eta'\pi$. This 
reflects in a larger forward-backward asymmetry of the $\eta'\pi$. 
Both channels present peaking structures in the $P$-waves at seemingly different masses.
For a a long time, the two structures were interpreted as two different states, the lighter one coupling mainly to $\eta\pi$ and the heavier one to $\eta'\pi$. However, the coupled-channel analysis in~\cite{Rodas:2018owy} showed that the data is consistent with the existence of a single exotic resonance. These conclusions have been confirmed by a recent independent analysis~\cite{Kopf:2020yoa}
and are supported by the latest lattice QCD computations~\cite{Woss:2020ayi}.

At higher invariant masses, the reaction is expected to be dominated by cross-channel Regge exchanges, 
which is consistent with the cross section peaking in the forward and backward directions, with the peaks shrinking with increasing $\eta^{(\prime)}\pi$ mass {\it cf.} Fig.~2 of Ref.~\cite{Adolph:2014rpp}. 
Since a forward-backward asymmetry arises from the interference between even and odd waves, the larger exotic $P$-wave in $\eta'\pi$ is consistent with the observed larger asymmetry. This connection between resonances and Regge exchanges can be formalized {\it via} dispersion relations, {\it e.g.} in the form of finite energy sum rules~\cite{Dolen:1967jr,Dolen:1967zz,Brower:1974yv}. 
Such relations can be used to constrain fits in the resonance region which, in combination with forthcoming high precision data from
GlueX~\cite{Austregesilo:2018mno,Gleason:2020mtb} and 
COMPASS~\cite{Ketzer:2019wmd},
 could lead to a more accurate determination of the exotic meson resonance parameters. A necessary step in this procedure is to fit the high mass region with analytical amplitudes that respect Regge asymptotic behavior. This is the main purpose of this work. 

The paper is organized as follows. In Section~\ref{sec:COMPASS}
we describe the COMPASS partial waves, the procedure to compute the
intensity distribution from them, and the main features of said
distribution. Section~\ref{sec:model}
describes the double-Regge model used to fit the
data. 
In Section~\ref{sec:truncation} we discuss
the consequences of truncating the partial wave expansion 
in the analysis
of the data and how it impacts the comparison to a given
model and the extraction of the dynamics.
In Section~\ref{sec:set_diag} we discuss what are the relevant contributions to the amplitudes, needed to reproduce the 
features of the angular and mass dependencies. 
Section~\ref{sec:results} describes our fitting strategy,
fit results, and comparison to data.
Section~\ref{sec:cPW} provides the connection between
the COMPASS partial waves and
the partial waves obtained from the double-Regge model.
Finally, in Section~\ref{sec:conclusion} we summarize our results. 
The kinematical description of the $\eta^{(\prime)}\pi$ reactions,
statistical analysis,
error propagation from the COMPASS partial waves to
the intensity distribution,
and other details and complementary information 
are left to the Appendices.

\section{COMPASS Intensities}\label{sec:COMPASS}

\begin{figure*}
    \includegraphics[width=0.85\textwidth]{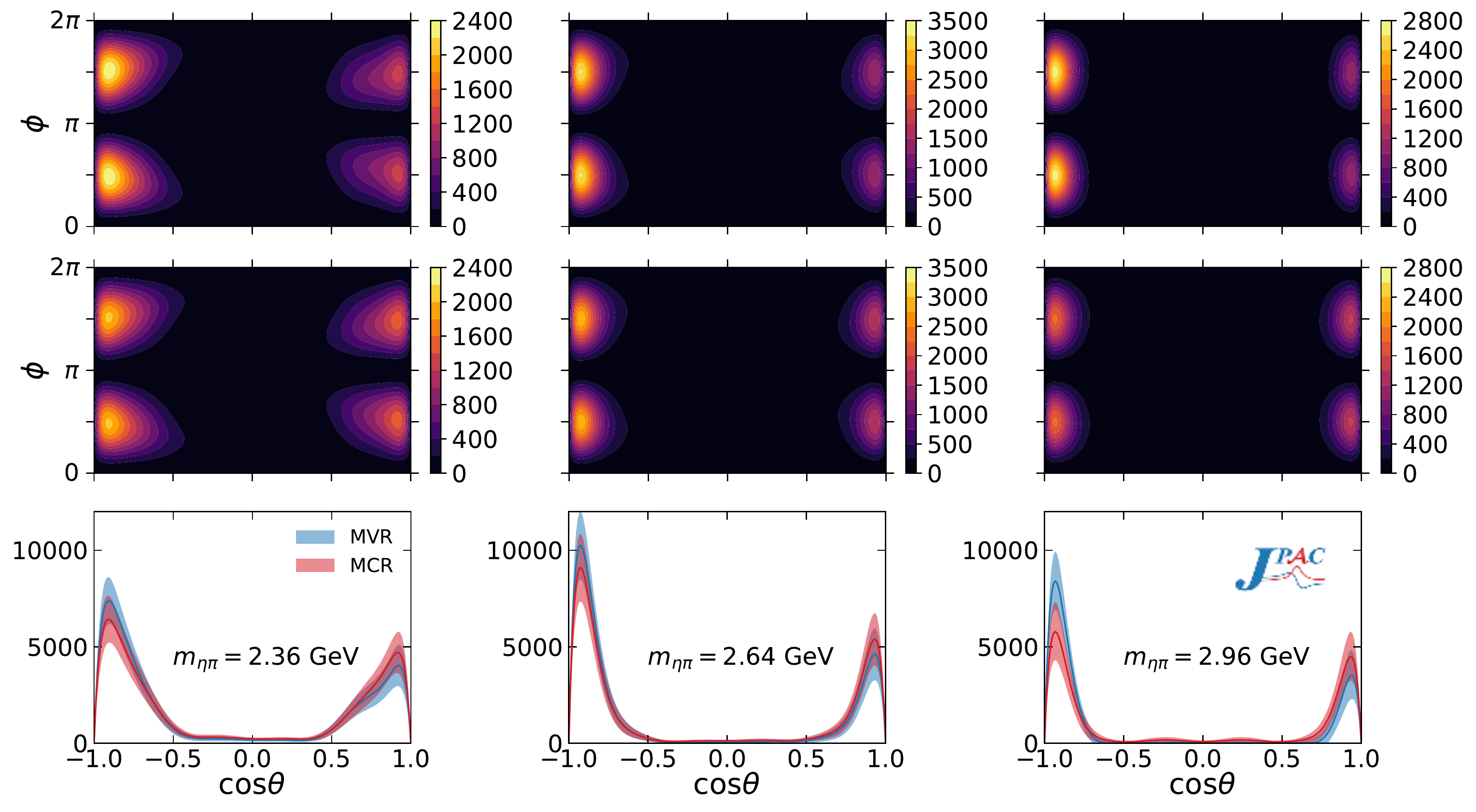} 
      \caption{$\cos \theta$ vs. $\phi$ expected value intensity $I(m_{\eta \pi},\Omega)$ density plot for MVR (upper) and MCR (center) of the COMPASS partial waves of the $\eta \pi$ channel for three fixed energies ($m_{\eta \pi}=2.36$, $2.64$, and $2.96\gev$). The lower row provides the $\phi$-integrated $I_\theta(m_{\eta \pi},\cos \theta)$ MVR (blue) and MCR (red) $1\sigma$ bands. Since the MVR does not propagate uncertainties, we show as MVR error bands the same computed from the MCR.}
      \label{fig:pieta_phi_theta}
\end{figure*}

\begin{figure*}
    \includegraphics[width=0.85\textwidth]{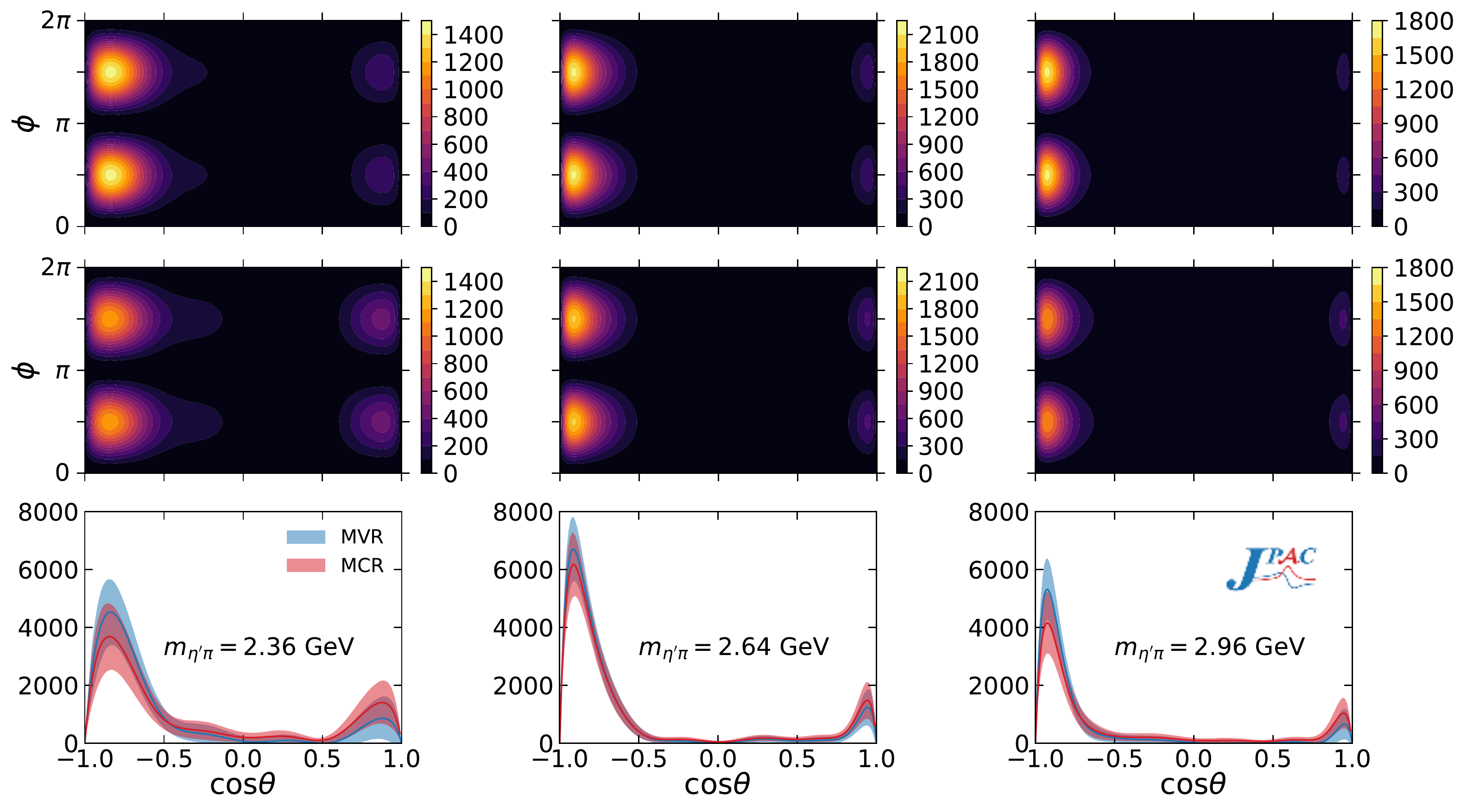} 
      \caption{Same as Fig.~\ref{fig:pieta_phi_theta} for $\eta' \pi$. We note that the $\eta' \pi$ backward peak is broader than the $\eta \pi$ one and that the forward-backward asymmetry is more pronounced.}\label{fig:pietap_phi_theta}
\end{figure*}
   
In this section we describe the data on reactions 
\begin{equation}
     \pi^-(q)+p(p_1)\to\eta^{(\prime)}(k_\eta)+\pi^-(k_\pi) +p(p_2)\, ,
    \label{reaction1}
\end{equation}
analyzed by COMPASS~\cite{Adolph:2014rpp}. 
The unpolarized cross sections for both reactions depend on five kinematical variables. These are, 
 for example, the total center of mass energy squared  $s=(q+p_1)^2$, the invariant mass of the  produced meson pair  
 $\m ^2 = m^2_{\eta^{(\prime)}\pi} = (k_\pi+k_\eta)^2$,
 the square of the momentum transfer between the target and the recoil  nucleon $t_p = (p_1-p_2)^2$, and the spherical angle $\Omega$ determining the direction of the  relative momentum between the two mesons in the rest frame of the pair. 
The COMPASS experiment operated with a fixed beam momentum of $191\gev$; in the analysis of~\cite{Adolph:2014rpp} $t_p$ was integrated  in the region
$t_p \in [-1.0,-0.1]\gevsq$. 
Furthermore, since there was no measurement of the initial flux, the normalization of the event distribution is unknown.
In the partial wave analysis of~\cite{Adolph:2014rpp}  the angular dependence of the  event distribution, {\it aka} intensity function, in bins of $\m$, was expanded in terms of angular functions,
\begin{figure}
    \includegraphics[width=\columnwidth]{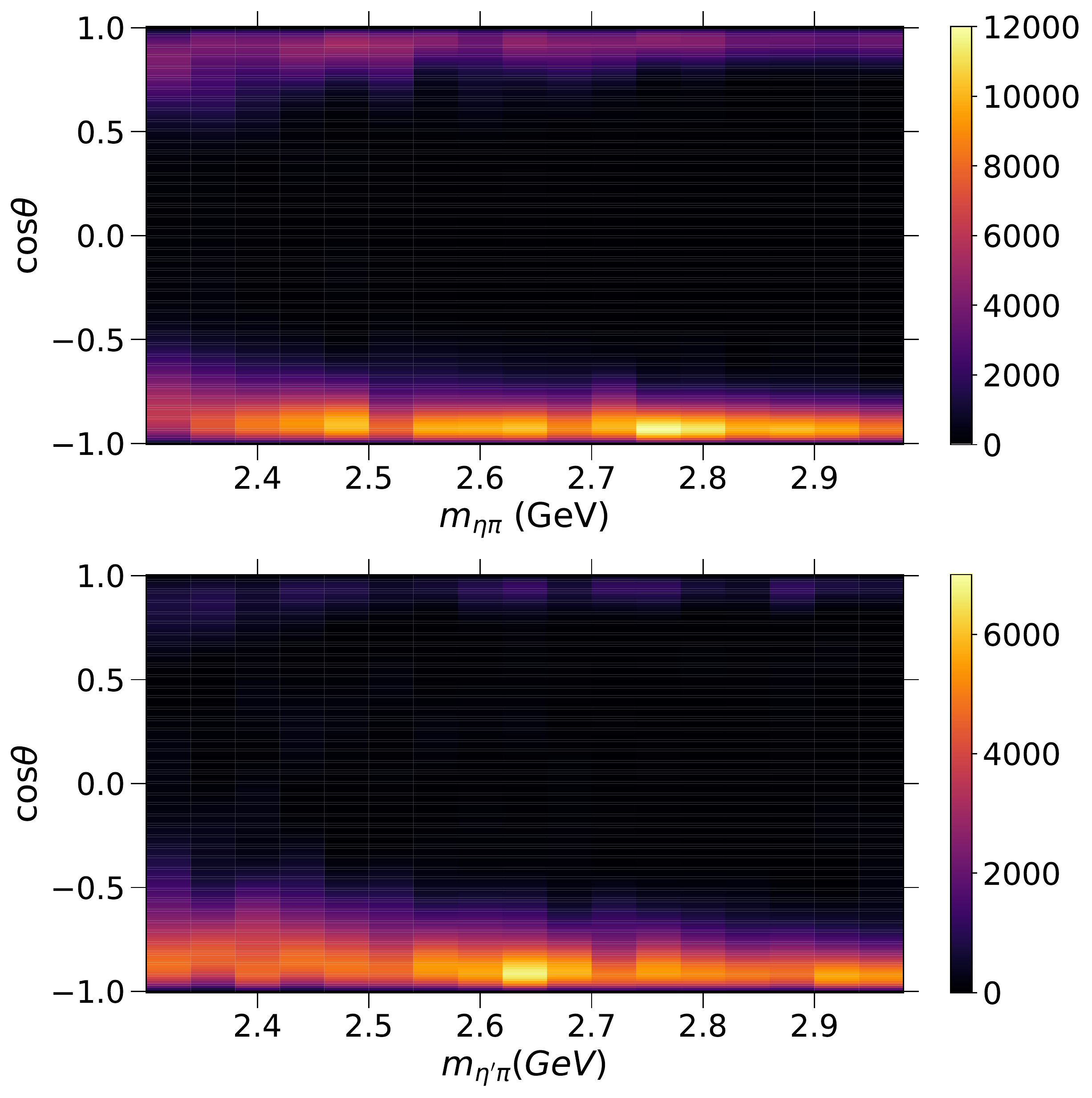} 
      \caption{Intensity $I_\theta(\m ,\cos\theta)$ 
      density distribution of the MVR from the 
      $\eta \pi$ (upper) and $\eta' \pi$ (lower)
      COMPASS partial waves.}\label{fig:compass_plot}
\end{figure}

\begin{figure}
    \includegraphics[width=\columnwidth]{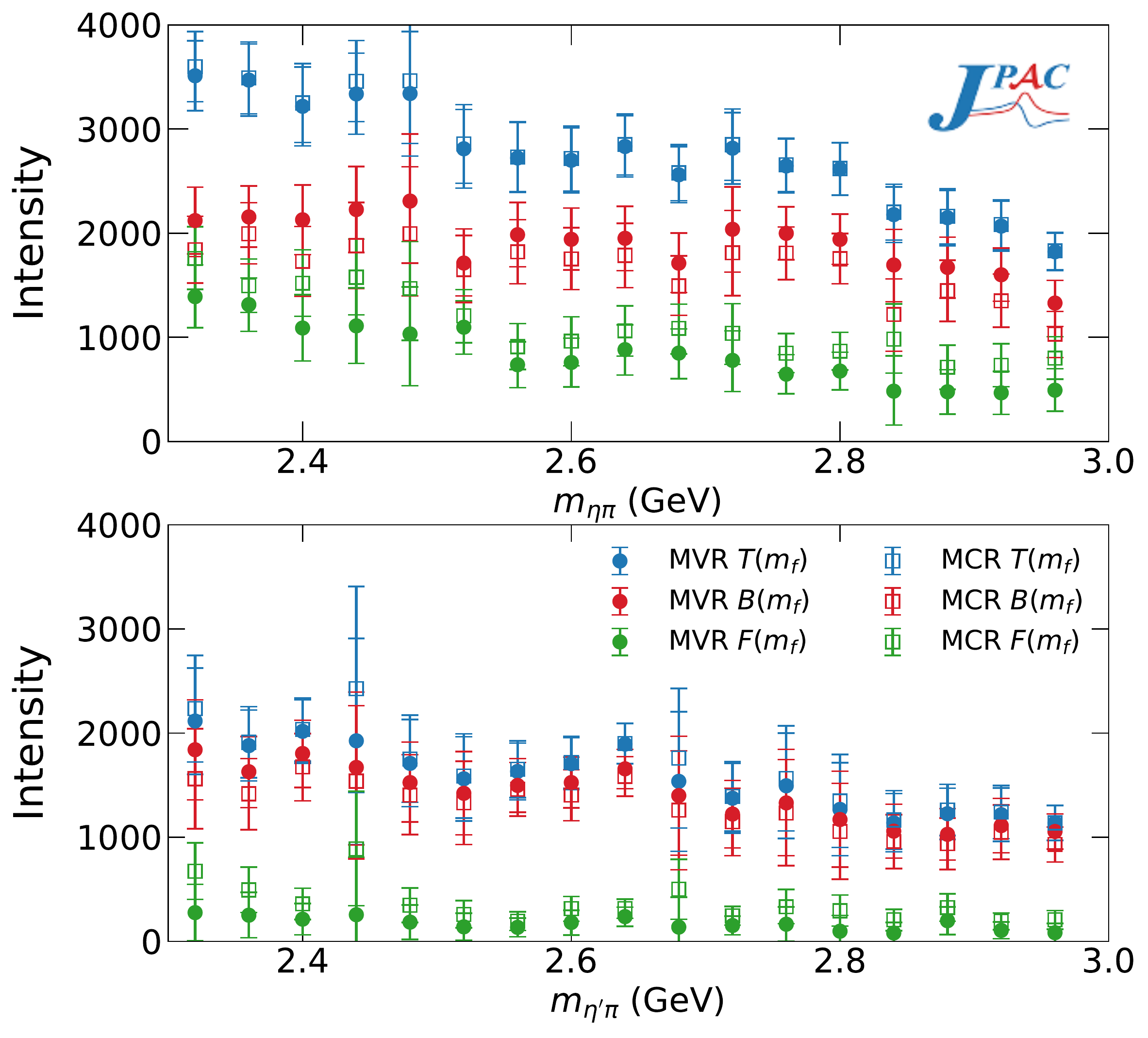} 
      \caption{Integrated forward $F(\m )$ (green), backward $B(\m )$ (red), and total $T(\m )=F(\m )+B(\m )$ (blue) intensities as defined in Eq.~\eqref{eq:fbintensities} for the MVR (full circles) and MCR (empty squares). Uncertainties in the MVR are taken from the MCR.
      We note that the slope of $F(\m )$ for both $\eta \pi$ and $\eta' \pi$ is steepper than for $B(\m )$.
      }\label{fig:fb_data}
\end{figure}

\begin{align} \label{eq:int_COMPASS}
    I(\m , \Omega) = \sum_{\epsilon = \pm} \left|\sum_{L,M} f^\epsilon_{LM}(\m ) \Psi^\epsilon_{LM}(\Omega) \right|^2 \, , 
\end{align}
 given by  $\Psi^{\epsilon =+}_{LM}(\Omega) = \sqrt{2} Y_L^M(\theta, 0) \sin M \phi$ and $\Psi^{\epsilon =-}_{LM}(\Omega) = \sqrt{2} Y_L^M(\theta, 0) \cos M \phi$, which are the real spherical harmonics with $\epsilon$ referred to as the reflectivity. 
The angular variables $\Omega \equiv (\theta,\phi)$ determine the direction 
of  the $\eta^{(\prime)}$ in the Gottfried-Jackson (GJ) frame (see Appendix \ref{sec:kin} for the axes orientation).
The complex functions $f^\epsilon_{LM}(\m )$ are obtained by fitting to the angular distributions for each energy bin $\m$. 
In the strict sense they are not partial waves, as they do not depend on the initial and final nucleon helicities. 
However, if a single helicity amplitude happens to dominate the reaction, the $f$'s can approach genuine partial waves. In general however, one should think of the $f$'s as defining an effective parametrization of the data at the amplitude level. Nevertheless, in the following we refer to the $f$'s as partial waves, as customary.

In practice, partial wave extraction requires the sum in Eq.~\eqref{eq:int_COMPASS}
to be  truncated. In the COMPASS analysis of $\eta \pi$, seven partial waves were used, $(L=1,\ldots,6; M=1)$ and $(L=2;M=2)$, while for the $\eta' \pi$ channel 
it was six partial waves, namely $(L=1,\ldots,6; M=1)$.
All the waves describing the $\eta^{(\prime)}\pi$ system have positive reflectivity $\epsilon = +$.  In the Regge asymptotic limit, reflectivity coincides with naturality of the exchange; at the nucleon vertex, the natural $\pom$ and $f_2$ are the dominant exchanges~\cite{Mathieu:2019fts}. 
A single negative reflectivity wave was included in the fit, $(L,M,\epsilon) = (0,0,-)$, that includes possible reducible backgrounds. It was found to contribute at the $0.5\%$ ($1.1\%$) level to the total $\eta\pi$ ($\eta'\pi$) intensity, and  will be neglected here. 

The partial waves
in Eq.~\eqref{eq:int_COMPASS} are written as
\begin{align} \label{eq:1} 
f_{LM}(\m ) = \sqrt{I_{LM}(\m )}\,\text{e}^{i\delta_{LM}(\m )}\, ,
\end{align}
where $I_{LM}(\m )$ are the partial wave intensities and the phases $\delta_{LM}(\m )$ are determined with respect to the phase of the $L=2,M=1$ wave, {\it i.e.}  $\delta_{21}(\m )\equiv 0$. In our analysis, we use the intensities and phases provided in the corrigendum to Ref.~\cite{Adolph:2014rpp}. The simplest way to compare the COMPASS results with a theoretical model would be to compare the partial waves. However, for reasons that will be discussed later in Section~\ref{sec:truncation}, we instead fit  
our amplitude model 
using an integral form of extended negative 
log-likelihood (ENLL) 
method~\cite{Lyons:1985vx,Barlow:1990vc,James:2006zz} to the intensity $I(\m , \Omega)$ reconstructed from the partial waves. There are two ways to reconstruct the $I(\m , \Omega)$ from the COMPASS partial waves. One approach is to  use the  mean values  of the intensity and phase at a given $\m $, and use Eq.~\eqref{eq:int_COMPASS} to obtain $I(\m , \Omega)$. We call this the mean value reconstruction (MVR). However, this method ignores the experimental uncertainties. The second method, which we refer to as MCR, uses Monte Carlo reconstruction. 
This is done by associating a probability distribution to the intensity and phase at each $\m $ independently.  
In doing so, instead of a single intensity value for each $(\m ,\phi,\cos\theta)$ point, we obtain a distribution. 
We can then compute the expected (mean) value of the intensity and its associated uncertainty at a given confidence level. The statistical errors are thus propagated from the partial waves to the intensity. The details on the MCR can be found in Appendix~\ref{sec:mcr}. What remains unknown, however, are the uncertainties associated to the systematics of the COMPASS fit and the correlations among partial waves. As a consequence, the intensities reconstructed using MVR and MCR differ.

In Figs.~\ref{fig:pieta_phi_theta} and~\ref{fig:pietap_phi_theta} we show the density plots of $I(\m , \Omega)$ at three fixed $\m $ as well as  the $\phi$-integrated distributions 
\begin{align}
I_\theta (\m ,\cos \theta) = 
\int_0^{2\pi} \diff\phi \, I(\m ,\Omega)\, . \label{eq:Itheta}
\end{align}

In Fig.~\ref{fig:compass_plot} we plot 
$I_\theta (\m ,\cos \theta)$ for 
$\m $ above $2.3\gev$, for a total of seventeen mass bins in each channel. 
This can be compared to the plot of the experimental data shown in Fig.~2 of Ref.~\cite{Adolph:2014rpp},
although we note that the data shown in the COMPASS paper
are not corrected for detector acceptance. Several features in Figs.~\ref{fig:pieta_phi_theta},~\ref{fig:pietap_phi_theta}, and~\ref{fig:compass_plot} are noteworthy:
\begin{enumerate}
    \item At fixed $\m $, the  intensity $I(\m ,\Omega)$ is periodic
    in $\phi$ with periodicity $2\pi$. Moreover,
    it presents a reflection symmetry 
    along the azimuthal angle $\phi$ with symmetry axis at $\phi=\pi$,
    {\it i.e.} $I(\m ,\theta,\phi)=I(\m ,\theta,2\pi-\phi)$ with $\phi \in [0,2\pi]$. Both facts stem from the definition of the intensity, Eq.~\eqref{eq:int_COMPASS};
    \item the intensity peaks in the forward $\cos\theta \sim 1$ and backward $\cos\theta \sim -1$ regions. In the forward region, most of the beam momentum is carried by the $\eta^{(\prime)}$, and in the backward region by the $\pi$. We call these clusters the ``fast-$\eta$'' and the ``fast-$\pi$'' regions, respectively;
    \item the backward (fast-$\pi$) peak is larger  than the forward (fast-$\eta$) peak, resulting in a forward-backward asymmetry. 
    This effect is more pronounced in  the case of the $\eta'\pi$ channel;
    \item the backward peak is broader in $\eta'\pi$  than in the
    $\eta \pi$; 
    \item both the forward and backward peaks become narrower as the invariant mass $\m $ increases;
    \item the MVR  intensities at backward peak are larger than those of the MCR, and in the small $\left|\cos\theta\right|$ region the intensity profile becomes smeared out in the MCR, so more structures are visible in the MVR in the region where intensities are low. Appendix~\ref{sec:mvr_vs_mcr} provides
    more insight on the differences between MVR and MCR.
\end{enumerate}
These features are typical of diffractive processes, indicating the dominance of double-Regge exchanges in the energy region $\m \gtrsim 2.3\gev$. 
In the $SU(3)$ flavor symmetric limit  
the $\pi$ and the octet $\eta_8$ are degenerate,
and so are the $a_2$ and $f_2$ Regge trajectories. 
Furthermore, if the $SU(3)$ singlet exchanges 
({\it e.g.} the $\pom$) are neglected, 
the forward and backward intensities are 
identical~\cite{Schwimmer:1970yk} for the production of the octet, and only even (nonexotic) waves contribute. 
Since $\eta'$ is dominated by the $SU(3)$ singlet
we expect the asymmetry to be larger for the production of $\eta'\pi$.
The broadness of the peaks is related to the relative
strength of the different double-Regge contributions
to the amplitudes and will be addressed in Sections~\ref{sec:set_diag} and~\ref{sec:results}.

To quantify the forward-backward asymmetry we define
\begin{subequations}
\begin{align}
F(\m ) \equiv   & \int_0^1 \diff\cos \theta \,
I_\theta(\m ,\cos\theta) \, , \label{eq:intensities_fwd} \\
B(\m ) \equiv  & \int_{-1}^0 \diff\cos \theta \, I_\theta(\m ,\cos\theta) \, , \label{eq:intensities_bwd} \\
A(\m ) \equiv & \,   \frac{F(\m ) - B(\m )}{F(\m ) + B(\m ) } \, ,
\label{eq:intensities_asy}
\end{align} \label{eq:fbintensities}
\end{subequations}
with $F(\m )$ and $B(\m )$ being the forward and backward intensities, respectively, and $A(\m )$ the forward-backward asymmetry.
Figure~\ref{fig:fb_data} shows
$F(\m )$, $B(\m )$, and their sum $T(\m )$ for both
MVR and MCR for the two channels.  
We find that the slope of $F(\m )$ is steeper 
than that of $B(\m )$.
These intensities show clearly the difference
between the MVR and the MCR, even though the total
intensity in the MVR and MCR are similar. 

\section{Double-Regge Model}\label{sec:model}

\begin{figure}
\includegraphics[width=0.35\textwidth]{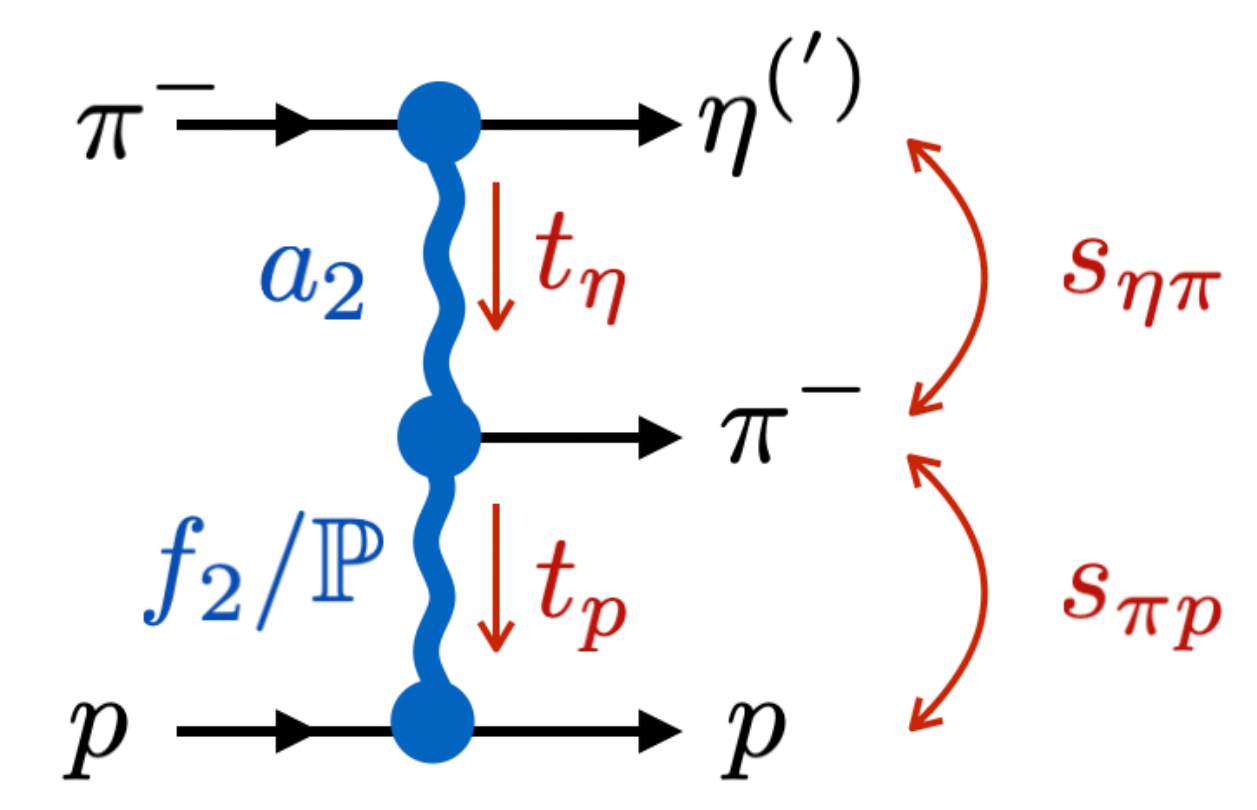} \\
\includegraphics[width=0.35\textwidth]{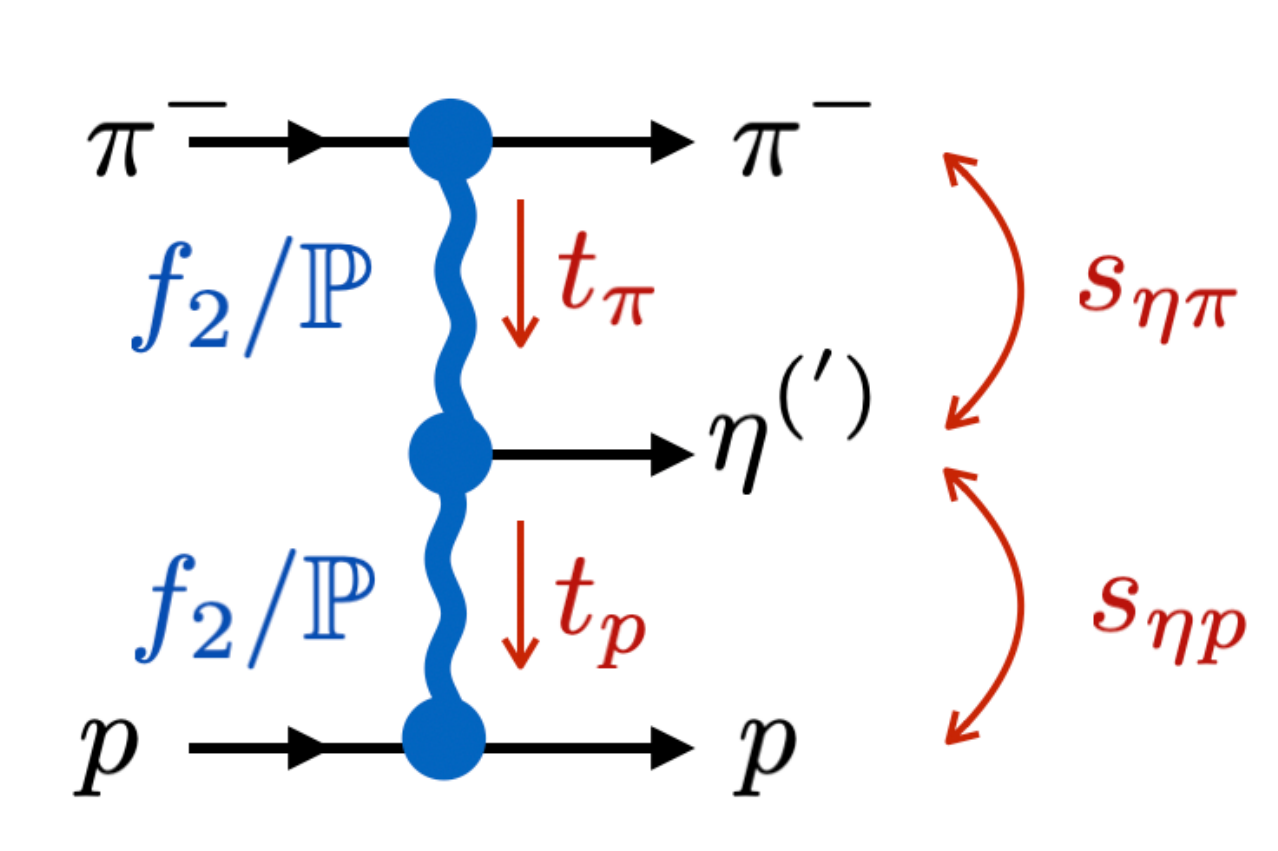}
\caption{Fast-$\eta$ (top) and fast-$\pi$ (bottom) amplitudes.}
\label{fig:exchanges}
\end{figure}

We present here a double-Regge exchange model for the reactions in Eq.~\eqref{reaction1}. 
Multi-Regge exchange formalism has been extensively studied theoretically in the past~\cite{Brower:1974yv, Drummond:1969ft,Bali:1967zz,Drummond:1969jv,Weis:1973vh,  Collins:1977jy}. 
An application of such formalism was 
presented in~\cite{Shimada:1978sx} for a similar reaction, 
two-pseudoscalar mesons production in $K^\pm$ and $\pi^\pm$ beam diffraction.
More recently the double-Regge exchange was used to describe the central meson production in the high energy proton-proton collisions ~\cite{Lebiedowicz:2019jru,Lebiedowicz:2020yre,Cisek:2021mju}.
We will adopt the same model and quote in this Section its main features. 

The fast-$\eta$ and fast-$\pi$ regions correspond to the fast-$\eta$ and fast-$\pi$ double-Regge exchange amplitudes depicted in Fig.~\ref{fig:exchanges}. 
The model assumes the dominance of leading Regge trajectories.
Although it is known that  daughter poles and cuts also contribute, for example to polarization observables~\cite{Mathieu:2015eia,Mathieu:2015gxa,Nys:2016vjz,Mathieu:2018mjw}, present data does not seem to be sensitive to subleading exchanges.

The top exchange is saturated by the $a_2$ trajectory for the fast-$\eta$ amplitude, and by the $f_2$ or $\pom$ trajectory for the fast-$\pi$ amplitude. The bottom exchange is either 
the $f_2$ or $\pom$ for both types of amplitude. It is common lore that, at COMPASS energies, the $\pom$ is the only relevant bottom exchange; however, this hypothesis is incompatible with data, as we will show in Section~\ref{sec:set_diag}. 

Consequently, the total amplitude $A_\text{Th}(\m,\Omega)$ is the sum of six possible double-Regge amplitudes 
\begin{align}
    \nonumber
    A_\text{Th}(\m,\Omega) &= c_{a_2\pom} \,A_{a_2 \pom} 
    + c_{a_2 f_2}\, A_{a_2 f_2}
    + c_{f_2\pom}\, A_{f_2 \pom} 
    \\
    & + c_{f_2f_2}\, A_{f_2 f_2}
    + c_{\pom\pom} \, A_{\pom \pom} 
    + c_{\pom f_2} \, A_{\pom f_2} \, ,
    \label{eq:ampTot}
\end{align}
where the $\{c\}$ are unknown
and will be fitted to data.
The intensity of the model is given by
\begin{align}
    I_\text{Th}(\m,\Omega) =  k(\m) \, |A_\text{Th}(\m,\Omega)|^2, \label{eq:model}
\end{align}
where $k(\m)= \lambda^\frac{1}{2}(\m^2,m^2_{\eta^{(\prime)}}, m_\pi^2)/(2\m)$ is the breakup momentum between the $\pi$ and the $\eta^{(\prime)}$, and  $\lambda(x,y,z) = x^2+y^2+z^2-2(xy+xz+yz)$
is the triangle function.

Regge amplitudes are expressed in terms of Lorentz invariants. In addition to $s$, $t_p$ and $\m$, as depicted in Fig.~\ref{fig:exchanges}, for the fast-$\eta$ and $\pi$ amplitudes, the GJ angles are related to the following Lorentz invariants 
\begin{subequations}  \label{invariantsdef}
\begin{align}
    \label{eq:invar_typeI}
    \text{fast-}\eta: &&
    t_\eta&=(q-k_\eta)^2, &
    s_{\pi p}&=(k_\pi+p_2)^2 \, ,
    \\
    \label{eq:invar_typeII}
    \text{fast-}\pi: &&
    t_\pi&=(q-k_\pi)^2, &
    s_{\eta p}&=(k_\eta+p_2)^2 \, .
\end{align}
\end{subequations}

There are only five independent variables. The fast-$\pi$ invariant $t_\pi$ and $s_{\eta p}$ can be expressed as linear combinations of the five fast-$\eta$ variables. Appendix~\ref{sec:kin} summarizes the relevant kinematical relations. 

The analytic structure is the same for all double-Regge amplitudes. The dependence in the momentum transferred $(t_\eta,t_p)$ for fast-$\eta$ and $(t_\pi,t_p)$ for fast-$\pi$ enters only via the trajectories $(\alpha_1,\alpha_2)$,  where $\alpha_1$ corresponds to the top exchange and $\alpha_2$ to the bottom one. 
Hence, for fast-$\eta$ amplitudes $\alpha_1 \equiv \alpha_{a_2}(t_\eta)$ and for fast-$\pi$ amplitudes $\alpha_1 \equiv \alpha_{f_2}(t_\pi)$ or $\alpha_1 \equiv \alpha_{\pom}(t_\pi)$. The bottom trajectory is $\alpha_2 \equiv \alpha_{f_2}(t_p)$ or $\alpha_2 \equiv \alpha_{\pom}(t_p)$ for both types depending on the bottom exchange.

Regge theory predicts the dependence in the invariant masses squared $(s_1,s_2)$ with $(s_1,s_2) = (s_{\eta\pi}, s_{\pi p})$ for the fast-$\eta$ amplitudes and $(s_1,s_2) = (s_{\eta\pi}, s_{\eta p})$ for the fast-$\pi$ amplitudes. Since the nucleons play a spectator role given the large total energy, their spins can be ignored. 
For five spinless particles with an odd number of pseudoscalars, the generic amplitude for a double-Regge exchange is~\cite{Brower:1974yv,Shimada:1978sx}
\begin{widetext}
\begin{align}
T(\alpha_1,\alpha_2 ; s_1,s_2) & = 
K\, \Gamma(1-\alpha_1)\,  \Gamma(1-\alpha_2) 
\frac{(\alpha' s_1)^{\alpha_1} (\alpha' s_2)^{\alpha_2}}{\alpha' s}  
\left[ 
\frac{\xi_1 \, \xi_{21}}{\kappa^{\alpha_1}}\, V(\alpha_1,\alpha_2,\kappa) + 
\frac{ \xi_2 \, \xi_{12}}{\kappa^{\alpha_2}} \, V(\alpha_2,\alpha_1, \kappa) 
\right] \, .
\label{eq:generic}
\end{align}
\end{widetext}

 The double-Regge limit corresponds to $s,s_1,s_2 \to \infty$ with $\kappa^{-1} \equiv  s/(\alpha' s_1 s_2)$ fixed, which
 is related to the cosine of the Toller angle~\cite{Brower:1974yv}. 
The variation of $\alpha'$ induces a smooth exponential dependence on the momentum transfer variables ($t_\pi$, $t_\eta$ and $t_p$), that effectively absorbs possible form factor contributions. In particular, since the COMPASS measurement is performed at integrated bottom exchange momentum, the data are not sensitive to $t_p$. Therefore our model Eq.~\eqref{eq:generic} does not require any additional momentum transferred dependence.
We have found that fitting simultaneously $\{c\}$ and $\alpha'$ does not lead to stable solutions, as the coefficients and the scale parameter are strongly correlated. Moreover, $\alpha'$ should be of the order of the hadronic scale, ${\cal O}(\nsgev^{-2})$.  
 We let $\alpha'$ vary in exploratory fits and found $\alpha' = 0.8\gev^{-2}$ to be the optimal choice.
The kinematical factor $K$ is detailed in Appendix~\ref{sec:kin}.
 
The presence of two symmetric terms in the bracket of Eq.~\eqref{eq:generic} is imposed from general considerations of the analytic structure of double-Regge amplitudes. The interested readers will find the technical details in Section~3.3 of Ref.~\cite{Brower:1974yv}.

The double-Regge amplitude of Eq.~\eqref{eq:generic} has poles for positive integer values of the trajectories $\alpha_1$ and $\alpha_2$, which are related to the spins of the physical particles in the $t$ channel. Since only poles with even signature $(-1)^J = +1$ can couple to $\eta\pi$ and $\pi\pi$, odd signature poles are removed by the signature factors
\begin{subequations}
\begin{align}
\xi_{n} & =\frac{1+e^{-i\pi\alpha_n}}{2}\, , \\
\xi_{nm}& =\frac{1+e^{-i\pi(\alpha_n-\alpha_m)}}{2} \, .
\end{align}
\end{subequations}

The vertex function $V(\alpha_1,\alpha_2,\kappa)$ is an analytic function of its arguments. 
Its most general form involves an infinite number of  
Reggeon-Reggeon-particle couplings and reduces to a polynomial in $\kappa^{-1}$ for integer $\alpha_{1}$ and $\alpha_2$~\cite{Drummond:1969ft}. In a dual model, all Reggeon-Reggeon-particle couplings are equal and the vertex simplifies to~\cite{Brower:1974yv,Shi:2014nea}
\begin{equation}
    V(\alpha_1,\alpha_2,\kappa)=\frac{\Gamma(\alpha_1-\alpha_2)}{\Gamma(1-\alpha_2)}\,_1F_1\left(1-\alpha_1,1-\alpha_1+\alpha_2,-\kappa\right) \, ,
    \label{eq:Vfun}
\end{equation}
where $_1F_1$ is the confluent hypergeometric function of the first kind.

As explained in Ref.~\cite{Shimada:1978sx}, the $V(\alpha_1,\alpha_2,\kappa)$ functions used in Eq.~\eqref{eq:generic} and defined in Eq.~\eqref{eq:Vfun} have poles at $\alpha_1-\alpha_2$ [and $\alpha_2-\alpha_1$ for $V(\alpha_2,\alpha_1,\kappa)$] equal to non-positive integers. However, these poles  cancel between the two terms in Eq.~\eqref{eq:generic}. For example, when $\alpha_2>\alpha_1$ the pole in the gamma function in Eq.~\eqref{eq:Vfun} cancels
out with the pole in the hypergeometric function from the second term Eq.~\eqref{eq:generic}. 

The six contributions in Eq.~\eqref{eq:ampTot} are obtained from the generic double-Regge amplitude in Eq.~\eqref{eq:generic} 
with the following substitutions
\begin{subequations}
\begin{align}
    A_{a_2 \pom} & = T(\alpha_{a_2}(t_\eta), \alpha_{\pom}(t_p) ; s_{\eta\pi}, s_{\pi p}) \, ,\\
    A_{a_2 f_2} & = T(\alpha_{a_2}(t_\eta), \alpha_{f_2}(t_p) ; s_{\eta\pi}, s_{\pi p}) \, ,
    \\
    A_{f_2 \pom} & = T(\alpha_{f_2}(t_\pi), \alpha_{\pom}(t_p) ; s_{\eta\pi}, s_{\eta p}) \, ,\\
    A_{f_2 f_2} & = T(\alpha_{f_2}(t_\pi), \alpha_{f_2}(t_p) ; s_{\eta\pi}, s_{\eta p}) \, ,
    \\
    A_{\pom \pom} & = T(\alpha_{\pom}(t_\pi), \alpha_{\pom}(t_p) ; s_{\eta\pi}, s_{\eta p}) \, ,\\
    A_{\pom f_2} & = T(\alpha_{\pom}(t_\pi), \alpha_{f_2}(t_p) ; s_{\eta\pi}, s_{\eta p}) \, .
\end{align}\label{eq:As}
\end{subequations}

Since the momentum transferred between the initial and final nucleon has been integrated over in the COMPASS analysis, we do not have access to the $t_p$ distribution. This distribution would allow us to discriminate between the bottom exchanges.
Since the amplitude decreases exponentially with $t_p$, we fix $t_p$ close to the COMPASS lowest limit, $t_p = -0.2\gevsq$.
Results are stable against small variation of this value.

Finally, we need to specify the Regge trajectories
\begin{subequations} \label{eq:trajectory}
\begin{align}
    \alpha_{a_2}(t)  & = 0.53 + 0.90\, t, \\
    \alpha_{f_2}(t)  & = 0.47 + 0.89\, t, \\
    \alpha_{\pom}(t) & = 1.08 + 0.25\, t,
\end{align}
\end{subequations}
where we adopted the standard parametrization for the $\pom$~\cite{Donnachie:1983hf} and the $f_2$~\cite{Oh:2003aw} trajectories. 
Phenomenologically, the $a_2$ trajectory is very similar to that of 
$\rho$, which is referred as exchange degeneracy (EXD)~\cite{Collins:1977jy,Girardi:1974wu}. Our model is thus entirely specified by the six real parameters $\{c\}$.
Each $\{c\}$ in  Eq.~\eqref{eq:ampTot} is a product of two  particle-Reggeon-particle couplings (top and bottom vertices) and one Reggeon-particle-Reggeon coupling (middle vertex). 
The particle-Reggeon-particle couplings could be extracted from quasi-two-body reactions~\cite{Irving:1977ea,Nys:2018vck}, but the Reggeon-particle-Reggeon couplings  
are largely unknown.
In principle, all couplings have residual dependence on $t$'s 
that cannot be disentangled. 
This prevents us from imposing further relations among the $\eta$ and $\eta'$ amplitude parameters.

\section{Partial wave truncation beyond the resonance region}
\label{sec:truncation}

\begin{figure*}
        \includegraphics[width=0.7\textwidth]{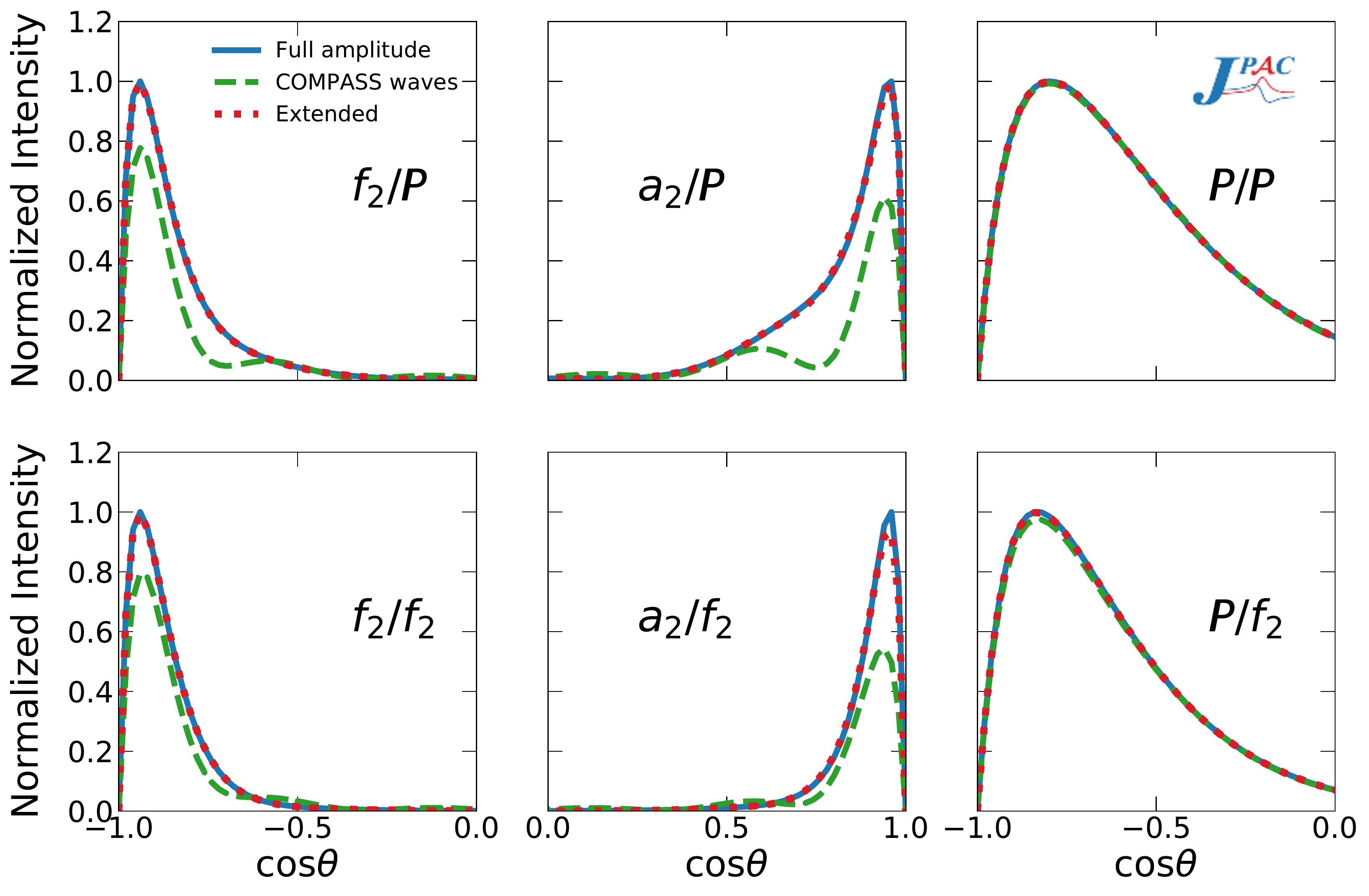}
      \caption{$I_\theta(m_{\eta\pi},\cos \theta)$ for individual amplitudes at $m_{\eta\pi}=2.64\gev$ for the $\eta\pi$ channel. 
      We compute the intensity of each individual 
      amplitude (solid blue) 
      normalized to its peak value. We compare it to two different partial wave truncations. 
      The dashed green curve adds up 
      the partial waves included in the COMPASS analysis, {\it i.e.} $(L;M) =$ $(1,\dots,6 ;1)$ and $(2;2)$. 
      The dotted red curve is obtained by extending the sum to $(L;M) = (1,\dots,10;1,2)$. One can appreciate that, while the COMPASS partial waves saturate the $\pom/\pom$ and $\pom/f_2$, they only account for $\sim 80\%$ of the $a_2/\pom$ peak intensity.
      } \label{fig:diags_eta}
\end{figure*}

\begin{figure*}
      \includegraphics[width=0.7\textwidth]{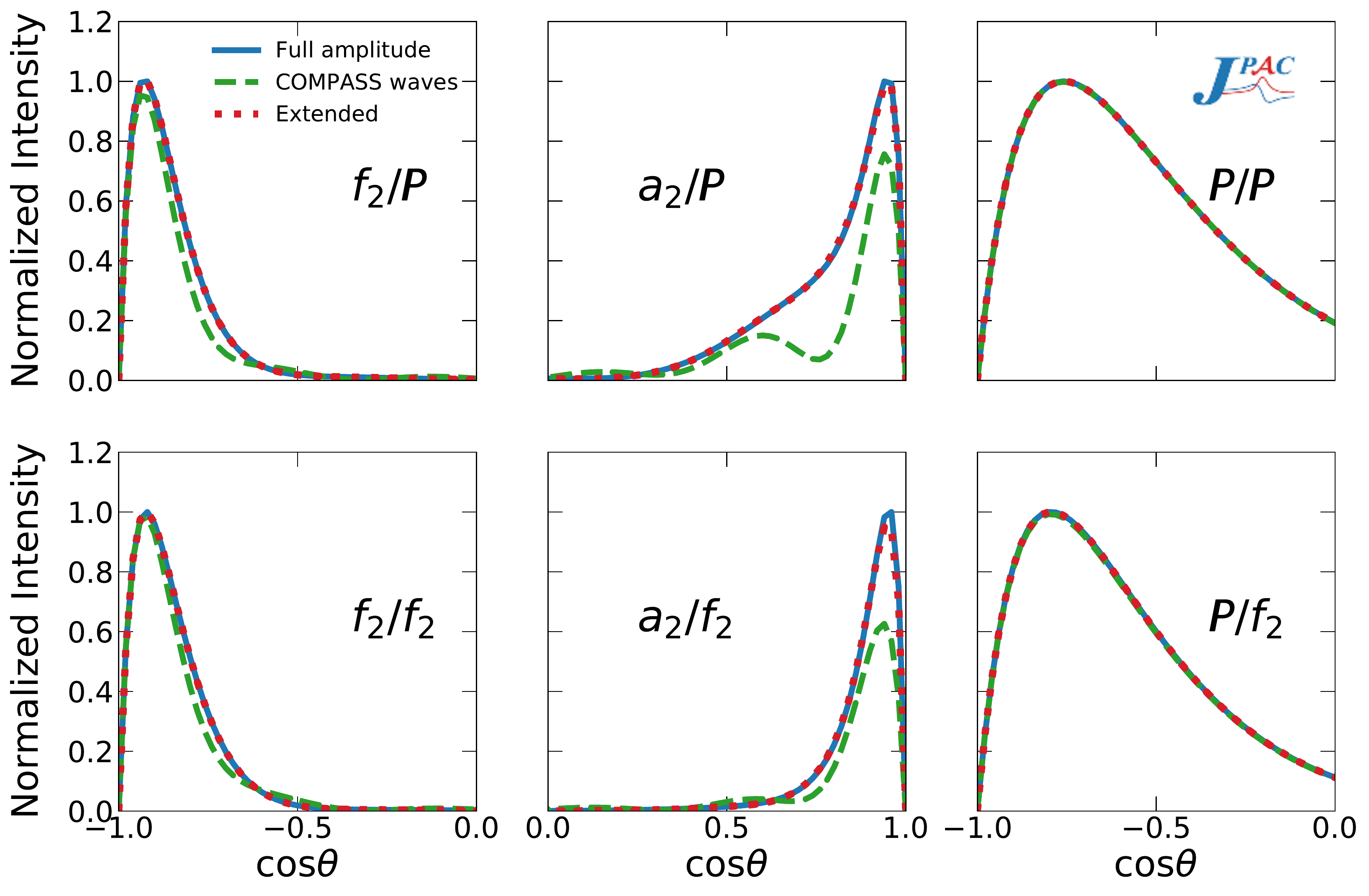}
      \caption{$I_\theta(m_{\eta' \pi},\cos \theta)$ for individual amplitudes at $m_{\eta' \pi}=2.64\gev$ for the $\eta' \pi$ channel. Conventions 
      are the same as in Fig.~\ref{fig:diags_eta}, except that here $(L;M)=(2;2)$ is not included in the COMPASS waves. Again, the COMPASS partial waves saturate the $\pom/\pom$ and $\pom/f_2$, but only account for $\sim 60\%$ of the $a_2/f_2$ peak intensity.} \label{fig:diags_etap}
\end{figure*}

COMPASS extracted partial waves under the assumption that only a limited number
of them contribute. This is justifiable in the resonance region, but as the invariant mass of the $\eta^{(\prime)}\pi$ system increases so does the number of relevant waves. Since the overall intensity decreases in the high energy region, the significance of higher waves ($L>6$) 
could not be established and, hence, they were neglected.

The Regge model developed in the previous Section is not based on a partial wave expansion and therefore implicitly includes all partial waves. One can thus study whether the approximation to truncate to $L\leqslant 6$ waves is appropriate for our model. 
In Fig.~\ref{fig:diags_eta} we show how the truncation  affects the total intensity in the $\eta \pi$ channel. We expand each amplitude into partial waves and then sum back only the ones considered in the COMPASS analysis. For example, at $m_{\eta\pi} = 2.64\gev$, the seven partial waves considered by COMPASS account for  $\sim80\%$ of the intensity at the peak for the $f_2/f_2$ exchange. 
At this $m_{\eta\pi}$, only for the $\pom/\pom$ and $\pom/f_2$
amplitudes this truncation adequately reproduces the intensity ($>99\%$ and $>97\%$, respectively). If we include the partial waves up to $L=10$ with $M=1$ and $M=2$, the intensity of the amplitudes is almost completely recovered ($>99\%$ for all amplitudes except for $a_2/f_2$, which is $>93\%$). In Fig.~\ref{fig:diags_etap} we show the same plots for the $\eta'\pi$ channel. In this case, the main disagreement happens in the forward peak ($a_2/\pom$ and $a_2/f_2$ amplitudes), where only between $60\%$ and $80\%$ of the peak strength is accounted for by the COMPASS partial waves. 

Thus, as mentioned earlier, in this energy region COMPASS waves should be considered as an effective parametrization of the data, rather than being directly compared with genuine partial waves from a model that contains an infinite number of waves.  However, given that they have been extracted under the constraint of summing up to the total intensity, we can reconstruct $I(\m,\Omega)$ from the partial waves 
using Eq.~\eqref{eq:int_COMPASS} with the two methods 
(MVR and MCR) explained in Section~\ref{sec:COMPASS}
and fit them with our model. 
In Section~\ref{sec:cPW}
we will discuss how the model partial waves compare to the 
$f$'s extracted from the data by COMPASS. 

\section{The minimal set of amplitudes}\label{sec:set_diag}

\begin{figure}
    \includegraphics[width=\columnwidth]{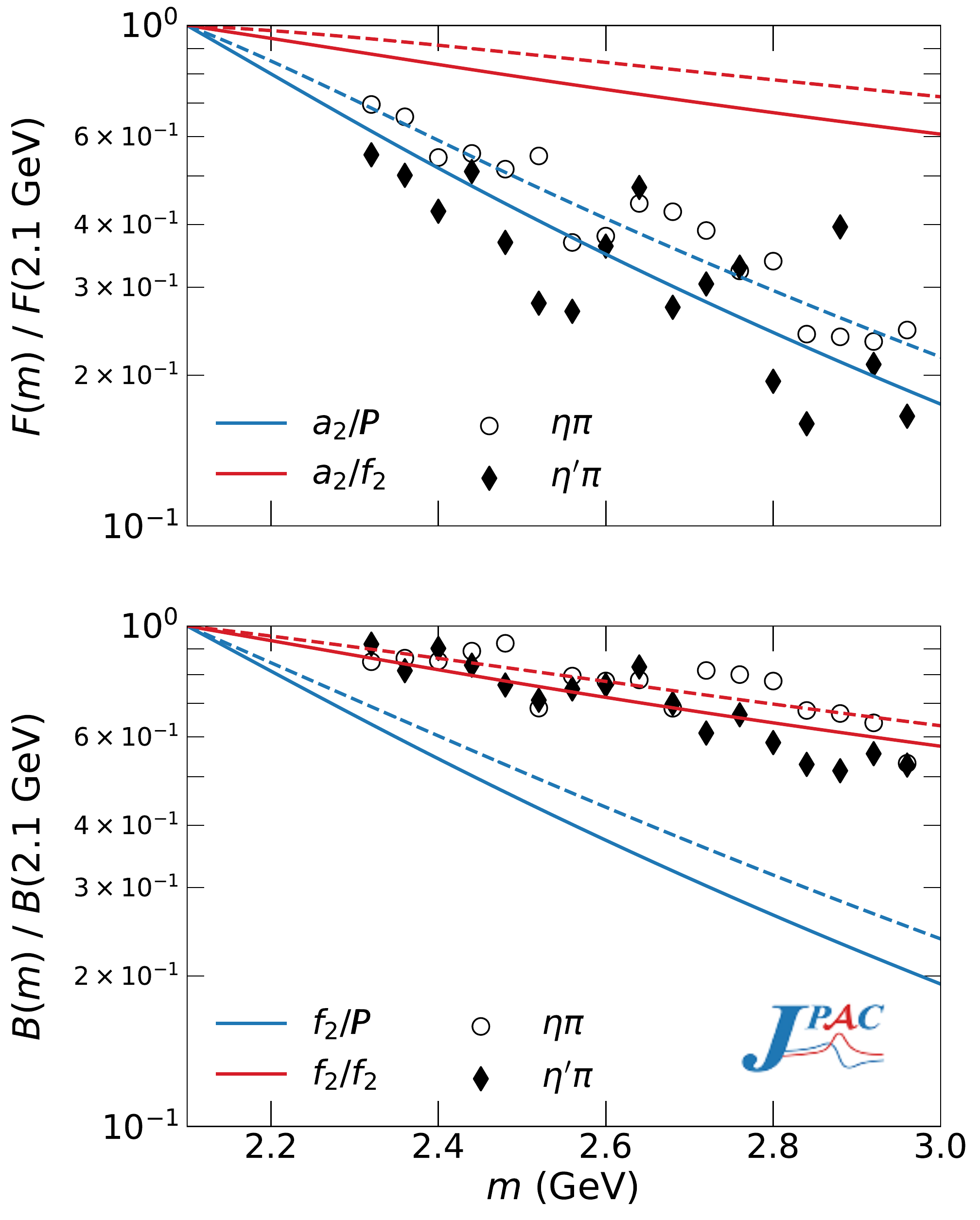}
      \caption{Forward (upper) and backward (lower) intensities as defined in Eq.~\eqref{eq:fbintensities}
      for the top-$a_2$ and top-$f_2$ amplitudes, respectively.
      Solid lines correspond to $\eta \pi$ and dashed
      to $\eta' \pi$.
      Each theoretical intensity is normalized to its value at $\m=2.1\gev$.
      In circles and diamonds we show the experimental data arbitrarily rescaled, as obtained by MVR.
}\label{fig:mpieta_slope}
\end{figure}

\begin{figure*}
    \includegraphics[width=0.8\textwidth]{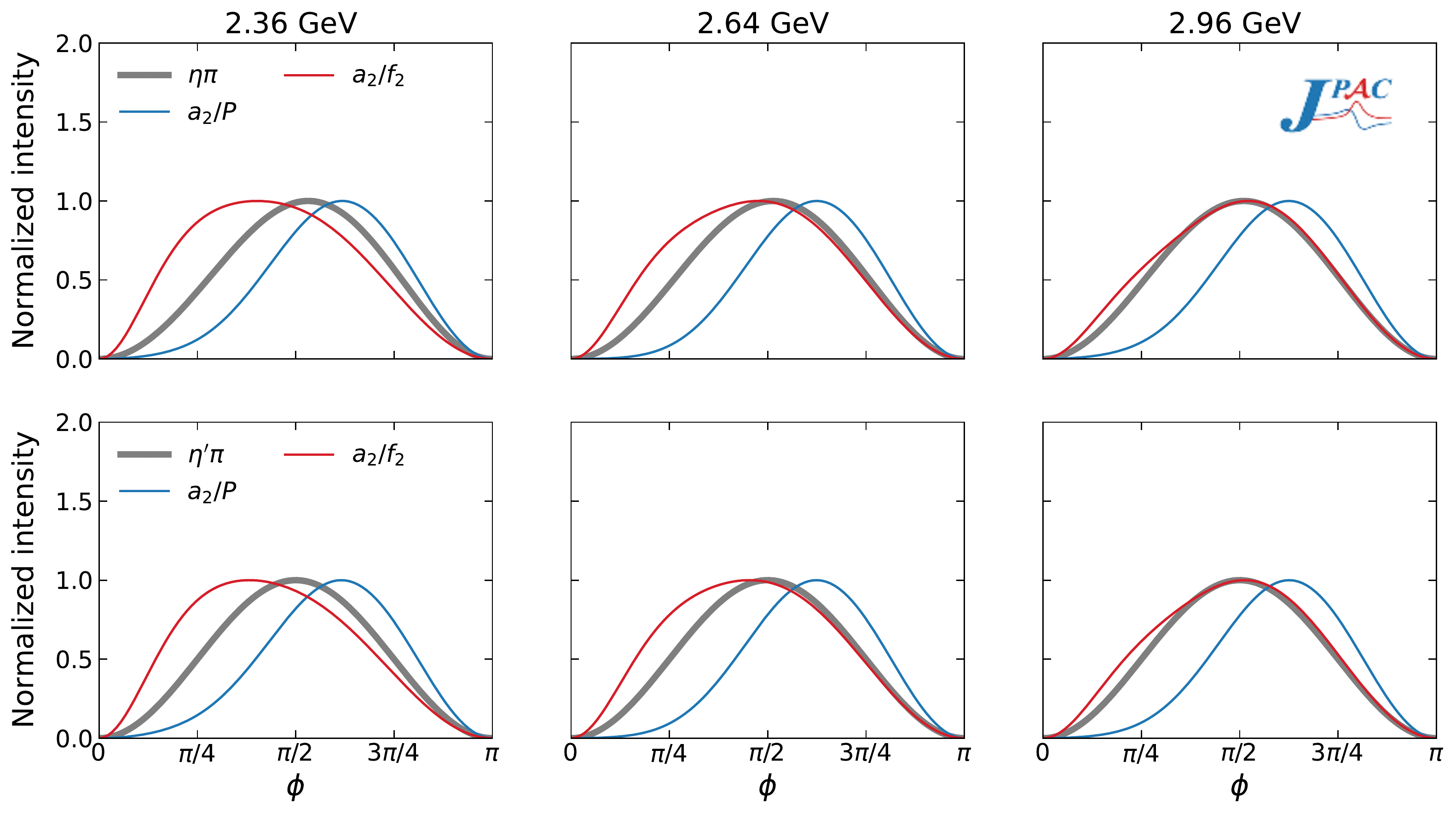} 
      \caption{Intensity distribution of $\phi$ from MVR (grey),
      for three fixed energies  
      and $\cos \theta = 0.866$ (close to the forward peak) 
      for $\eta \pi$ (upper row) and
      $\eta' \pi$ (lower row) compared
      to the same distributions for the $a_2/\pom$ (blue) and $a_2/f_2$ (red) amplitudes.
      Each distribution is normalized to its peak value.
      Due to the reflection symmetry shared by the model and the intensity, we only show $\phi\in[0,\pi]$.
}\label{fig:phi_fixedtheta_forward}
\end{figure*}

\begin{figure*}
    \includegraphics[width=0.8\textwidth]{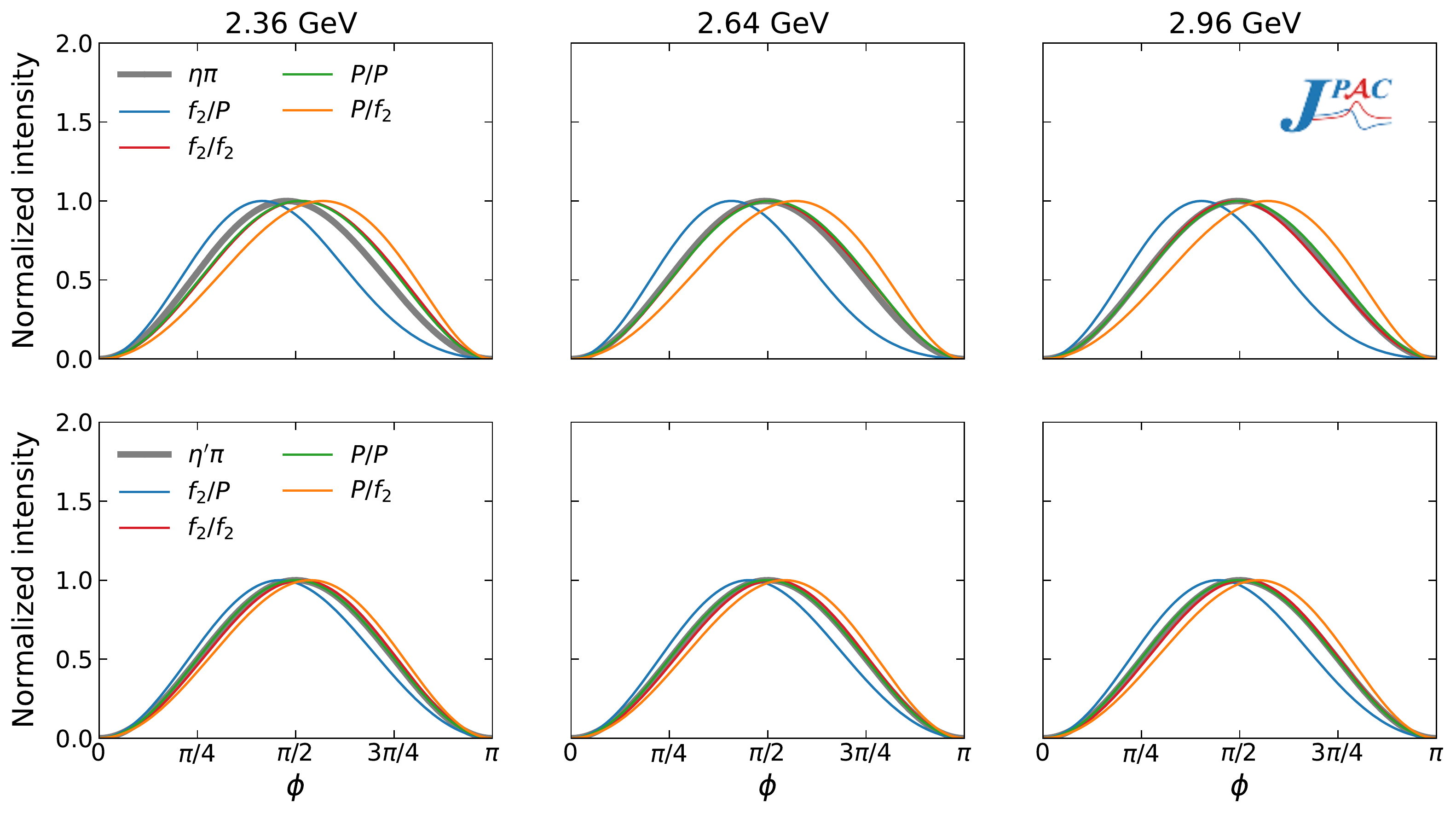} 
      \caption{Intensity distribution of $\phi$ from MVR (grey),
      for three fixed energies  
      and $\cos \theta = -0.866$ (close to the backward peak) 
      for $\eta \pi$ (upper row) and
      $\eta' \pi$ (lower row)  compared
      to the same distributions for the $f_2/\pom$ (blue), $f_2/f_2$ (red), $\pom/\pom$ (green), and $\pom/f_2$ (orange) amplitudes.
      Each distribution is normalized to its peak value. 
      }\label{fig:phi_fixedtheta_backward}
\end{figure*}

Our model described in Section~\ref{sec:model} is completely determined by the six coefficients of the double-Regge exchange amplitudes. As a first approach, we fitted the intensity with all the six parameters unconstrained. 
However, those fits did not lead to a unique solution, 
sometimes having coefficients compatible with zero. The error estimation was unreliable. In order to make the fits to reach stable solutions, we have to restrict the parameters to at most four.
Consequently, to establish which amplitudes must be neglected or included in the fits, in this Section we compare the angular and mass dependencies of the individual exchanges to the experimental ones from MVR shown in Figs.~\ref{fig:pieta_phi_theta}--\ref{fig:fb_data}. Conclusions are identical for MCR.

In the $SU(3)$ limit, the event distribution of $\eta\pi$ becomes symmetric in $\cos \theta$. At the amplitude level, this is manifested via EXD, meaning that the parameters of the $a_2$ and $f_2$ Regge poles are equal, including the couplings, 
$c_{a_2\pom} \simeq - c_{f_2\pom}$ and
$c_{a_2f_2} \simeq - c_{f_2f_2}$.\footnote{The minus sign is due to the kinematic factor $K$ being odd under permutation of the $\pi$ and $\eta$ momenta.} 
Deviations from the EXD relation are manifested in the nonvanishing forward-backward asymmetry. The $\cos \theta$ dependence is correlated to the $t_\pi$ or $t_\eta$ dependence arising from the top exchange trajectory. We thus expect the amplitudes with the same top exchange 
to  have similar $\cos \theta$ behavior. Both $a_2/f_2$ and $a_2/\pom$ amplitudes will be collectively denoted as top-$a_2$ amplitudes, and similarly for the top-$f_2$ and top-$\pom$ amplitudes. By EXD, we expect that both top-$a_2$ and -$f_2$ matter. 

As shown in Figs.~\ref{fig:diags_eta} and~\ref{fig:diags_etap}, the top-$a_2$ and top-$f_2$ amplitudes produce a narrow forward and a narrow backward peak, respectively. 
The top-$\pom$ amplitudes produce a wider backward peak, which is due to the smaller slope of the $\pom$ trajectory. 
Given that the widths of the experimental backward peaks in $\eta\pi$ and $\eta'\pi$ are similar to what is expected from the top-$f_2$ exchange,  we conclude that the $f_2/f_2$ and/or $f_2/\pom$ amplitudes should account for most of the backward intensity. The residual contribution from the $\pom/\pom$ and $\pom/f_2$ amplitudes may be needed to further widen the peak. In particular, the top-$\pom$ contributions might be necessary for the $\eta' \pi$ channel. 

We next investigate the mass dependence of the top-$a_2$ and top-$f_2$ amplitudes. 
In Fig.~\ref{fig:fb_data} we find that $F(\m)$ is steeper than $B(\m)$. 
The slope of the distribution is determined by the slopes of the trajectories in Eq.~\eqref{eq:trajectory} of both top and bottom exchanges, once
the angular variables have been integrated over.
The $\m$ dependence for individual amplitudes in Fig.~\ref{fig:mpieta_slope} shows this effect. A steeper slope of the intensity is observed when the bottom exchange is $\pom$.
Hence, the steeper $F(\m)$ favors a 
bottom-$\pom$, while the flatter $B(\m)$ a bottom-$f_2$.
Consequently, both the $a_2/\pom$ and $f_2/f_2$
amplitudes should be included. 

Another important feature is the  $\phi$ dependence.
In Fig.~\ref{fig:phi_fixedtheta_forward} we compare the $\phi$ dependence in the forward region of $I(\m,\Omega)$ and the top-$a_2$ amplitudes. 
We see that a single amplitude cannot reproduce the experimental distributions at all $\m$.
Therefore, we will include both $a_2/f_2$ and $a_2/\pom$  amplitudes in our fits. 

In Fig.~\ref{fig:phi_fixedtheta_backward} we make the same comparison for the backward region.
The $f_2/\pom$ and $\pom/f_2$ amplitudes do not peak at the correct position at any $\m$. On the other hand, the $f_2/f_2$  and the $\pom/\pom$ match better the data.  
As explained above, the $f_2/f_2$ amplitude is already favored by the observed $B(\m)$ slope.

In conclusion, the minimal set of amplitudes (MIN) common to both channels, consists of $a_2/\pom$, $a_2/f_2$, and $f_2/f_2$.

 Additionally, as discussed earlier,  we may extend this  set in order to take into account the width of the backward peak. 
In particular, the $\eta' \pi$ 
  peak is broader than predicted by the $f_2/f_2$ amplitude. 
Including the $\pom/f_2$ would help.
However, it may disrupt the $\phi$ distribution as shown in Fig.~\ref{fig:phi_fixedtheta_backward}. 
An option would be to include both $\pom/f_2$ and
$f_2/\pom$. 
However, as stated earlier in this Section, including more than four amplitudes makes the fits unstable. For this reason we do not include the $\pom/f_2$ amplitude in any fits.

Therefore, we are left with two options to broaden the backward peak: either $\pom/\pom$ or $f_2/\pom$. The $\pom/\pom$ amplitude allows to broaden the backward peak without  affecting much the $\phi$ dependence. It also 
would make the backward peak broader than
the $f_2/\pom$ exchange. The $f_2/\pom$ exchange shifts the $\phi$ distribution to peak below $\pi/2$, but by interfering with $f_2/f_2$ this shift may be reduced. Hence, we explore adding either the $f_2/\pom$ or $\pom/\pom$ amplitudes to the MIN set for fitting the intensities.

To summarize, the sets of amplitudes we explore are: 
\begin{itemize}
    \item[(i)] MIN, that includes the $a_2/\pom$, $a_2/f_2$ and $f_2/f_2$ amplitudes, {\it i.e.} parameter 
    set $\{c_{a_2\pom},\, c_{a_2f_2},\, c_{f_2f_2}\}$;
    \item[(ii)] MIN$+f/\pom$, with parameter set $\{c_{a_2\pom},\, c_{a_2f_2}, \, c_{f_2\pom},\, c_{f_2f_2}\}$;
    \item[(iii)] MIN$+\pom/\pom$, with parameter set $\{c_{a_2\pom},\, c_{a_2f_2},\, c_{f_2f_2},\, c_{\pom\pom}\}$.
\end{itemize}

\section{Results}\label{sec:results}
\subsection{Extended negative log-likelihood fit}

\begin{table*}
\caption{ENLL and fit parameters $\{c\}$ (in appropriate \nsgev units)
for both MVR and MCR. 
For MVR, $\mathcal L$ corresponds to the best fit found, while for MCR the value and error of $\mathcal L$ 
correspond to the mean value 
and dispersion  of the best ENLL for each pseudodataset.  
The $\mathcal{L}$ distributions are
depicted in Fig.~\ref{fig:ellh}, while the parameter distributions 
are discussed in Appendix~\ref{sec:mcrdistributions}.
}
\begin{ruledtabular}
\begin{tabular}{c|c|cc|cc|cc}
Channel    &  &    \multicolumn{2}{c|}{MIN} &   \multicolumn{2}{c|}{MIN$+f/\pom$} &   \multicolumn{2}{c}{MIN$+\pom/\pom$} \\ 
& &    MVR &   MCR &   MVR & MCR &   MVR & MCR \\   
\hline
&$\mathcal{L}\times 10^{-4}$&$-22.8$  & $-21.9\pm 0.9$& $-22.7$  & $-22.0\pm 0.9$&$-22.8$  & $-22.1\pm 0.8$\\
          &$c_{a_2\pom}$  &     $0.29$& $0.42\pm0.03$&     $0.28$&  $0.40\pm0.04$&      $0.29$&    $0.36\pm0.04$\\
          &$c_{a_2f_2}$   &     $3.67$&   $3.3\pm0.4$&     $3.70$&    $3.4\pm0.4$&      $3.59$&      $3.8\pm0.4$\\
$\eta  \pi$&$c_{f_2\pom}$  &\textemdash&   \textemdash&    $-0.20$& $-0.30\pm0.05$& \textemdash&     \textemdash \\
          &$c_{f_2f_2}$   &   $-11.82$& $-11.0\pm0.3$&    $-8.99$&     $-6.6\pm0.7$&    $-10.86$&    $-8.9\pm0.4$ \\
          &$c_{\pom\pom}$ &\textemdash&   \textemdash&\textemdash&    \textemdash&    $0.0073$& $0.0135\pm0.002$\\
\hline
&$\mathcal{L}\times 10^{-4}$&$-11.7$  & $-10.9\pm 1.0$&$-11.7$  & $-11.0\pm 1.0$&$-11.8$  & $-11.4\pm 1.0$\\
&$c_{a_2\pom}$ &  $0.16$ & $0.37\pm0.07$&     $0.16$& $0.34\pm0.05$&$0.19$ &   $0.35\pm0.05$ \\
          &$c_{a_2f_2}$   &     $1.50$&   $0.4\pm0.6$&     $1.51$&   $ 0.7\pm0.5$&$1.22$ &      $0.6\pm0.5$ \\
 $\eta' \pi$ & $c_{f_2\pom}$ &\textemdash &  \textemdash & $-0.21$ & $-0.29 \pm 0.03$ & \textemdash & \textemdash \\
           &$c_{f_2f_2}$   &   $-11.42$& $-11.0\pm0.5$&    $-7.73$&   $-5.5\pm0.7$&$-9.01$ &     $-7.1\pm0.6$ \\
           &$c_{\pom\pom}$ &\textemdash&              \textemdash&\textemdash&               \textemdash& $0.012$&$0.018\pm0.002$ \\
\end{tabular}
\end{ruledtabular}\label{tab:parameters}
\end{table*}

The contribution of each amplitude in a given set (MIN, MIN$+f/\pom$, and MIN$+\pom/\pom$) is determined  
by fitting the MVR and the MCR distributions   for each $\eta^{(\prime)}\pi$ channel independently. 

We first discuss how to fit the MVR. In each mass bin, the intensity $I(\m,\Omega)$ depends 
on two angles $\Omega=(\phi,\cos \theta)$. The continuous variables prevent us from using a standard $\chi^2$ fit. Besides, we need to take into account the fact that the total intensity is a fixed quantity.
Hence, we use an integral form of the extended negative log-likelihood function (ENLL)~\cite{Lyons:1985vx,Barlow:1990vc}:
\begin{align}
    \mathcal{L}(\{c\}) =& \sum_{i}
    \int \diff \Omega \, \left[ I_\text{Th}(\m_i,\Omega|\{c\}) \right. \nonumber \\
    &\, \left. -I_\text{Exp}(\m_i,\Omega)\log I_\text{Th}(\m_i,\Omega|\{c\}) \right], \label{eq:likelihood}
\end{align}
where the experimental intensity $I_\text{Exp}(\m,\Omega)$ is the fitting objective function computed using Eq.~\eqref{eq:int_COMPASS} with MVR, and the theoretical intensity $I_\text{Th}(\m,\Omega|\{c\})$ 
is computed from  Eq.~\eqref{eq:model}.
The experimental distributions are fitted simultaneously in 15 bins of $\{\m\}$, in the range $2.38 < \m < 2.98\gev$. Varying slightly this interval leaves the results unchanged. 
We minimize $\mathcal{L}$ using {\sc MINUIT}~\cite{minuit} to obtain the $\{c\}$ parameters weighting each theoretical amplitude. 

We note that the ENLL makes the total intensity of the model 
as close as possible to the
total intensity of data.
We remind the normalization of data is unknown, thus the $\{c\}$ cannot be directly compared to normalized couplings. The overall sign of the amplitude is also undetermined, so we fix $c_{a_2\pom}$ to be positive. As said above, we expect $c_{f_2\pom}$ to be negative.
The best fits found are reported in 
Table~\ref{tab:parameters}.
Local minima that do not follow the sign expectations were found, 
although with worse $\mathcal{L}$ than the reported best fit values.

We note that the absolute value of the parameters is not a measure of the importance of any given amplitude
contribution, because the $A$'s 
in Eq.~\eqref{eq:As} have largely different magnitudes. In particular, 
bottom-$\pom$ are much larger than bottom-$f_2$.

Fitting the MCR is more challenging. 
Each pseudodataset $j$ is fitted using Eq.~\eqref{eq:likelihood}, obtaining an independent  set of parameters $\{c\}^j$. 
We estimate the expectation value of the parameters
 by averaging over $N=10^4$ fits and the uncertainties from the
appropriate quantiles. This number of pseudodatasets allows us to obtain the probability distribution of each parameter and the correlations with $\sim 1\%$ statistical uncertainty (more details Appendix~\ref{sec:mcr}).
  
\subsection{Fit results}\label{sec:fitresults}

\begin{figure} 
    \includegraphics[width=\columnwidth]{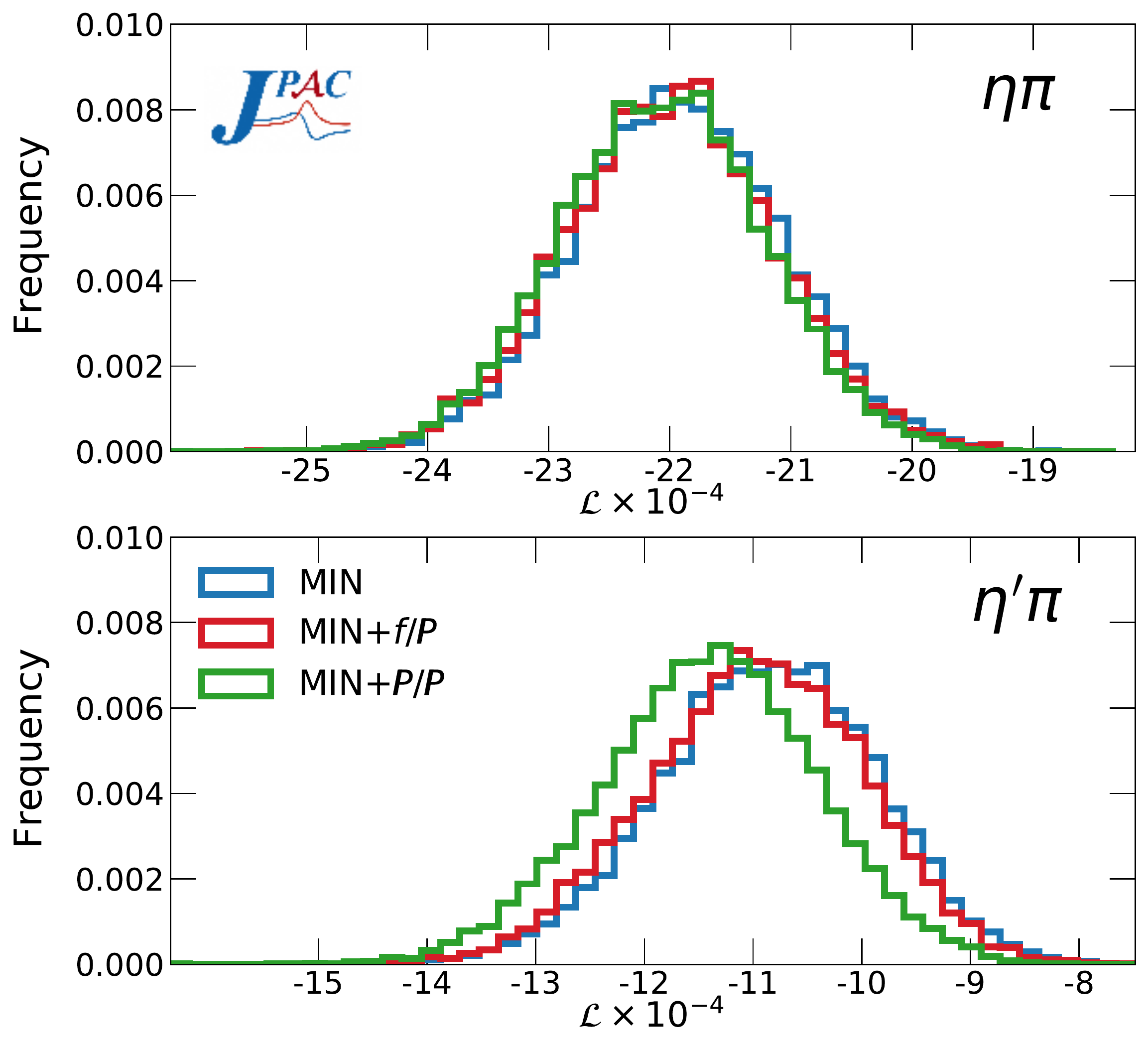} 
      \caption{ENLL $\mathcal L_\text{MCR}$ distributions for the $10^4$ bootstrap fits with the three models to the MCR
      distributions for $\eta \pi$ (upper) and $\eta' \pi$ (lower) channels.}
      \label{fig:ellh}
\end{figure}

Table~\ref{tab:parameters}
gives the value of the ENLL for the three models fitted to the MVR and MCR for both channels, as well as the resulting fit parameters. The distribution of the ENLL for the MCR fits is shown in Fig.~\ref{fig:ellh}.
We see that all the models have similar ENLL, with a nonsignificant preference for MIN$+\pom/\pom$ fit, in particular for the $\eta' \pi$ channel.
In Appendix~\ref{sec:statanalyses} we analyze this difference more systematically and conclude that, statistically, there is indeed a preference for the MIN$+\pom/\pom$ model for the  $\eta'\pi$ channel. 

In Appendices~\ref{sec:mvr_vs_mcr} and~\ref{sec:mvrfits} we compare MVR and MCR observables and fits, respectively, finding that MCR fits are more reliable. Here we summarize the results of the MCR fits and leave the MVR fit results for Appendix~\ref{sec:mvrfits}.

\subsubsection{$\eta \pi$ MCR fits} 
\label{sec:etapimcr}

\begin{figure*}
    \includegraphics[width=0.8\textwidth]{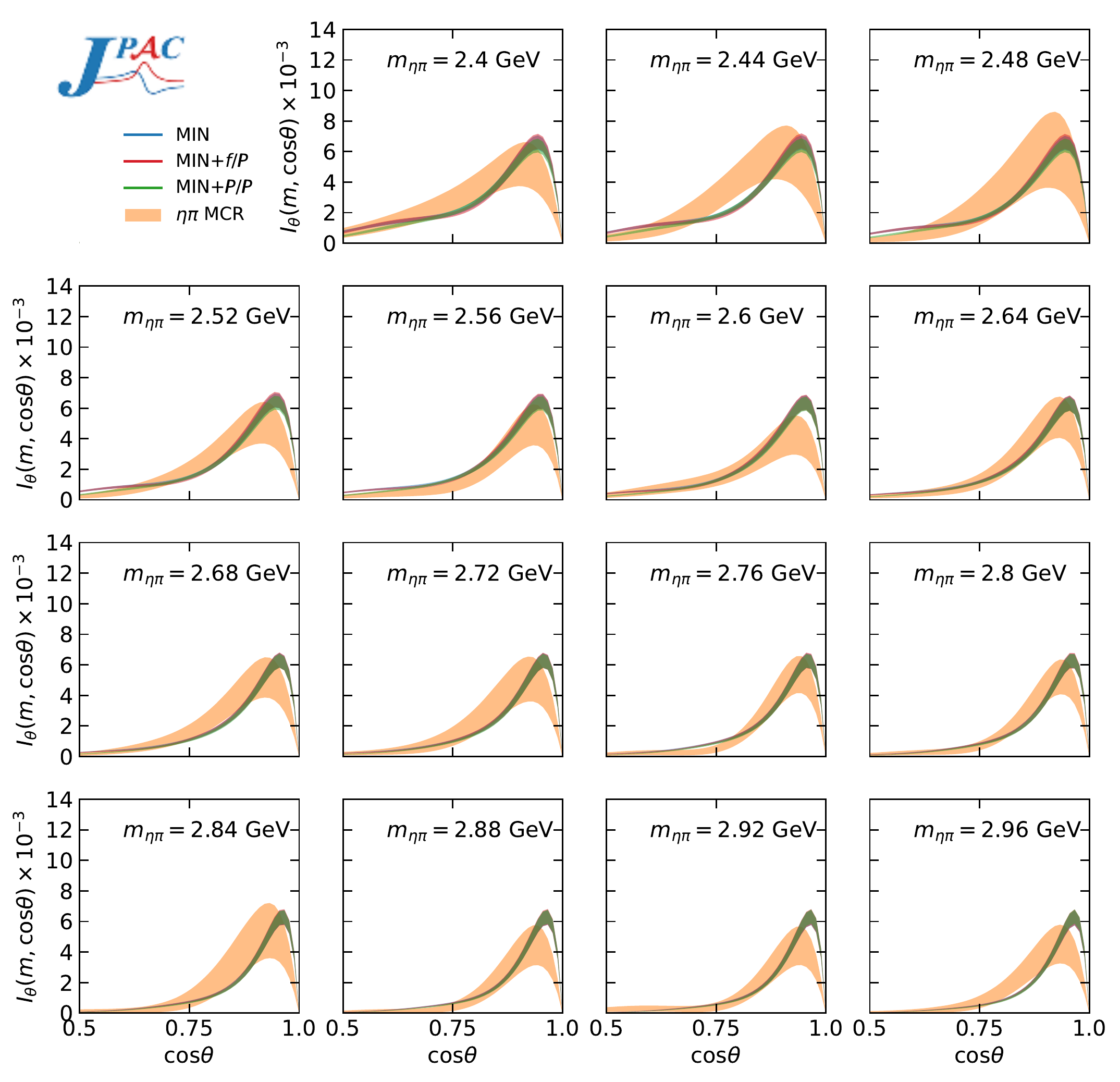}
    \caption{Experimental $I_\theta(m_{\eta \pi},\cos\theta)$ from MCR (orange) for the forward region compared to the MIN (blue), MIN$+f/\pom$ (red) and
    MIN$+\pom/\pom$ (green) fits.
    Bands correspond to the $68\%$ confidence level.
    The three model curves mostly overlap.}
     \label{fig:pieta.theta.fwd}
\end{figure*}

\begin{figure*}
    \includegraphics[width=0.8\textwidth]{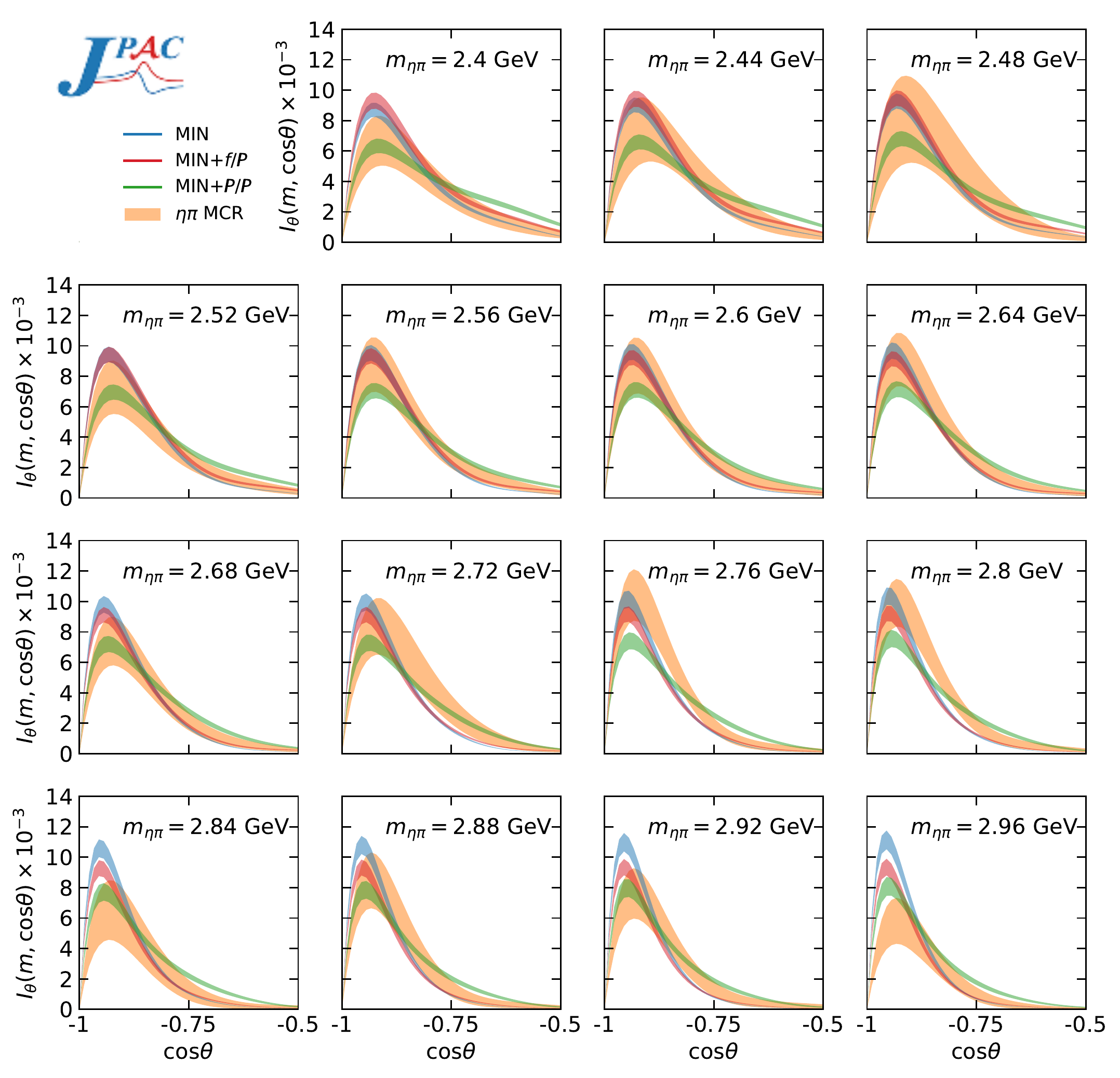}
      \caption{Same as Fig.~\ref{fig:pieta.theta.fwd} for the 
      $\eta \pi$ backward region. Model differences are now apparent.}
\label{fig:pieta.theta.bwd}
\end{figure*}

\begin{figure*}
    \includegraphics[width=0.8\textwidth]{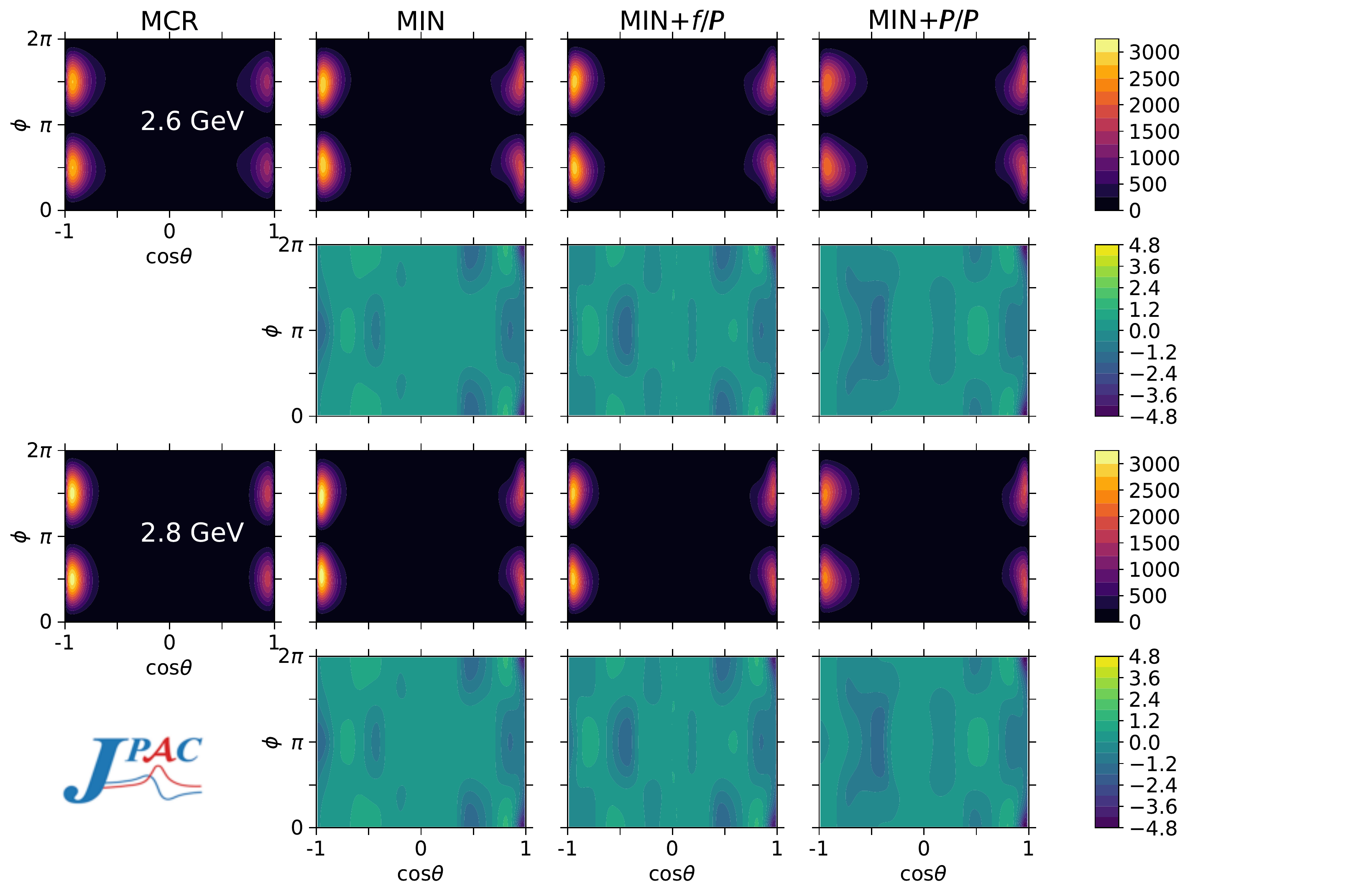}
    \caption{Examples of $\eta \pi$ density plots for $\bar{I}_\text{Exp}(m_{\eta \pi},\Omega)$, 
    $\bar{I}_\text{Th}(m_{\eta \pi},\Omega)$, and 
    $D(m_{\eta \pi},\Omega)$ defined in Eq.~\eqref{eq:difIEIT}.
    The upper two rows displays refer to $m_{\eta\pi}=2.6\gev$,
while the lower ones to $m_{\eta\pi}=2.8\gev$.
     In the second and fourth rows $D(m_{\eta \pi},\Omega)$ shows where the main discrepancies
between theory and experiment appear.
Most of its structures are located in the small and intermediate $\left|\cos \theta\right|$ regions,
where the models are expected to be less accurate. Such differences are not large. 
The largest discrepancies between theory and experiment at the 
$\cos \theta=1$, $\phi=0,2\pi$
corners. This is a consequence of the exact zero that occurs in both experiment and theory, that makes $D(m_{\eta \pi},\theta,\phi = 0)$ indeterminate.\label{fig:pieta.phitheta}}
\end{figure*}

\begin{figure*}
    \includegraphics[width=0.8\textwidth]{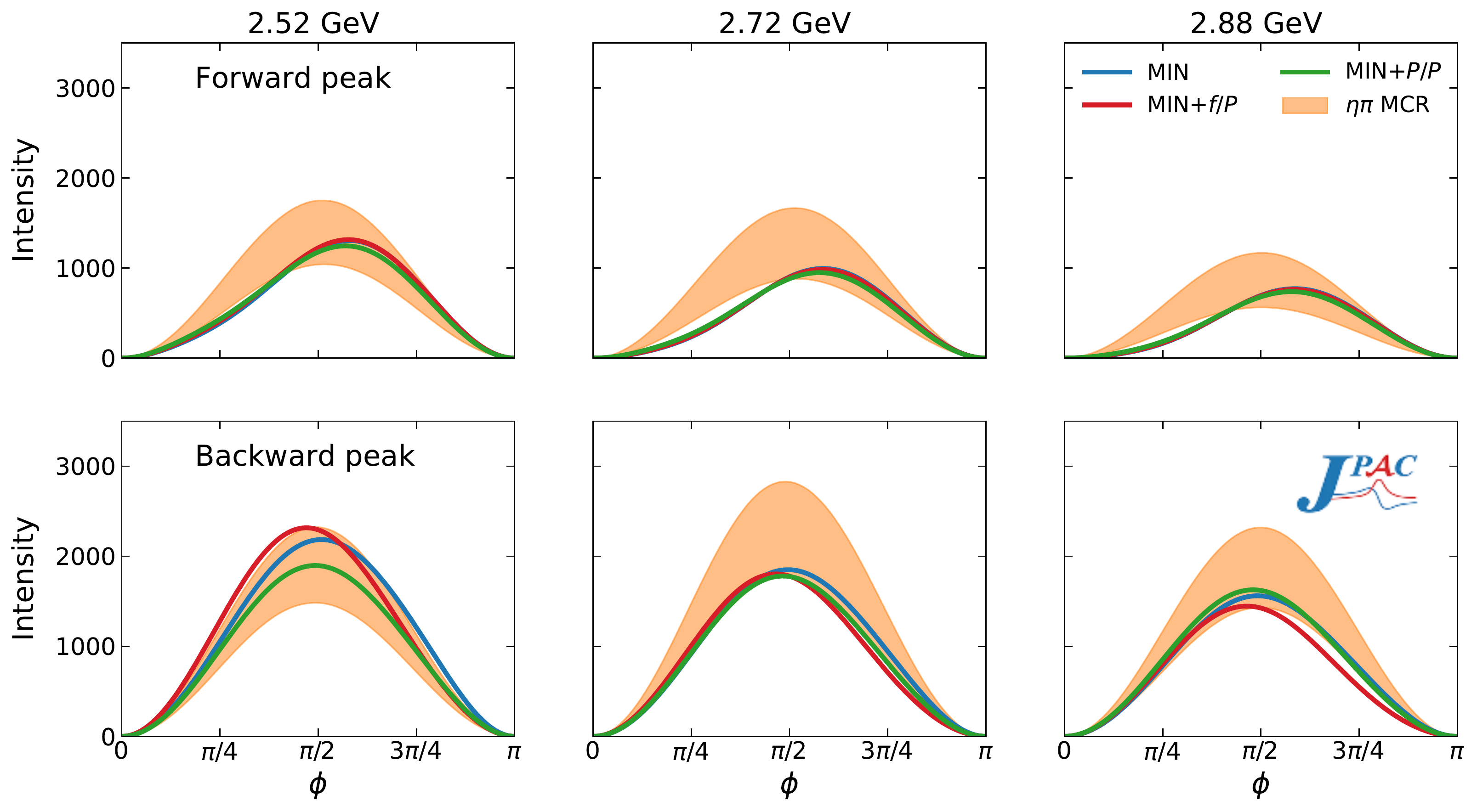} 
      \caption{
      $\phi$-dependence of the
      experimental $I(m_{\eta \pi},\Omega)$ from MCR (orange)  compared to the MIN (blue), MIN$+f/\pom$ (red) and
      MIN$+\pom/\pom$ (green) fits.
     at three fixed energies for
      $\cos \theta = 0.866$ (close to the forward peak, upper row) and
       $\cos \theta = -0.866$ (close to the backward peak, lower row). 
      Due to the reflection symmetry shared by the model and the intensity, we only show $\phi\in[0,\pi]$.
}\label{fig:phi_fixedtheta_pieta}
\end{figure*}

The three models give consistent values for $c_{a_2\pom}$ and $c_{a_2f_2}$, providing
almost identical descriptions of the forward peak. 
This can be appreciated in Fig.~\ref{fig:pieta.theta.fwd},
where the experimental
$I_\theta(m_{\eta \pi},\cos \theta)$ MCR
for the fast-$\eta$ region is compared
to the three models 
for all the fitted $m_{\eta \pi}$ bins. 
The three models agree very well.
Figure~\ref{fig:pieta.theta.bwd} shows the same results
for the fast-$\pi$ region.
Here the differences among models can be appreciated. 
As expected from Fig.~\ref{fig:diags_eta}, 
the MIN+$\pom/\pom$ provides a wider peak, 
and, since the normalization is fixed in a ENLL fit,   
the maximum intensity at the peak is smaller than the MIN and
MIN+$f/\pom$ results.
The latter two fits are similar, with their 
 uncertainty bands overlapping,  
except in the highest $m_{\eta \pi}$ bin. 
We note that for some energies the MIN+$\pom/\pom$
provides a better description  of the experimental
distribution, 
while for others the MIN and MIN+$f/\pom$ fits look better.

Further insight can be obtained by examining the three-dimensional $I(m_{\eta \pi},\Omega)$ distributions.
We define 
\begin{align}
D (m_{\eta\pi},\Omega) = 
\frac{ \bar{I}_\text{Exp}(m_{\eta\pi},\Omega) - \bar{I}_\text{Th}(m_{\eta\pi},\Omega)} 
{\sqrt{ \left[ \Delta I_\text{Exp}(m_{\eta\pi},\Omega)  \right]^2 
+ \left[ \Delta I_\text{Th}(m_{\eta\pi},\Omega) \right]^2 }} \, ,
\label{eq:difIEIT}
\end{align}
where $\bar I$ and $\Delta I$ are the mean and dispersion of the experimental and theoretical distributions as obtained from MCR.\footnote{We do not take into account the fact that both the theoretical and experimental distributions are evaluated out of the same pseudodatasets, and therefore correlated. However, this still gives a qualitative description of the discrepancy between theory and experiment. }
This quantifies point-by-point
how similar the MCR and the theoretical distributions are.
Figure~\ref{fig:pieta.phitheta} shows these distributions for the $\eta \pi$
channel  
 in two mass bins. 
Figure~\ref{fig:phi_fixedtheta_pieta} provides the $\phi$ distribution at the
backward and forward peaks for three energies.
Comparing the model and data distributions, one concludes that 
the former  has more structure in the forward region. The experimental peak has two 
symmetric blobs in $\phi$ while the theory is rather asymmetric.  As
shown in Fig.~\ref{fig:phi_fixedtheta_forward}, this is due to the asymmetry in $\phi$ of the top-$a_2$ amplitudes. 
We remind that the symmetry of the experimental $\phi$ distribution is exact and stems from Eq.~\eqref{eq:int_COMPASS}.
This symmetry is not imposed in the model, and is approximately reached by having both top-$a_2$ amplitudes interfering. 
All the three models peak at roughly the correct $\phi=\pi/2$ and $3\pi/2$. The situation is different for the backward peak. The MIN fit peaks 
slightly below (above) the experimental value of $\phi=3\pi/2$ ($\phi=\pi/2$). Hence, we do not favor the MIN model, {\it i.e.}\ the $f_2/f_2$ amplitude is not enough to reproduce the $\phi$ dependence of the fast-$\pi$ region. 

From the bootstrap fits, we can study 
the parameter distributions and their 
correlations, summarized in Appendix~\ref{sec:mcrdistributions}.
The correlations confirm that the
fast-$\pi$ and fast-$\eta$ amplitudes 
are essentially independent.
The parameters are generally well determined and exhibit Gaussian behavior, except the $\pom/\pom$ coefficient $c_{\pom\pom}$ that has a bimodal distribution.

Finally, although including a bottom-$\pom$ amplitude is necessary to describe the backward region, data do not show a clear preference for either MIN$+f/\pom$ or MIN$+\pom/\pom$. Since the two models point to different values for the  $f_2/f_2$ coupling, the latter cannot be determined unambiguously either.
Currently, we do not have enough precision in the data to determine the contributions from the individual exchanges in the backward region. 

\subsubsection{$\eta' \pi$ MCR fits}\label{sec:etaprimepimcr}

\begin{figure*}
    \includegraphics[width=0.8\textwidth]{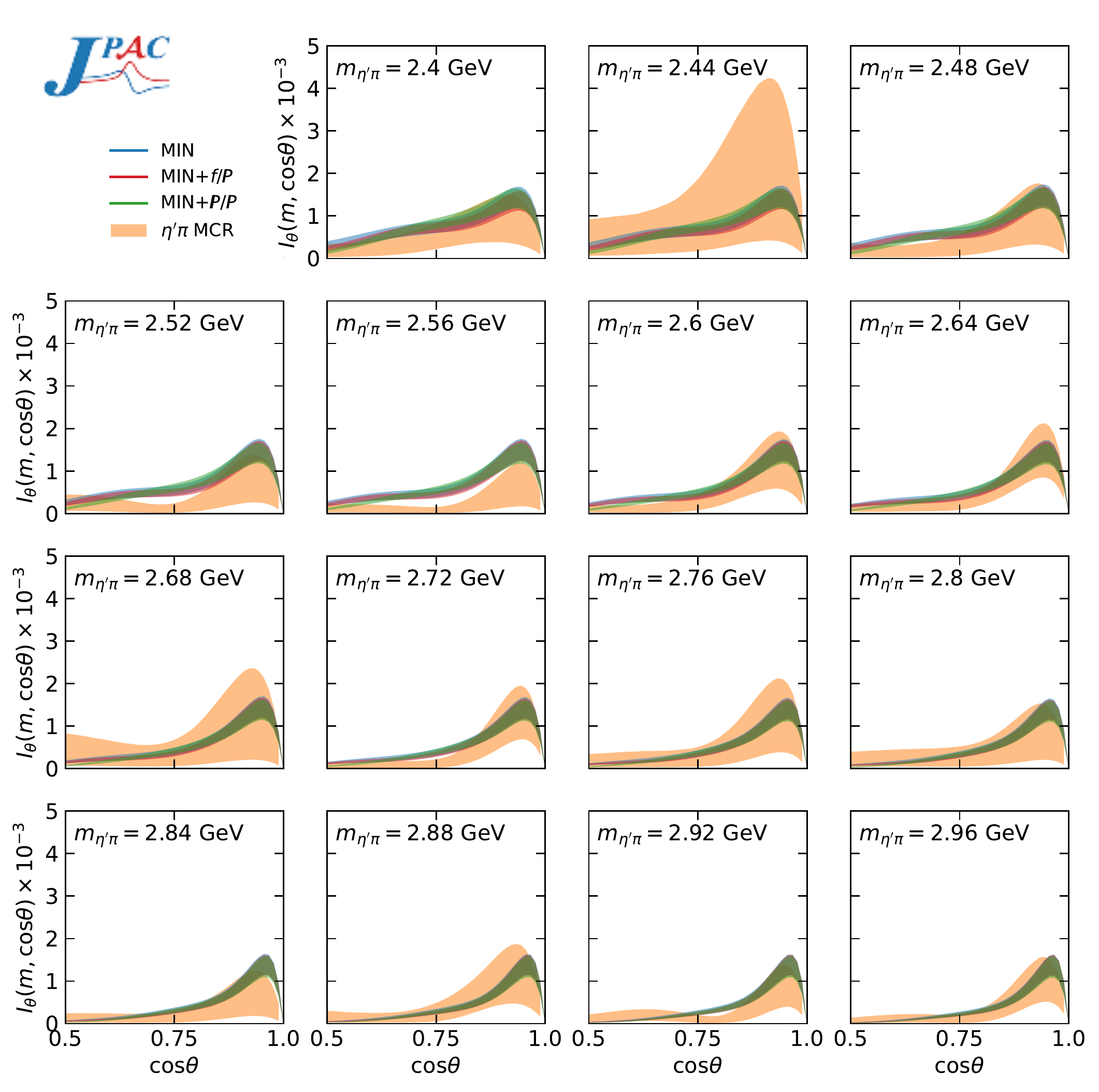}
      \caption{Same as Fig.~\ref{fig:pieta.theta.fwd} for 
      $\eta' \pi$.}
     \label{fig:pietap.theta.fwd}
\end{figure*}
\begin{figure*}
    \includegraphics[width=0.8\textwidth]{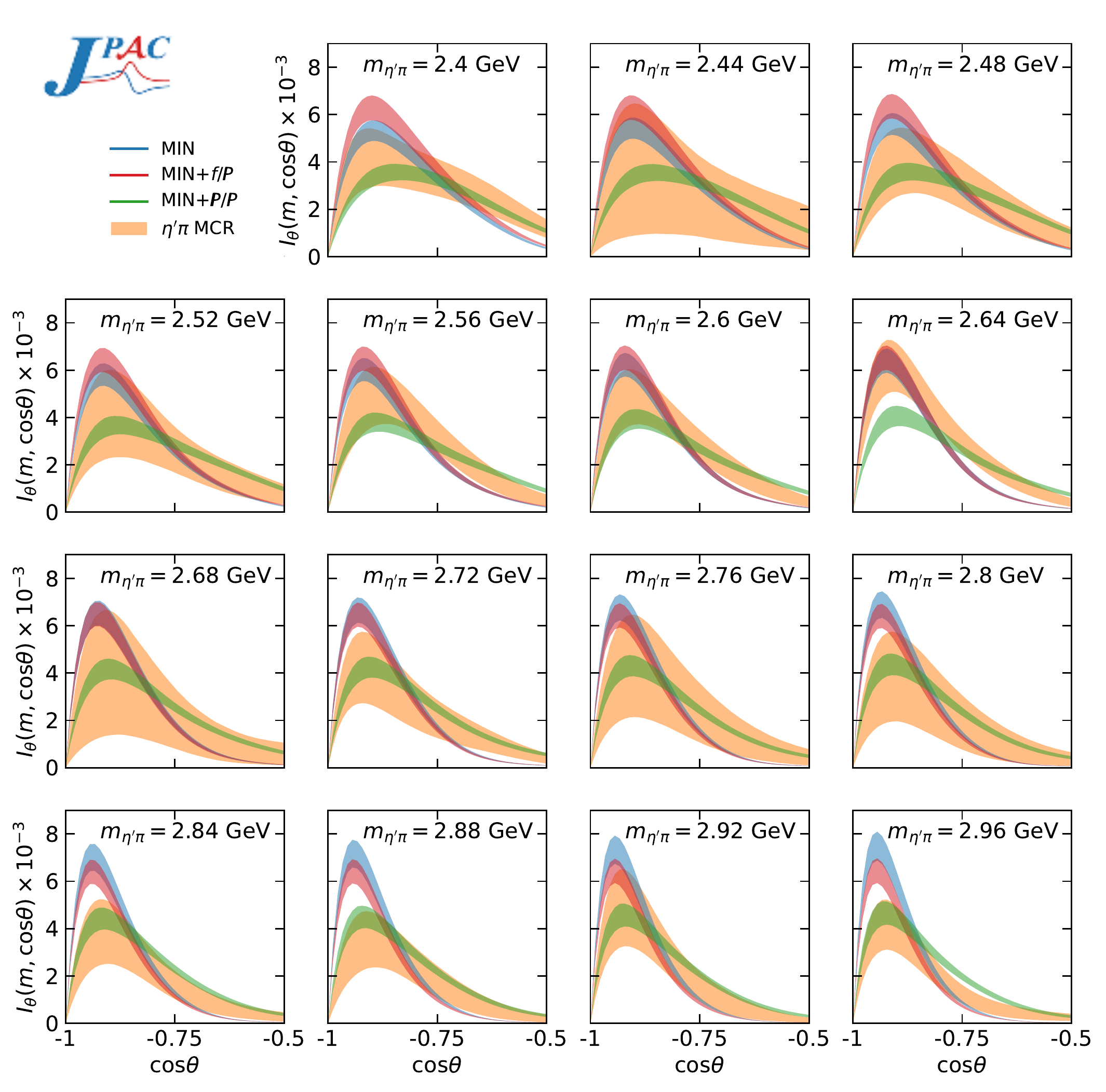}
      \caption{Same as Fig.~\ref{fig:pieta.theta.bwd} for 
      $\eta' \pi$.} \label{fig:pietap.theta.bwd}
\end{figure*}

\begin{figure*}
    \includegraphics[width=0.8\textwidth]{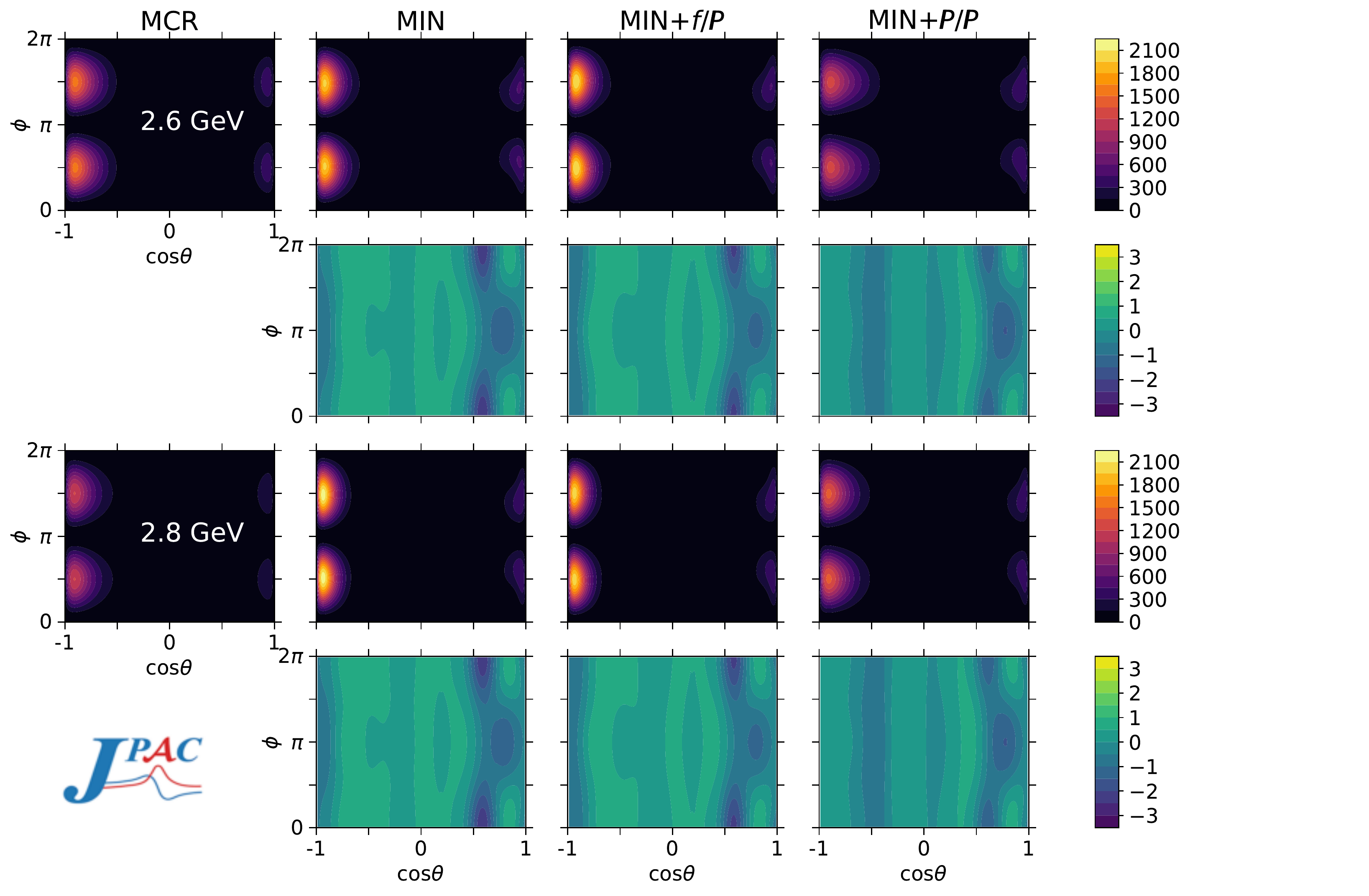}
    \caption{Same as Fig.~\ref{fig:pieta.phitheta} for the $\eta' \pi$
    channel.} \label{fig:pietap.phitheta}
\end{figure*}

\begin{figure*}
    \includegraphics[width=0.8\textwidth]{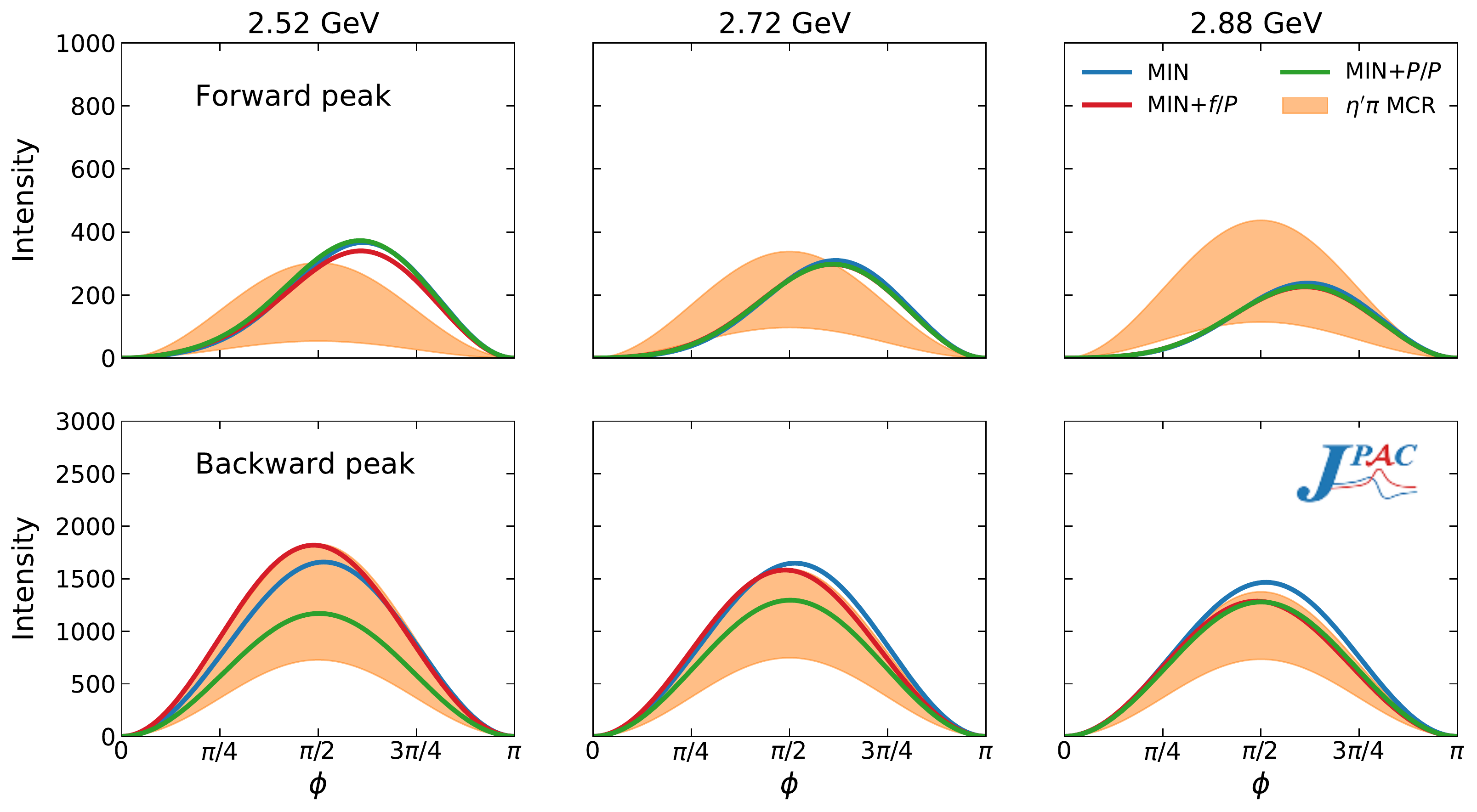} 
      \caption{
       Same as Fig.~\ref{fig:phi_fixedtheta_pieta}
       for $\eta'\pi$.
      }\label{fig:phi_fixedtheta_pietap}
\end{figure*}

The fit parameters for the three models
are presented in the MCR columns of Table~\ref{tab:parameters}. 
As for $\eta \pi$, the three models give consistent values for 
$c_{a_2\pom}$ and $c_{a_2f_2}$, providing
almost identical descriptions of the forward peak. However, 
$c_{a_2f_2}$ is compatible with zero at a $2\sigma$ level. 
This suggests larger level of 
EXD  breaking in the $\eta' \pi$ channel.
Figures~\ref{fig:pietap.theta.fwd} and~\ref{fig:pietap.theta.bwd},
compare the experimental
$I_\theta(m_{\eta' \pi},\cos \theta)$ with the three models in the forward and backward regions, respectively. 
The models completely agree in the forward region, while
the MIN+$\pom/\pom$ provides a wider backward peak, in better
agreement with the data. 

The three-dimensional distributions 
for $\eta'\pi$ are shown in Figure~\ref{fig:pietap.phitheta} 
for the three models and MCR at $m_{\eta'\pi}=2.60$ and $2.80\gev$.
Figure~\ref{fig:phi_fixedtheta_pietap} provides the $\phi$ distribution at the
backward and forward peaks for three energies.
As for $\eta\pi$, the MIN does not peak at the correct value of $\phi$ in the backward region.
Results and conclusions are qualitatively similar to the $\eta\pi$ channel.
In particular, the preference for a MIN$+\pom/\pom$ is clear. This points to a large affinity of $\eta'$ to gluons as discussed in the literature~\cite{Bass:2018xmz}.

\subsubsection{Forward and backward intensities
and asymmetry}

\begin{figure}
    \includegraphics[width=\columnwidth]{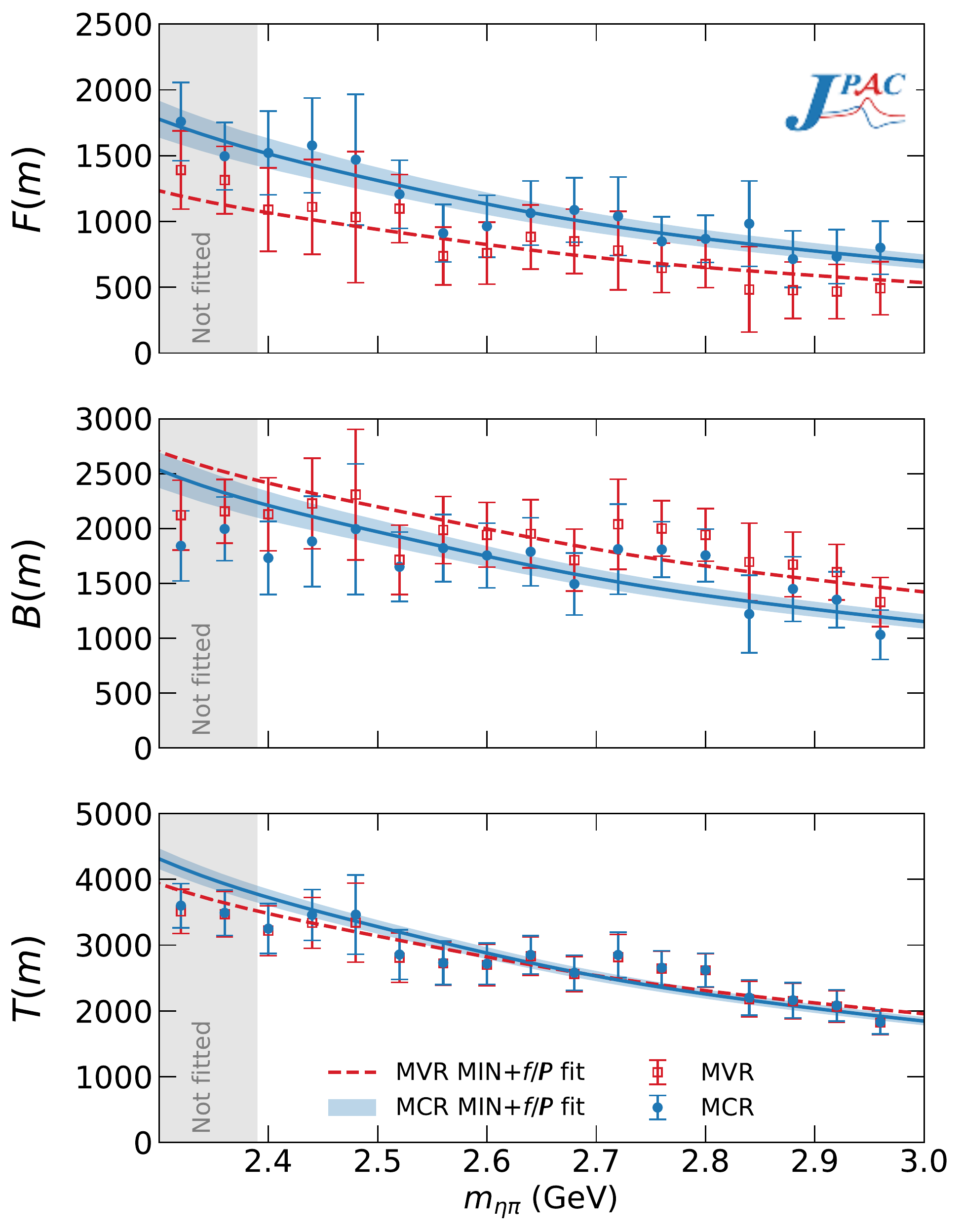}
      \caption{Forward
      (upper), backward
      (center), and total
      (lower) intensities for the $\eta \pi$ channel as defined in Eq.~\eqref{eq:fbintensities} for the MCR and MVR and their respective MIN$+f/\pom$ fits.
      We show as experimental MVR error bars the same computed from the MCR.}
      \label{fig:pieta.fb.fit}
\end{figure}

\begin{figure}
    \includegraphics[width=\columnwidth]{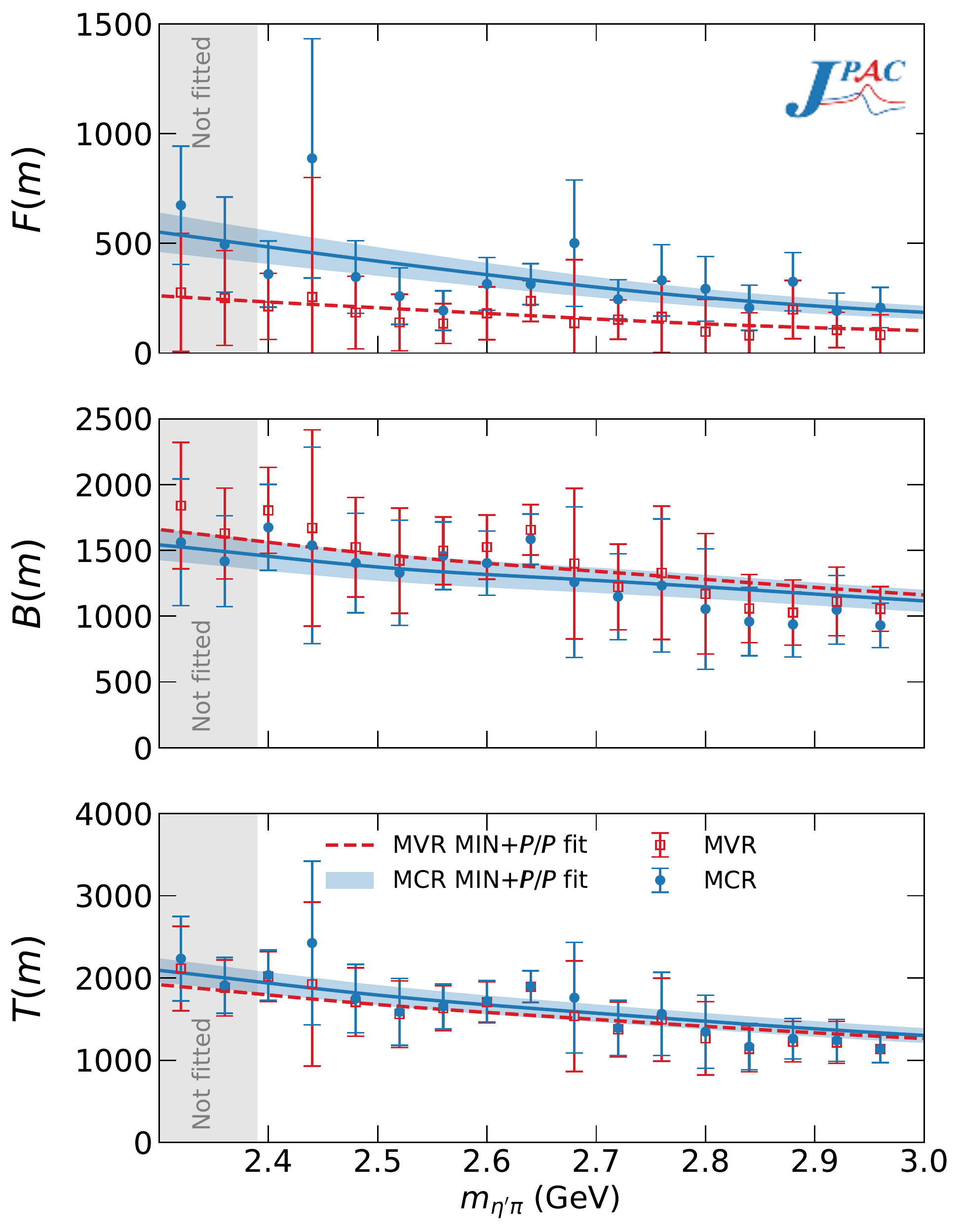}
      \caption{Same as Fig.~\ref{fig:pieta.fb.fit} for the $\eta' \pi$
      data and the MIN$+\pom/\pom$ model.}
      \label{fig:pietap.fb.fit}
\end{figure}

\begin{figure*}
    \begin{tabular}{cc}
    \includegraphics[width=0.45\textwidth]{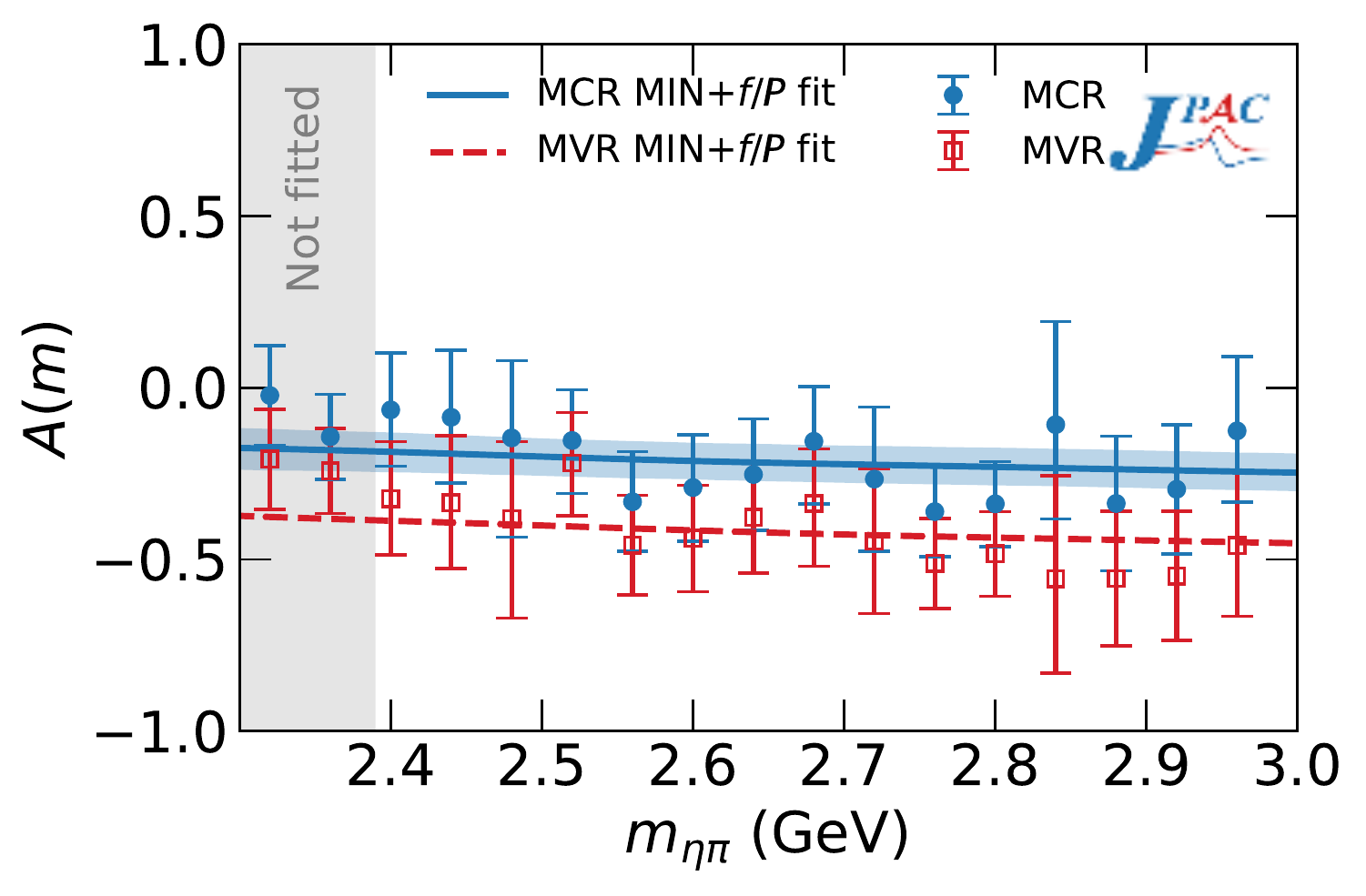} &
    \includegraphics[width=0.45\textwidth]{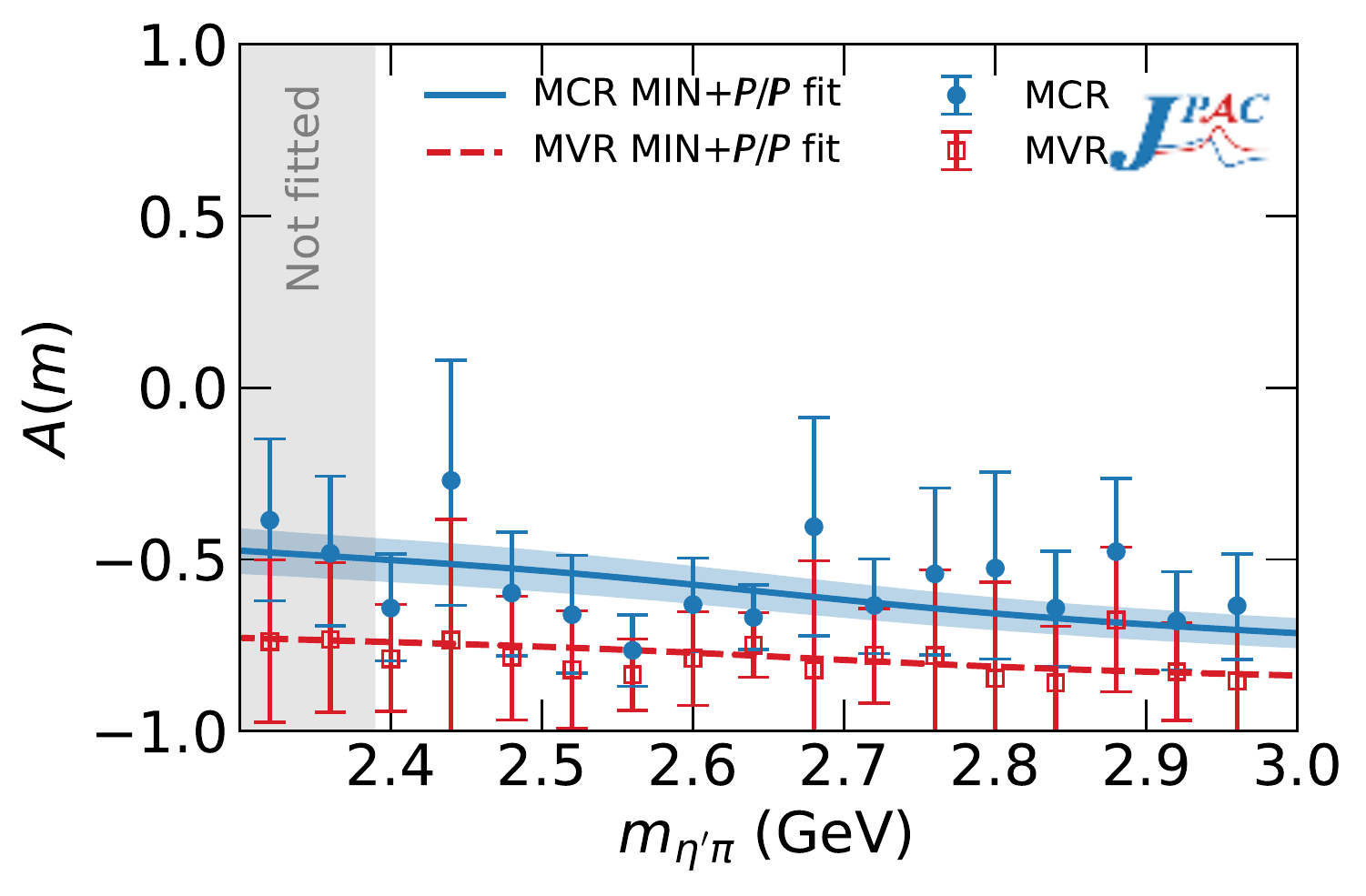}
    \end{tabular}
      \caption{Forward-backward intensity asymmetry as defined in Eq.~\eqref{eq:fbintensities} for $\eta \pi$ (left) and $\eta' \pi$ (right) for both MCR and MVR.
      We show as experimental MVR error bars the same computed from the MCR.
      }
      \label{fig:asy.fit}
\end{figure*}

Figures~\ref{fig:pieta.fb.fit} and~\ref{fig:pietap.fb.fit}
show  the forward, backward,
and total intensities. 
We show that 
MIN+$f/\pom$ for $\eta\pi$ 
and MIN+$\pom/\pom$ for $\eta' \pi$  
reproduce all the intensities rather  well. 

We also note that the integrated forward intensity 
is systematically larger for the MCR.
The opposite is true for 
the the backward one, which is
systematically larger for the MVR.
Consequently, the forward-backward asymmetry $A(\m)$,
defined in Eq.~\eqref{eq:intensities_asy},
is, in absolute value, larger for the MVR than for the MCR, as shown
in Fig.~\ref{fig:asy.fit}.
The existence of the asymmetry is a consequence of the
odd (exotic) partial waves contribution.
Taking into account the uncertainties in the 
partial waves makes the asymmetry less acute, 
but it is still sizeable and negative
for both channels.
The asymmetry is larger for the $\eta'\pi$ reaction, 
making this channel appropriate to search for hybrid candidates~\cite{Dudek:2011bn,Meyer:2015eta,Bass:2018xmz}.

\section{Constrained partial wave analysis}
\label{sec:cPW}

\begin{figure*}
    \includegraphics[width=0.8\textwidth]{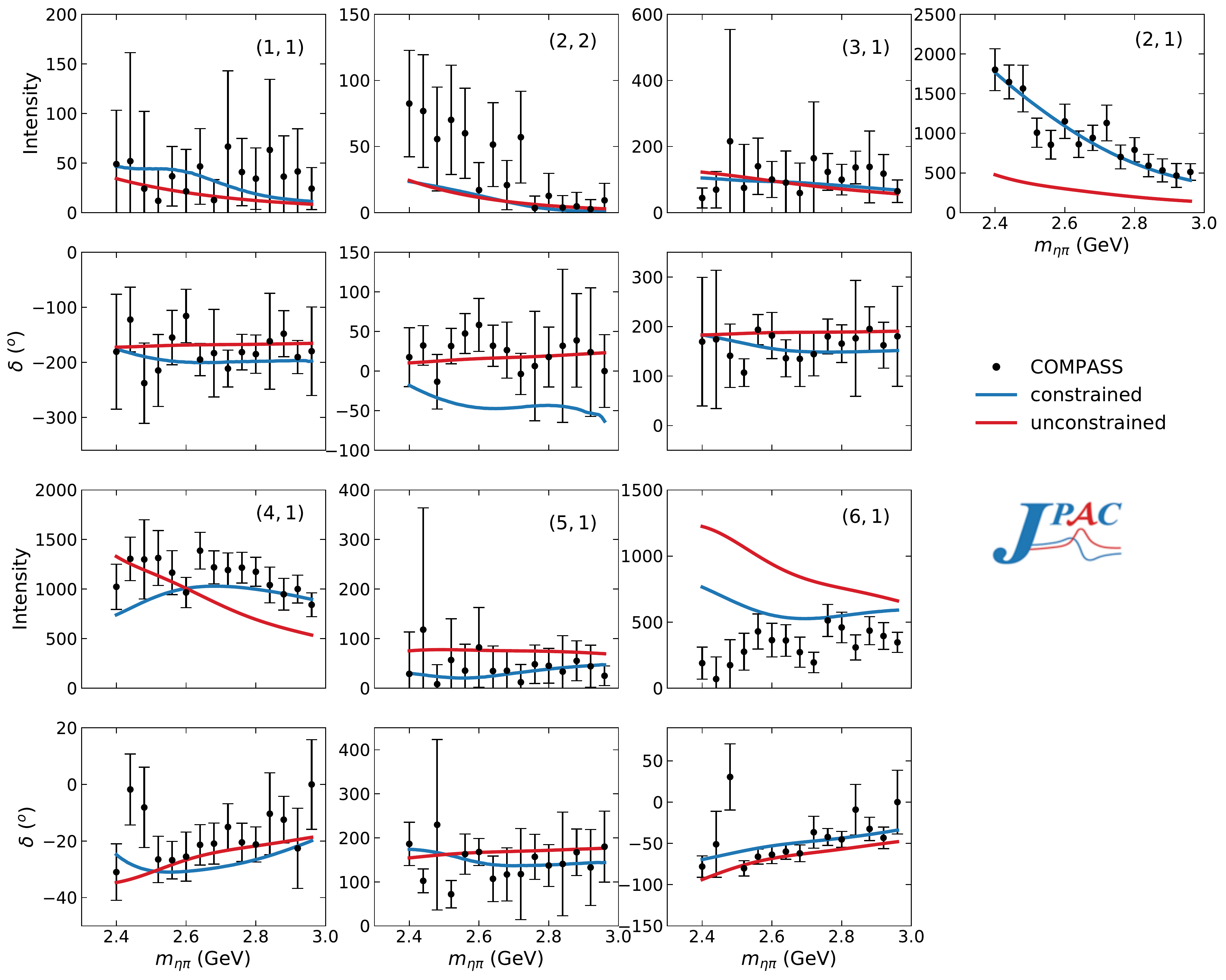}
    \caption{Constrained (red) and unconstrained (blue) $\eta\pi$ partial waves from Eqs.~\eqref{eq:mpw} and~\eqref{eq:likelihoodcpw}
    for the MIN+$f_2/\pom$, compared with the COMPASS data (black) . 
    Partial waves intensities are labeled as $(L,M)$
    and the corresponding phases $\delta$ are presented below them. In this convention, $\delta_{21}\equiv 0$.}
      \label{fig:cPW.etapi}
\end{figure*}

\begin{figure*}
    \includegraphics[width=0.8\textwidth]{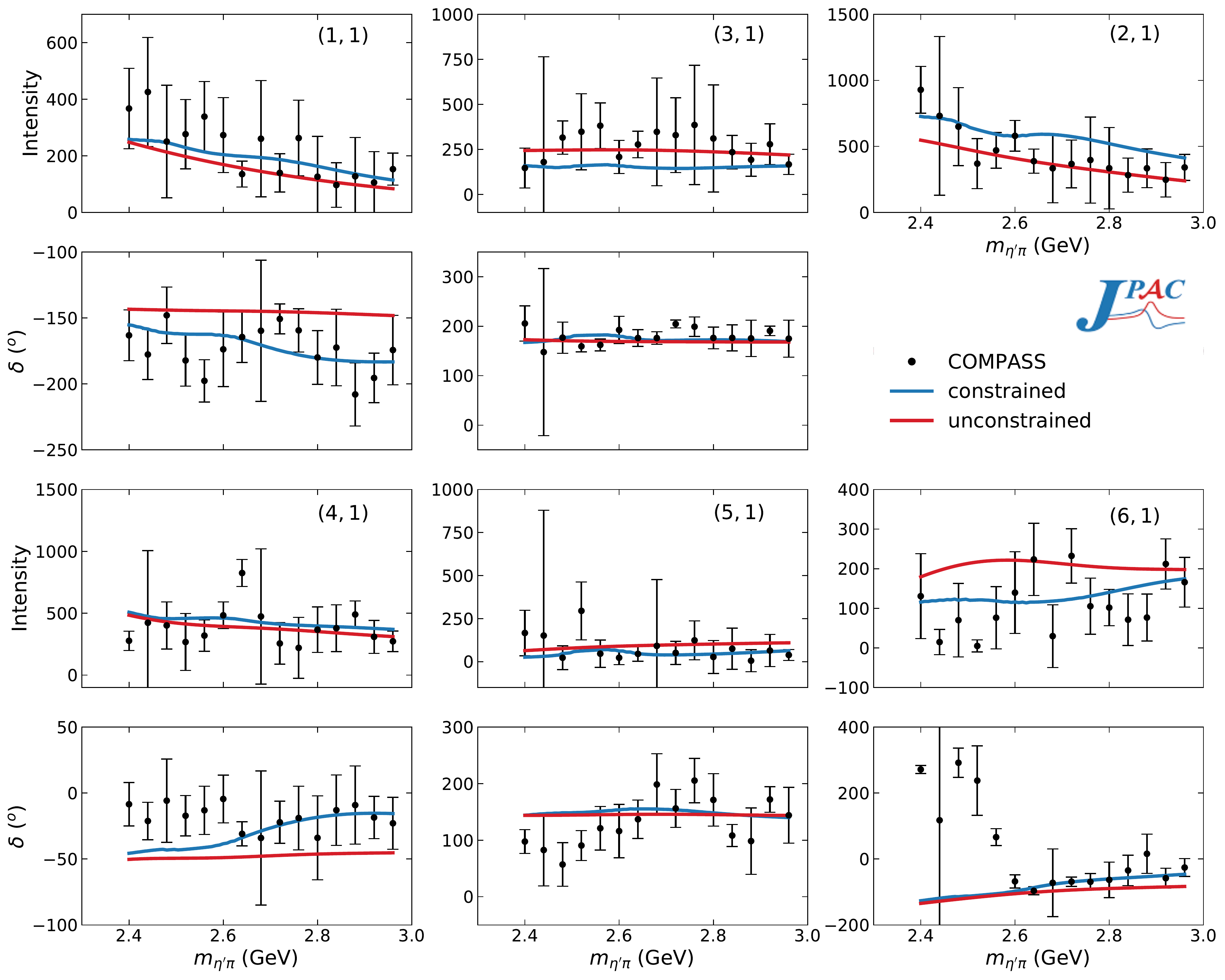}
    \caption{Same as Fig.~\ref{fig:cPW.etapi} for the $\eta' \pi$ system and the MIN+$\pom/\pom$ fit.}
    \label{fig:cPW.etappi}
\end{figure*}

As discussed earlier in Section~\ref{sec:truncation}, the  
$\left[f^+_{LM}(\m)\right]_\text{Exp}$ 
amplitudes extracted by COMPASS are 
not exactly genuine partial waves. 
It is rather a parametrization that minimizes the ENLL 
estimator used to fit the actual event distributions. 
The ENLL fit makes a finite set of amplitudes
reproduce the total intensity.  
Hence, any contribution from higher partial waves 
gets redistributed into the set included in the fit.
Our model contains an infinite number of partial waves,
which leads to a mismatch between the model partial waves and the
COMPASS ones.
The comparison can still be done if
we project the model onto partial waves applying the same constrained procedure implemented by COMPASS. For simplicity, we consider $I_\text{Th}(\m,\Omega)$ of Eq.~\eqref{eq:model}, as obtained from MVR.
We follow the conventions in Eq.~\eqref{eq:int_COMPASS}.
For each energy bin $\m_i$,
we extract the constrained partial waves (cPW) 
 by minimizing the ENLL estimator
\begin{align}
    \mathcal{L}(\{f_i\})=&
    \int \diff \Omega \, \left[ I_\text{cPW}(\m_i ,\Omega\,|\{f_i\}) \right. \nonumber \\
    &\left. -I_\text{Th}(\m_i,\Omega)\log I_\text{cPW}(\m_i,\Omega\, |\{f_i\}) \right]\, ,
    \label{eq:likelihoodcpw}
\end{align}
where $I_\text{cPW}$ is given by
\begin{align} \label{eq:int_COMPASSpw}
    I_\text{cPW}(\m_i, \Omega\,|\{f_i\}) = \left|\sum_{L,M} 
    \left[f^+_{LM}(\m)\right]_\text{cPW} \Psi^{+}_{LM}(\Omega) \right|^2\, .
\end{align}
The $L,M$ employed are 
the same truncated set as COMPASS. 

The truncated set of partial waves suffers from the problem of discrete ambiguities, the so-called Barrelet zeros~\cite{Barrelet:1971pw}.
The intensity $I_\text{cPW}(\m_i, \Omega|\{f_i\})$ with $L=1,\dots,6$; $M=1$ is identical for $2^5$ different sets of parameters $\{f_i\}$, leading to 32 different minima of the $\mathcal{L}(\{f_i\})$ function.
These are exact degeneracy for $\eta'\pi$.
While the presence of $M=2$ in $\eta \pi$ 
resolves the exact degeneracy of the solutions.
However, these solutions remain as nearly-indistinguishable local minima due to the small size of the $M=2$ components for $\eta\pi$. We select the solution $\{f_i\}$ the closest to the COMPASS $f^+_{LM}(\m_i)$ values.

The unconstrained partial waves (uPW) are computed using
 \begin{align}
 \left[f^+_{LM}(\m)\right]_\text{uPW} = \sqrt{k(\m)} \int \diff\Omega \, 
  \, A_\text{Th}(\m,\Omega)
 \, \Psi^{+}_{LM}(\Omega)\, , \label{eq:mpw}
 \end{align}

The cPW, uPW and COMPASS waves
are shown in Figs.~\ref{fig:cPW.etapi} and~\ref{fig:cPW.etappi}.
As anticipated, the cPW agree with the COMPASS data very well,
while the uPW can be quite different.
Indeed, the truncation of uPW to the COMPASS set would reduce the integrated intensity  to $86\%$  for $\eta\pi$ and $95\%$ for $\eta'\pi$.
This is in agreement with the expectations from 
Figs.~\ref{fig:diags_eta} and~\ref{fig:diags_etap},
where truncation effects were shown to be critical for $\eta\pi$.
The dominant $(2,1)$ intensity
in the $\eta \pi$ channel is noteworthy.
The unconstrained wave is very small compared to the COMPASS one,
but the cPW matches the data, showing how the
truncation makes low-lying partial waves to absorb
the intensity of higher waves.

\section{Summary and conclusions}
\label{sec:conclusion}

We studied the COMPASS data on the 
$\pi^-p\to\eta^{(\prime)}\pi^- \,p$ reactions
for $\m > 2.38\gev$ where the dynamics is expected to be dominated by Regge phenomenology. We considered a double-Regge model composed of up to six amplitudes that account for the possible top/bottom Regge exchanges. In particular, we included  $a_2/\pom$ and $a_2/f_2$ to describe the fast-$\eta$ (forward) region, and $f_2/\pom$, $f_2/f_2$, $\pom/\pom$, $\pom/f_2$ for the fast-$\pi$ (backward) region.

The COMPASS collaboration reported 
partial waves extracted from data 
under the assumption that only seven (six) 
partial waves contributed to the $\eta \pi$ ($\eta' \pi$) channel. This is justifiable in the resonance region, {\it i.e.} $m_{\eta^{(\prime)}\pi} \lesssim 2\gev$. For higher energies, the number of relevant partial waves increases. 
Our Regge model is not based on a partial wave expansion and therefore implicitly includes all partial waves. Truncating to the set of waves used
by COMPASS is not appropriate for this energy region, as in our model the discarded higher partial waves amount to a nonnegligible contribution to the intensities. 
Nevertheless, we reconstructed the total intensities 
from the COMPASS partial waves and fitted with our double-Regge model. 
We found that the $\eta \pi$ intensity can be well
described with four amplitudes, $a_2/\pom$, $a_2/f_2$, $f_2/f_2$, and either $f_2/\pom$ or $\pom/\pom$. The inclusion of either bottom-$\pom$ amplitude is necessary to describe the forward region, but the data do not show a clear preference for either $f_2/\pom$ or $\pom/\pom$ amplitudes. For this reason, we could not disentangle the contributions from the individual exchanges in the backward direction.  
In  the $\eta'\pi$ channel, we  found that the best model to reproduce the data consists of $a_2/\pom$, $a_2/f_2$, $f_2/f_2$, and $\pom/\pom$ amplitudes. 
The $\pom/\pom$ contribution is necessary to describe
the data and points to a large gluon affinity of the $\eta'\pi$ system, potentially related to the existence of hybrid mesons. This is also consistent with the observed breakdown of exchange degeneracy between $a_2$ and $f_2$ in $\eta'\pi$ production. 

The importance of the bottom-$f_2$ exchange,
as shown in the slope of the integrated backward intensity,
contradicts the common lore that, at COMPASS energies, the $\eta^{(\prime)}\pi$
pairs are produced via $\pom$ exchange only, at least for this range of $m_{\eta^{(\prime)}\pi}$.

A consequence of having an amplitude model that  contains an infinite number of partial waves is that these cannot match the truncated waves from COMPASS. 
To bridge this apparent contradiction, we performed a constrained partial wave analysis of the model, using 
the same procedure as COMPASS. We found that these waves indeed agree well with the COMPASS ones.
This proves the importance of studying the full amplitude rather than a truncated partial wave decomposition once the double-Regge regime is reached.

\begin{acknowledgments}
We thank COMPASS Collaboration for numerous discussions.
\L.B. acknowledges the financial support and hospitality of the Theory Center at  Jefferson Lab and Indiana University.
This work was supported by 
Polish Science Center (NCN) Grant No.~2018/29/B/ST2/02576,
PAPIIT-DGAPA (UNAM, Mexico) Grant No.~IN106921,
CONACYT (Mexico) Grant No.~A1-S-21389,
Ministerio de Educaci\'on, Cultura y Deporte (Spain)
Grant No.~PID2019-106080GB-C21,
and U.S. Department of Energy Grants 
No.~DE-AC05-06OR23177 and No.~DE-FG02-87ER40365.
V.M. is a Serra H\'unter fellow and acknowledges support 
from the Community of Madrid through 
the Programa de Atracci\'on de Talento Investigador 2018-T1/TIC-10313.
A.P. has received funding from the European Union's Horizon 2020 research and innovation programme under the Marie Sk{\l}odowska-Curie grant agreement No.~754496.
\end{acknowledgments}

\appendix
\section{Kinematics}
\label{sec:kin}
\begin{figure}
    \centering
    \includegraphics[width=0.45\textwidth]{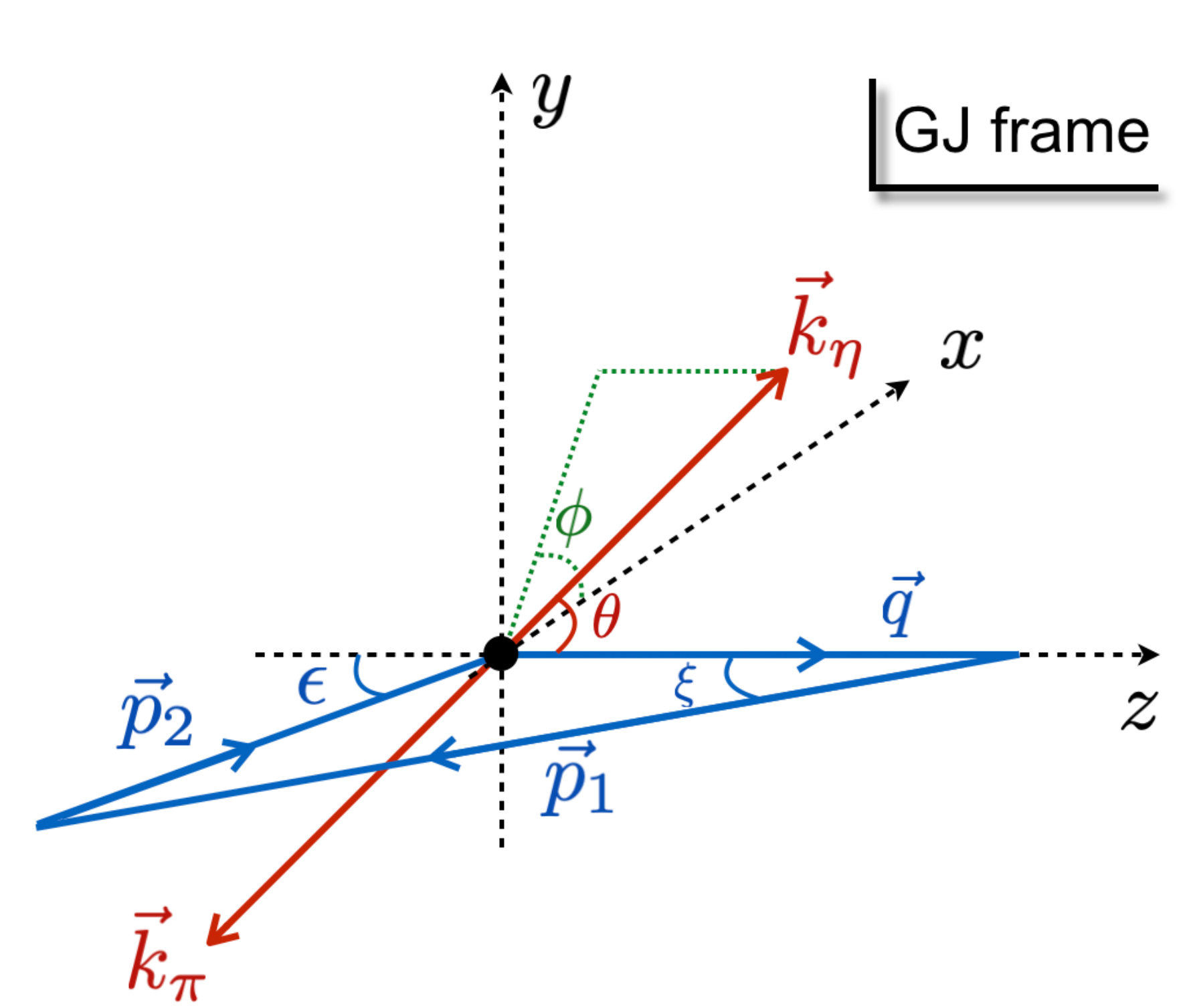}
    \caption{GJ 
    reference frame.
    See text in Appendix~\ref{sec:kin} for the definition of the 
    variables and vectors.}
    \label{fig:GJframe}
\end{figure}

The momenta of the reaction in Eq.~\eqref{reaction1} in the GJ frame are represented in Fig.~\ref{fig:GJframe}. In this frame, 
\begin{subequations}
\begin{align}
    q & = (E_q, \mathbf{q}) \, ,& 
    p_{1,2} & = (E_{1,2}, \mathbf{p}_{1,2}) \, , \\
    k_{\eta} & = (E_{\eta}, \mathbf{k}) \, , &
    k_{\pi} & = (E_{\pi}, -\mathbf{k}) \, .
\end{align}
\end{subequations}

In the $\eta\pi$ center-of-mass, the $\hat{z}$-axis is along the beam and the $\hat{y}$-axis, perpendicular to the production plane, is parallel to $\mathbf{q}\times\mathbf{p_2}$.

The particle energies can be expressed in terms of invariants as 
\begin{subequations}
\begin{align}
    E_q &=(m^2_{\eta\pi}-t_p+m_\pi^2)/2\m_{\eta\pi}, \\
    E_1 &=(s-m_p^2+t_p-m_\pi^2)/2m_{\eta\pi}, \\
    E_2 &=(s-s_{\eta\pi}-m_p^2)/2m_{\eta\pi}, \\
    E_\pi &=(m^2_{\eta\pi}+m_\pi^2-m_\eta^2)/2m_{\eta\pi}, \\
    E_\eta& =(m^2_{\eta\pi}+m_\eta^2-m_\pi^2)/2m_{\eta\pi}.
\end{align}
\end{subequations}
$\xi$ and $\epsilon$ are the angles between the beam and the target, and between the beam and the recoil, respectively.
They are given by
\begin{subequations}
\begin{align}
    2 |\mathbf{q}||\mathbf{p}_1| \cos \xi & = s- 2\, E_q\, E_1-m_p^2-m^2_\pi \, , \\
    2 |\mathbf{q}||\mathbf{p}_2| \cos \epsilon & = s- 2\, E_q\, E_2-m_p^2-m^2_{\eta\pi} +t_p \, ,
\end{align}
\end{subequations}
where $t_\eta$ and $s_{\pi p}$ are defined in Eq.~\eqref{eq:invar_typeI}. The energies and the nucleon angles depend only on $s$, $m^2_{\eta\pi}$ and $t_p$. The polar and azimuthal angles of the $\eta$ must then depend on the remaining independent invariants. Indeed we obtain
\begin{subequations}
\begin{align}
    t_\eta = &\, m_\pi^2+m_\eta^2- 2 E_q E_\eta + 2 |\mathbf{q}||\mathbf{k}| \cos\theta \, ,
    \\ \nonumber
    s_{\pi p}  = & \, m_\pi^2+m_p^2+ 2 E_2 E_\pi \\
    & \, - 2 |\mathbf{p}_2||\mathbf{k}| \left( \sin\epsilon \sin\theta \cos \phi  + \cos \epsilon \cos \theta\right)  \, .
\end{align}
\end{subequations}
The invariants $t_\pi$ and $s_{\eta p}$ needed in fast-$\pi$ amplitudes and defined in Eq.~\eqref{eq:invar_typeII} are related to the other Mandelstam variables by
\begin{subequations}
\begin{align}
 s_{\eta p} & = s-s_{\pi p}-m^2_{\eta\pi} + m_\eta^2 + m_\pi^2 + m_p^2, \\
 t_\pi & = t_p- t_\eta- m^2_{\eta\pi}+ m_\eta^2 + 2m_\pi^2.
\end{align}
\end{subequations}

The kinematic function $K$ that appears 
in Eq.~\eqref{eq:generic} is given by
\begin{align} \nonumber
    K & = \varepsilon_{\alpha\beta\gamma\delta} (q+p_2)^\alpha (q-p_2)^\beta (k_\eta+k_\pi)^\gamma (k_\eta-k_\pi)^\delta  \\
    &= 4 m_{\eta \pi}|\mathbf{q}| |\mathbf{k}| | \mathbf{p}_2| \sin \epsilon \sin\theta \sin \phi \, .
\end{align}

\section{MCR and bootstrap}
\label{sec:mcr}
\begin{figure}
    \includegraphics[width=\columnwidth]{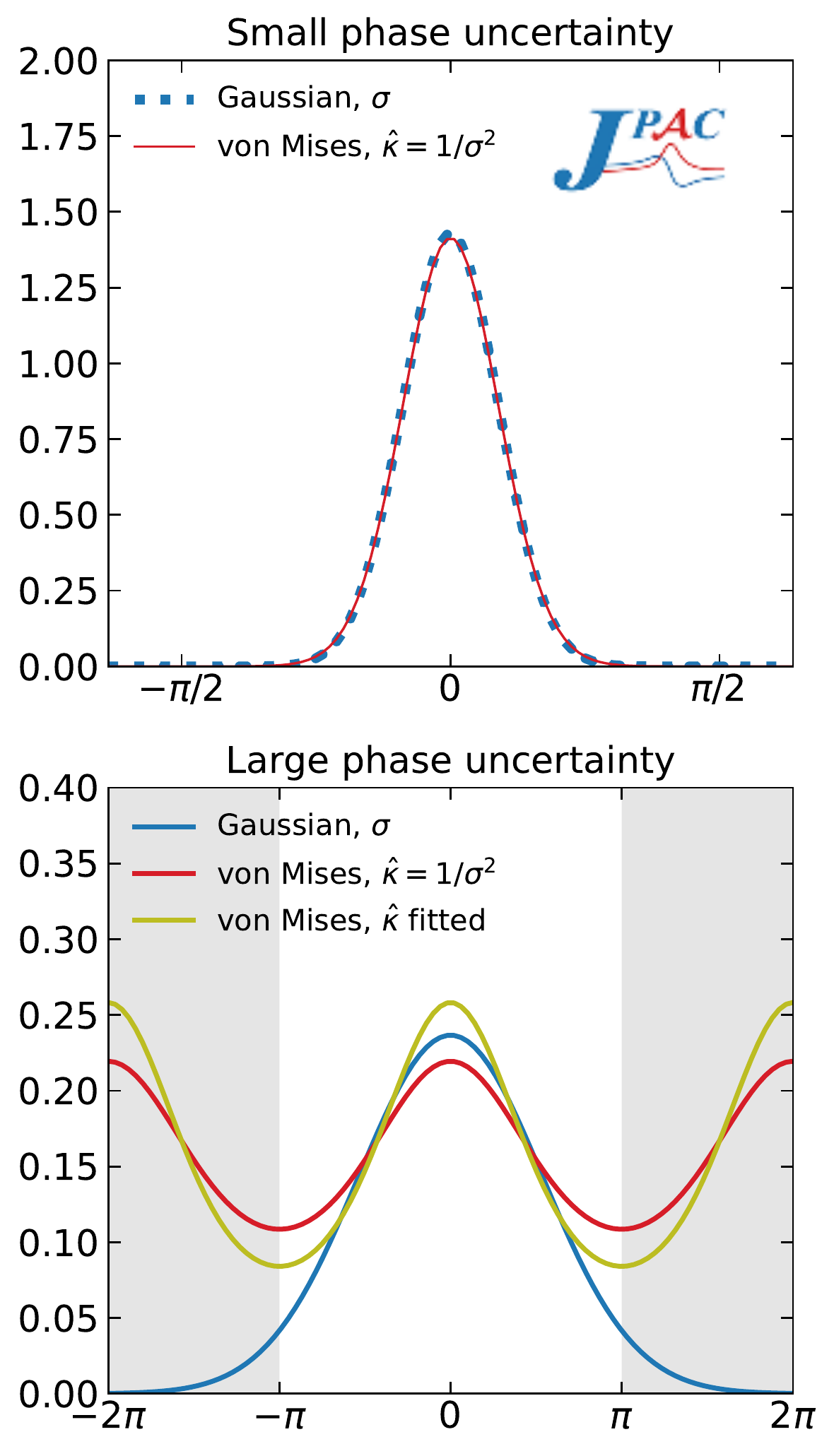} 
      \caption{Comparison between the Gaussian and von Mises distributions for small (top)
      and large (bottom) phase shift uncertainties, centered in $\mu = 0$. 
      In the upper plot the Gaussian distribution has $\sigma=0.28$ (dotted blue), 
      and von Mises has $\hat{\kappa} = 1/\sigma^2 = 12.82$ (solid red).
      In the lower plot the Gaussian has $\sigma=1.69$ (solid blue, one of the largest error in COMPASS datasets), 
      von Mises has $\hat{\kappa} = 1/\sigma^2=0.35$ (solid red),
      and another von Mises has $\hat{\kappa}=0.56$ (solid gold) obtained by
      fitting to the Gaussian distribution.
      The grey bands
      hide the region outside the $[-\pi,\pi]$ range.}
      \label{fig:vonMises}
\end{figure}

We describe how the MCR is performed, and consequently the
bootstrap fit to it. The first step is to associate a probability distribution to each value of intensity and phase shift.
For the intensities, we associate 
to each data point a Gaussian distribution
\begin{align}
f(x|\mu,\sigma) = \frac{1}{\sqrt{2\pi}\, \sigma}
\text{e}^{ -\frac{1}{2}\left( \frac{x - \mu}{\sigma}  \right)^2} \, ,
\end{align}
with $\mu$ equal to the mean value and $\sigma$ the uncertainty reported by COMPASS. For phase shifts, we use the von Mises distribution, {\it i.e.} the periodic equivalent to the Gaussian distribution, 
\begin{align}
f(x|\mu,\hat{\kappa}) = \frac{1}{2\pi I_0(\hat{\kappa})}\text{e}^{\hat{\kappa} \cos \left( x - \mu\right)} \, ,
\end{align}
where $I_0(\hat{\kappa})$ is the modified Bessel function of order $0$. The $\mu$ parameter is set to the mean value of each phase shift. The concentration parameter $\hat{\kappa}$ is the reciprocal measurement of the dispersion. 
If the uncertainty is small, the Gaussian distribution with $\sigma$ equal to the experimental uncertainty is almost equal to the von Mises distribution with $\hat{\kappa} = 1/\sigma^2$,
as shown in Fig.~\ref{fig:vonMises}. For larger values of the uncertainty, the
Gaussian and the von Mises distribution with $\hat{\kappa} = 1/\sigma^2$ are quite different, and we decided to refit the concentration parameter to the Gaussian distribution. We believe this gives a better description of the actual experimental uncertainties.
Nevertheless, we computed the MCR with the three options for the phase shifts distributions 
and we did not find relevant differences among the results.

We are assuming that the uncertainties are statistical only, that data at different energy bins are statistically independent, and that the partial waves intensities and phases are uncorrelated. Although this is clearly not true, the correlation information is not available from the COMPASS analysis. This assumption likely leads to an overestimation of the error bands. There is an additional caveat: some of the uncertainties are large enough to make the intensities negative, which is unphysical. Hence, if a resampling provides a negative intensity for a given $\m$, we set that particular intensity to zero. Alternatively, we consider a negative intensity in Eq.~\eqref{eq:int_COMPASS} instead of setting the value to zero. We found the difference to be negligible in the resulting MCR distributions and fits.

Once the distributions are chosen, we resample them $N=10^{4}$ times to achieve a precision of $\sim 1\%$ in the extracted distributions. 
For each resampled pseudodataset we can compute the value of the intensity $I(\m,\Omega)$, and we can fit it to obtain the corresponding parameters $\{c\}$ (bootstrap fit)~\cite{EfroTibs93,Landay:2016cjw,Perez:2014jsa}. Then, the expected value (mean) of each observable and the associated uncertainty ($16\%$ and $84\%$ quantiles to obtain the $68\%$ error bands) can be computed. 

\section{MVR vs. MCR observables}
\label{sec:mvr_vs_mcr}
\begin{figure}
\begin{tabular}{c}
    \includegraphics[width=\columnwidth]{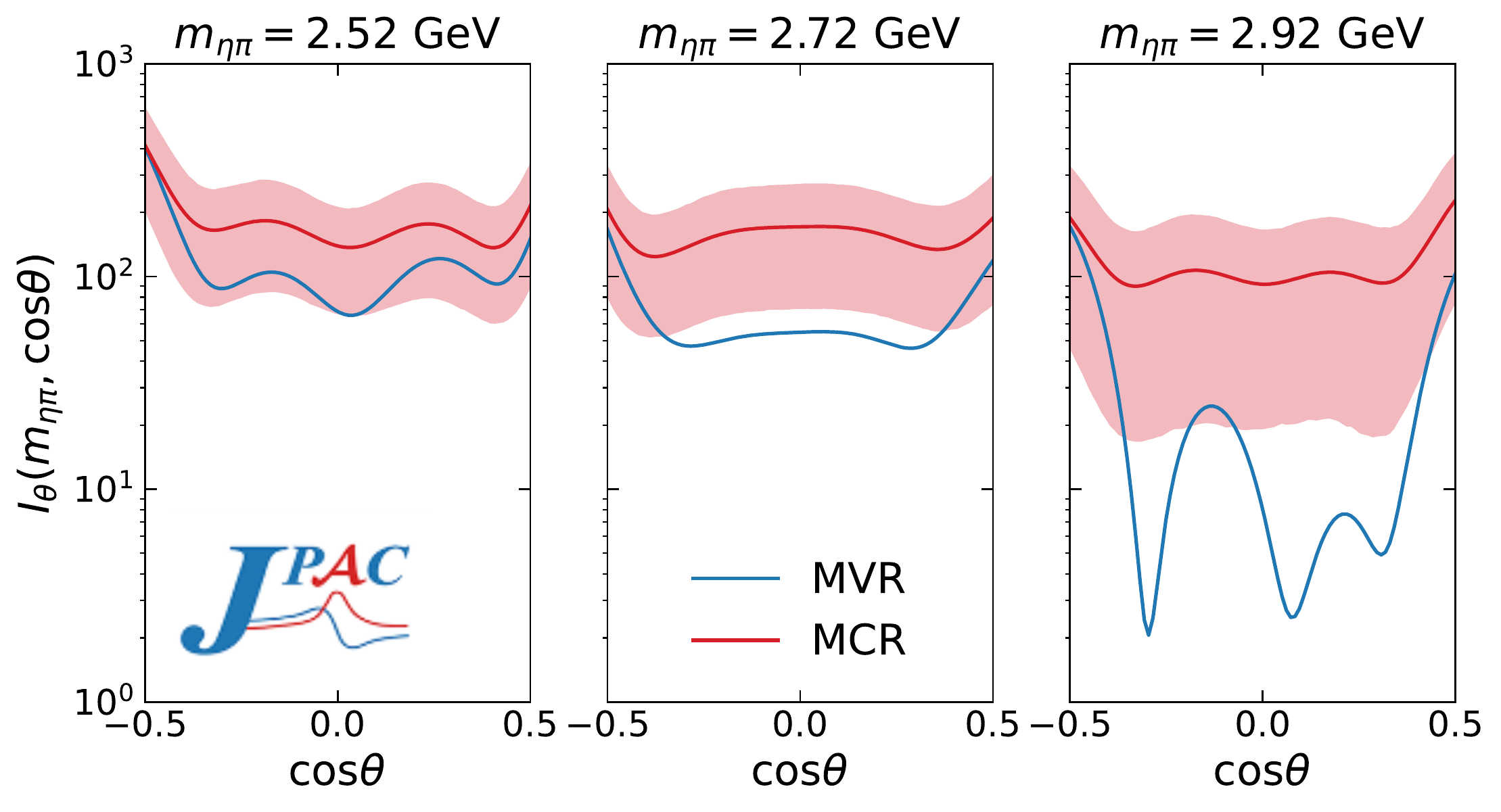} \\
    \includegraphics[width=\columnwidth]{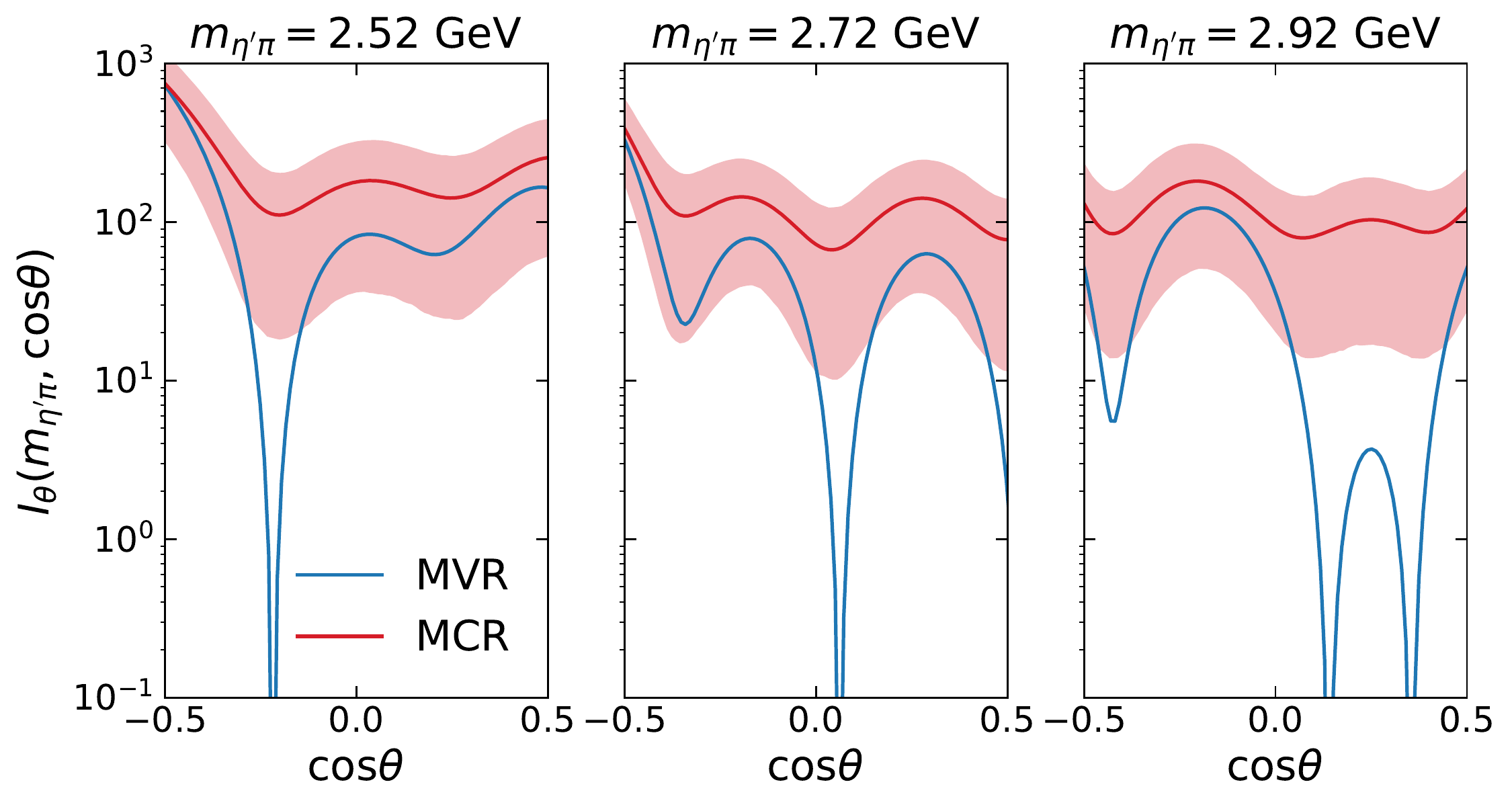}
    \end{tabular}
    \caption{MCR (red) vs. MVR (blue) for the $I_\theta(\m,\cos \theta)$
    observable in the $\eta \pi$ (top) and 
    $\eta' \pi$ (bottom) channels and the small $\cos\theta$ region, $-0.5\leq\cos \theta \leq 0.5$, for three $\m$ energies. 
    }
  \label{fig:pieta.mvrmvr}
\end{figure}

In Section~\ref{sec:COMPASS} we mentioned that
there were meaningful differences between 
the MVR and MCR.
In particular, for  $I_\theta(\m,\cos \theta)$
the forward peak is smaller for the MVR than for the MCR. 
This can be noticed in both
Fig.~\ref{fig:pieta_phi_theta} and~\ref{fig:pietap_phi_theta}.
It can also be noticed in the integrated forward intensity of
Figs.~\ref{fig:pieta.fb.fit} and~\ref{fig:pietap.fb.fit}, where $F(\m)$
is systematically larger for the MCR.
The opposite is true for 
the backward peak and $B(\m)$, which are
systematically larger for the MVR.
Consequently, the forward-backward asymmetry $A(\m)$
is, in absolute value, larger for the MVR than for the MCR, as shown
in Fig.~\ref{fig:asy.fit}.
The total intensity $T(\m)$ is very similar for both MCR and MVR,
as displayed in Fig.~\ref{fig:fb_data}.

The difference between MVR and MCR
is also apparent if we inspect the 
small $\left|\cos\theta\right|$ 
region.
The inclusion of the uncertainties and the calculation
of expected values in the MCR leads to
a smearing that flattens the experimental distribution.
This effect is shown
in Fig.~\ref{fig:pieta.mvrmvr}.
Fits to the MVR would try to match these structures that 
are mostly killed by uncertainties.
However, removing the small $\left|\cos\theta\right|$
region from the fits is not feasible,
given that the total intensity is an important experimental
constraint to ENLL fits.
Consequently, we consider the physics extracted from 
the MCR fits more reliable.

\section{Statistical analyses of the likelihood}
\label{sec:statanalyses}
\begin{figure*}
    \begin{tabular}{cc}
    \includegraphics[width=0.48\textwidth]{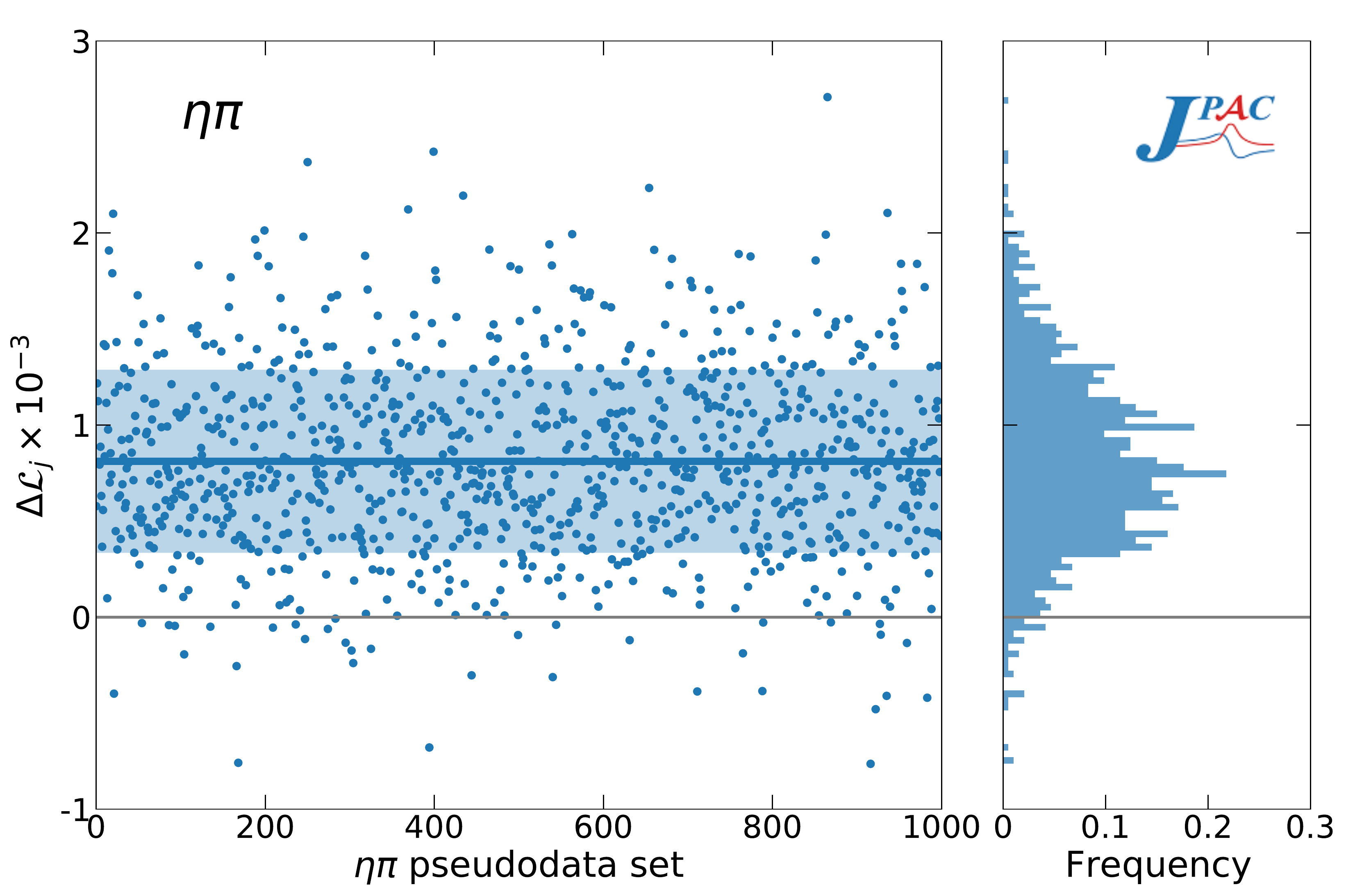} &
    \includegraphics[width=0.48\textwidth]{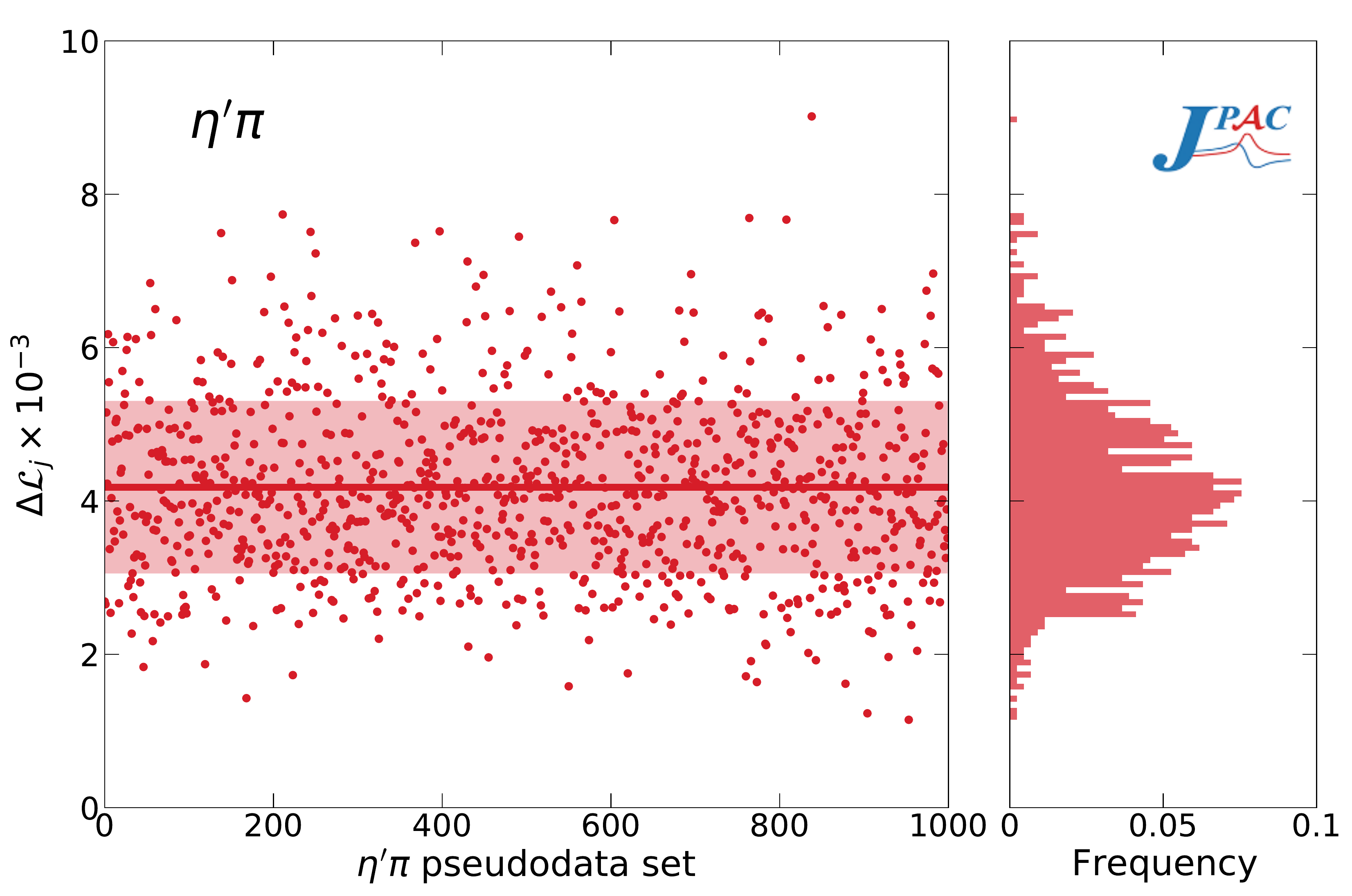}
    \end{tabular}
      \caption{Extended negative log-likelihood difference
      $\Delta {\mathcal L}_j = \mathcal{L}_j(\text{MIN}+f/\pom) - \mathcal{L}_j(\text{MIN}+\pom/\pom)$ for $10^3$ MCR $\eta \pi$ (left, blue) $\eta' \pi$ (right, red) resamples. 
      The colored line is the mean of the distribution and
      the band represents the $68\%$ confidence level. While for $\eta'\pi$ the distribution sits safely above $\Delta {\mathcal L}_j = 0$ (gray line), for $\eta\pi$ this happens only for $96.5\%$ of the pseudodatasets, corresponding to a $2\sigma$ preference. It is worth noticing that the agreement with the naive difference of the mean values of the ENLL in Table~\ref{tab:parameters}.
      }
      \label{fig:ellhdiff}
\end{figure*}

As mentioned in Section~\ref{sec:fitresults},
in ENLL fits the comparison of models with different number of parameters is nontrivial (see for example~\cite{Pilloni:2016obd}).
However, we can compare easily MIN$+f/\pom$ and MIN$+\pom/\pom$.
In Table~\ref{tab:parameters}
we saw that the MIN$+\pom/\pom$ fit
was slightly better than the MIN$+f/\pom$ 
for both MVR and MCR, 
although it did not look significant. 
The quantitative comparison between the two models can be addressed by
checking if, for each pseudodataset,
one of the models is systematically better. 
In doing so, we build $10^3$ resampled datasets
for each channel, we fit each dataset $j$ with both models.
Then we compute the difference between the two ENLL.
Figure~\ref{fig:ellhdiff} shows the result of this exercise.
We find that for $\eta' \pi$, MIN$+\pom/\pom$
 is better
systematically and significantly than MIN$+f/\pom$.
For $\eta \pi$, this happens for $96.5\%$ of the resampled datasets, which only results in a $2\sigma$ preference.
Moreover, this preference is not conclusive
as it stems from Fig.~\ref{fig:pieta.theta.bwd}, where the
addition of either $f_2/\pom$ or $\pom/\pom$ amplitudes 
is favored by different values of $m_{\eta\pi}$. 
Therefore, it is likely that the inclusion of both 
could be necessary.

\section{MVR vs. MCR fits}
\label{sec:mvrfits}
The fit parameters to the MVR for the three models in both channels are presented in the MVR columns of Table~\ref{tab:parameters}. No error is provided as none can be reliably computed.

The differences between the MVR and the MCR fits
can be read from the fitted parameters
in Table~\ref{tab:parameters}.
In the $\eta \pi$ channel,
we notice that $c_{a_2\pom}$, 
is larger for the MCR while the $c_{a_2f_2}$, 
is very similar for both MVR and MCR fits. 
This happens systematically for all of the MIN, MIN$+f/\pom$, and  MIN$+\pom/\pom$ models.
In $\eta' \pi$ find the same behavior
for $c_{a_2\pom}$.
However, $c_{a_2f_2}$ is very different for 
the MVR and the MCR fits.
For the MCR, the three models provide results compatible within uncertainties.
Moreover, the $a_2/f_2$ exchange is compatible with zero within a $2\sigma$ confidence level, signaling a large EXD breaking.
For the MVR fits, the $a_2/f_2$ amplitude is larger than for the MCR and 
yields very similar results for the MIN and MIN$+f/\pom$ models.
The MIN$+\pom/\pom$ $c_{a_2f_2}$ is slightly smaller. This is due to the non-negligible 
interference between
the top-$a_2$ amplitudes and the $\pom/\pom$ contribution in the central region.
However, our double-Regge model is based on leading Regge poles and is reliable at small scattering angles, not in the central region where corrections (cuts and daughters) are expected. Hence, any correlation or interplay among the amplitudes in the central region cannot be trusted.

Hence, MVR and MCR fits provide a very different balance 
between the two fast-$\eta$ amplitudes.
For the fast-$\pi$ ones,
we find that the $c_{f_2f_2}$ parameter is 
larger for the MVR fits than for the MCR ones for both channels, as expected.
The asymmetry is negative for both $\eta^{(\prime)}\pi$ channels, as shown in Fig.~\ref{fig:asy.fit}.
Taking into account the uncertainties with MCR makes the asymmetry less acute, but sizeable.

\section{Parameter distributions and correlations}
\label{sec:mcrdistributions}
\begin{figure*}
    \includegraphics[width=0.98\textwidth]{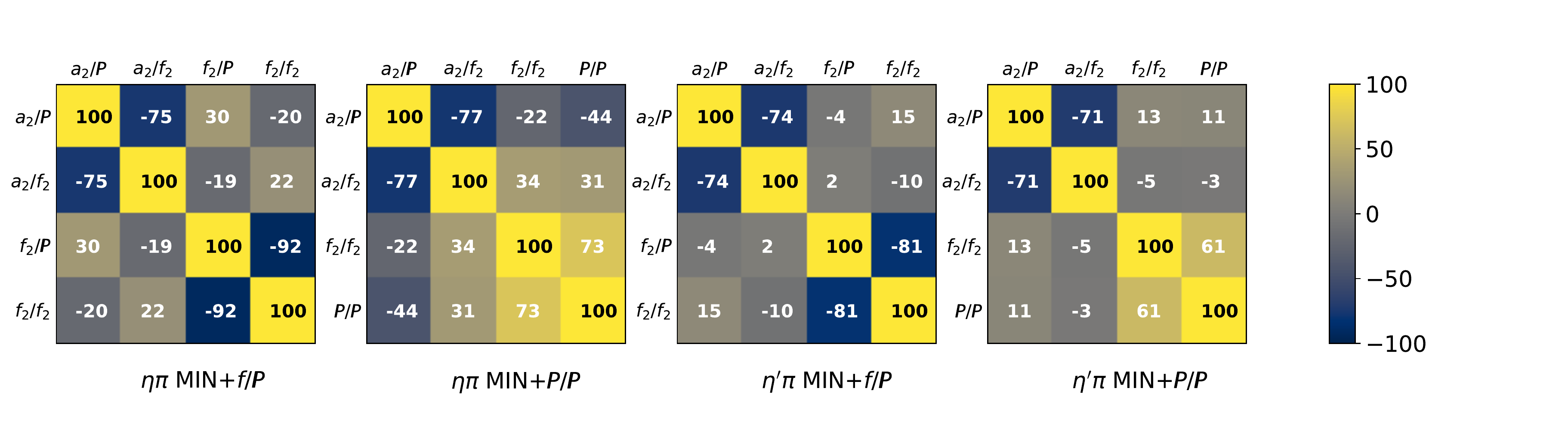}
      \caption{Correlation plots for the fit parameters 
      of the MIN$+f/\pom$ and MIN$+\pom/\pom$ models for
      both $\eta\pi$ and $\eta'\pi$ channels.}
      \label{fig:correlation}
\end{figure*}
\begin{figure*}
    \begin{tabular}{c}
    \includegraphics[width=0.81\textwidth]{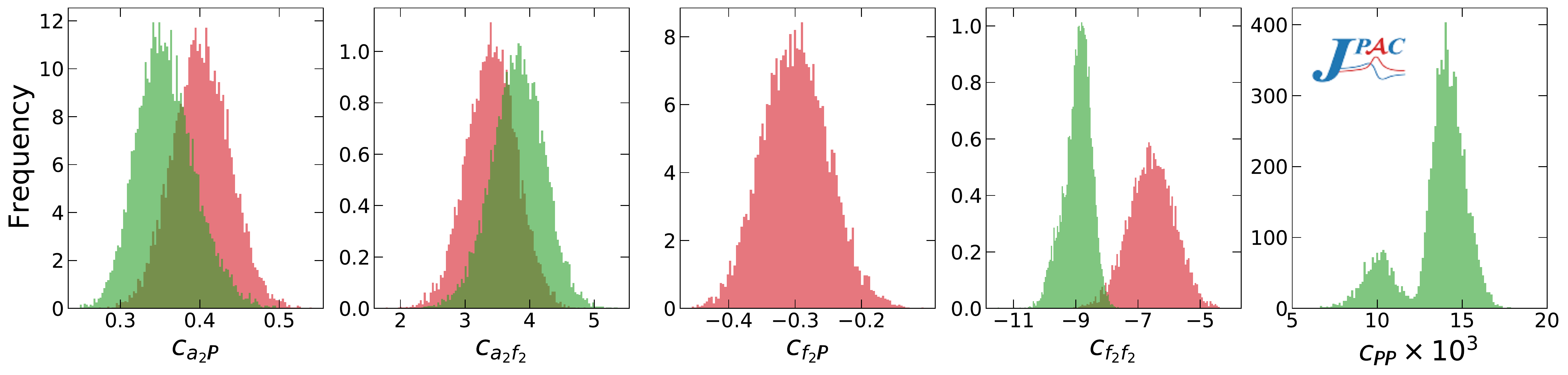} \\
    \includegraphics[width=0.8\textwidth]{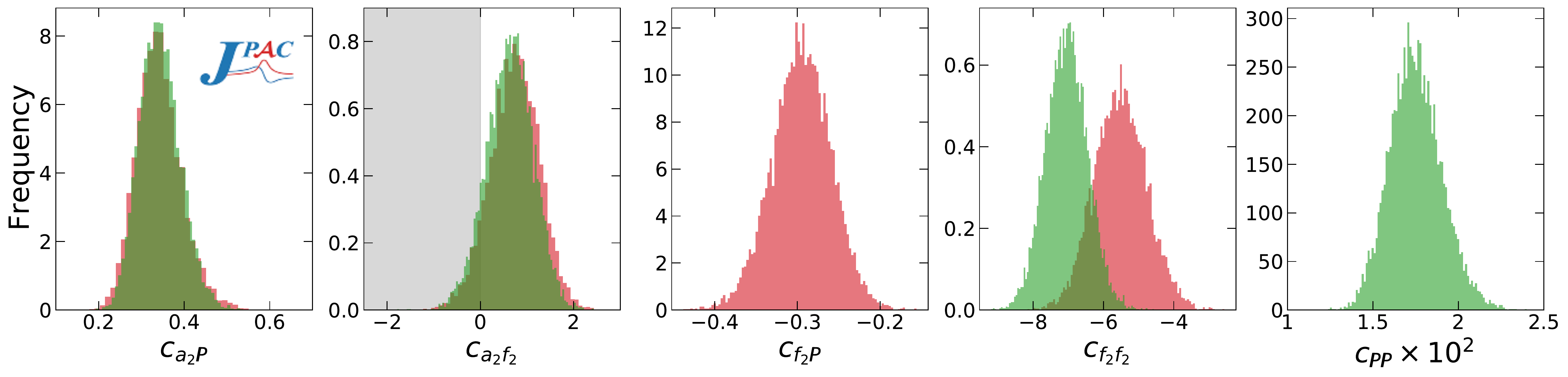}
    \end{tabular}
    \caption{MCR fit parameter distributions for $\eta \pi$ (top row) and $\eta' \pi$ (bottom row) ,
     for the MIN+$f/\pom$ (red) and the MIN+$\pom/\pom$ (green) models.
    The grey area hides the region where $c_{a_2f_2}$ is negative.}
      \label{fig:pieta.histograms}
\end{figure*}

The bootstrap fits to the MCR provide 
the parameter distributions and their 
correlations. We show the correlation matrices
in Fig.~\ref{fig:correlation}
and the fit parameter distributions in
Fig.~\ref{fig:pieta.histograms}
for both MIN$+f/\pom$ and MIN$+\pom/\pom$ fits to both channels.
We do not show the results for the MIN fits as they do not give
an appropriate description of the $I(\m,\Omega)$ distributions, see Sections~\ref{sec:etapimcr} 
and~\ref{sec:etaprimepimcr}.

From the correlation matrices the substantial independence of fast-$\pi$ and fast-$\eta$
amplitudes is
apparent for both models and channels. 

For the $\eta \pi$ channel,
the parameters for the MIN$+f/\pom$ model
are well determined and exhibit a Gaussian behavior.
However, this does not happen for the MIN$+\pom/\pom$ model,
where the $\pom/\pom$ amplitude parameter is not well determined
and presents a bimodal distribution.
The fit to the MVR using the MIN$+\pom/\pom$ model presents
a single and isolated absolute minimum, so the appearance
of a two peak structure in the fit to the MCR is entirely due
to the inclusion of the uncertainties. The $c_{\pom\pom}$ parameter cannot be well determined.
The $c_{a_2\pom}$ and $c_{a_2f_2}$ distributions for mostly overlap in both models. 
For $c_{f_2f_2}$, the
distributions barely overlap,
as a consequence of the differences between the 
$f_2/\pom$ and $\pom/\pom$ amplitudes.

For the $\eta' \pi$ channel,
all the parameters for both models
are well determined and exhibit a nice Gaussian behavior.
The $c_{a_2f_2}$ distribution is compatible with zero at a $2\sigma$
level for both models, indicating that it is possible
that the associated amplitude vanishes.
This would mean a large violation of the EXD between the
$a_2$ and $f_2$ Regge poles.
Given that both the statistical analysis
in Appendix~\ref{sec:statanalyses}
and the comparison to the data in Section~\ref{sec:etaprimepimcr}
favor the MIN$+\pom/\pom$,
we find that the contribution of all four amplitudes
($a_2/\pom$, $a_2/f_2$, $f_2/f_2$, and $\pom/\pom$)
to the $\eta'\pi$ process
can be well established with reasonable uncertainties.

\bibliographystyle{apsrev4-2.bst}
\bibliography{quattro}
\end{document}